\documentclass{article}
\usepackage[left=1.2in,right=1.2in,top=2in,bottom=2in]{geometry}

\usepackage{amsthm,amsmath}
\usepackage{amssymb}
\usepackage{graphicx}
\usepackage{enumitem}
\usepackage{comment}
\usepackage{array}
\usepackage{subcaption}
\usepackage{multirow}
\usepackage{tabularx}
\usepackage{arydshln}
\usepackage{booktabs}
\usepackage{authblk}
\usepackage{adjustbox}
\usepackage{setspace}
\usepackage{xcolor}

\DeclareMathOperator*{\argmax}{arg\,max}
\newcommand{\cN}{\mathcal{N}}
\newcommand{\e}{\varepsilon}
\newcommand{\h}{\hat}

\newcommand{\vb}{{\boldsymbol b }}
\newcommand{\norm}[1]{\left\lVert#1\right\rVert}
\newcommand{\ignore}[1]{}

\newcommand{\vt}{\boldsymbol{t}}
\newcommand{\vg}{\boldsymbol{g}}

\newcommand{\vX}{\boldsymbol{X}}
\newcommand{\vY}{\boldsymbol{Y}}

\newcommand{\vbeta}{\boldsymbol{\beta}}
\newcommand{\vu}{\boldsymbol{u}}

\newcommand{\vSigma}{\boldsymbol{\Sigma}}
\newcommand{\vXf}{\boldsymbol{X}^\text{f}}
\newcommand{\vXr}{\boldsymbol{X}^\text{r}}
\newcommand{\vXs}{\boldsymbol{X}^\text{s}}
\newcommand{\vXg}{\boldsymbol{X}^\text{g}}
\newcommand{\TS}{\texttt{TS}}

\newcommand{\Proj}{\text{Proj}}
\newcommand{\toinv}[1]{{#1}|_{\text{1st}}}
\providecommand{\keywords}[1]{\textit{Keywords: #1}}
\RequirePackage[colorlinks,citecolor=blue,urlcolor=blue]{hyperref}

\newcommand{\red}[1]{#1}

\title{Joint Latent Class Trees:
A Tree-Based Approach to Modeling Time-to-event and Longitudinal Data}

\date{}
\author{Ningshan Zhang}
\author{Jeffrey S. Simonoff}
\affil{Technology, Operations, and Statistics Department, Leonard N.
Stern School of Business, New York University}

\begin{document}
\maketitle
{\abstract{
        \red{
In this paper, we propose a semiparametric, tree based joint latent class modeling 
approach (JLCT) to model the joint behavior of longitudinal and time-to-event data.
Existing joint latent class modeling approaches are parametric and can suffer from high
computational cost. The most common parametric approach,
the joint latent class model (JLCM), further restricts analysis to using
time-invariant covariates in modeling survival risks and latent class memberships.
Instead, the proposed JLCT is fast to fit, and can use time-varying covariates in
all of its modeling components.  We demonstrate the prognostic value of 
using time-varying covariates, and therefore the advantage of JLCT over JLCM 
on simulated data.  We further apply JLCT to the PAQUID data set and 
confirm its superior prediction performance and orders-of-magnitude speedup
over JLCM.}
}}

\bigbreak
\noindent \keywords{Biomarker; Conditional independence; Recursive partitioning;
Survival data; Time-varying covariates.}

%--------------------
\section{Introduction}
\label{sec:intro}

Clinical studies often collect three types of data on each patient:
the time to the event of interest (possibly censored),
the longitudinal measurements on a continuous response
(for example, some sort of biomarker viewed as clinically important),
and an additional set of covariates (possibly time-varying) about the patient.
These studies then focus on analyzing the relationship between the
time-to-event and the longitudinal responses, using the additional covariates.
\ignore{A common approach is to jointly model the time-to-event by a survival model,
while modeling the longitudinal responses using a linear mixed-effects model,
with both the survival and the linear mixed-effects models potentially
making use of the additional covariates.}
In particular, clinical studies are often interested in the idea of uncovering 
latent classes among the population of patients,
which can be used to describe disease progression in clinical
studies \cite{lin2002latent,garre2008joint,proust2009joint}.
The idea of latent classes is partially motivated by the 
fact that many diseases have different stages; 
examples include dementia \cite{perneczky2006mapping},
AIDS \cite{longini1989statistical}, cancer \cite{dicker2006principles}, 
and chronic obstructive pulmonary disease (COPD) \cite{antonelli2003gold}.
Once those stages are identified, one can design personalized medicine
that, for a particular patient, changes with membership in different stages \cite{hajiro2000stages}.
\red{From this point of view, incorporating the time-varying nature of latent 
class membership into any modeling strategy is crucial, and any approach 
that does not allow for time-varying membership of latent classes cannot 
possibly be informative in the implementation of what is necessarily dynamic 
personalized medicine treatment.}
The desire to find latent classes in order to design personalized treatment
also extends to other contexts as well.

Currently the clinical definitions of stages of a disease consist
of using diagnostic findings (such as biomarkers) to produce clusters of patients.
In this paper, we focus on an alternative approach to find data-dependent 
and possibly time-varying latent classes, which uncovers \red{potentially} 
meaningful stages by modeling the biomarker trajectories and
survival experiences together. This is known as the joint latent class problem.

The joint latent class model (JLCM) \cite{lin2002latent,proust2009development,proust2009joint,proust2017estimation} 
is the most common approach for constructing latent classes that 
account for the joint behavior of time-to-event and longitudinal data.
JLCM assumes that the population of patients consists 
of multiple latent classes, and models the latent class
memberships by a multinomial logistic regression model.
\red{The key assumption of JLCM is that}, conditioning on her latent class membership,
a patient's time-to-event and longitudinal responses are independent.
JLCM further assumes that the
latent classes are homogeneous and thus patients within a latent class
follow the same survival and longitudinal model. 
\red{The assumed latent classes could potentially represent clinically
meaningful subpopulations, but such a connection requires further clinical research.}

JLCM is a parametric approach, 
which estimates the parameters for all modeling components via maximizing
the log-likelihood function. However, this can be prohibitively slow for 
large scale data. In addition, software implementations of JLCM \cite{proust2017estimation} 
cannot use time-varying covariates in its latent class membership and survival models, 
which can greatly reduce its prediction performance.
In Appendix~\ref{app:jlcm}, we give a brief introduction to JLCM,
and discuss at length its strengths and weaknesses;
see also \cite{zhangsimonoff}.

A nonparametric approach that addresses some of these issues would be desirable,
and tree-based approaches \cite{breiman1984classification,friedman2001elements} 
are natural candidates.
Tree-based methods are powerful modeling tools in statistics
and data mining,
especially because they are fast to construct,
able to uncover nonlinear relationships between covariates,
and intuitive and easy to explain.
In this work, we propose the joint latent class tree (JLCT) method for the 
joint latent class problem,
which consists of two steps. The first and primary
step of JLCT is to provide a tree-based
partitioning that uncovers meaningful latent classes of the population.
JLCT, like JLCM, is based on the key assumption that conditioning on
latent class membership, time-to-event and longitudinal responses are independent.
JLCT therefore looks for a tree-based partitioning such that
within each estimated latent class defined by a terminal node,
the time-to-event and longitudinal responses display a lack of association.
In the second step, we assign each observation to a latent class
(i.e.\ terminal node of the constructed tree), 
and independently fit any type of survival and
longitudinal models, using the class membership information.
\red{Since these time-to-event and longitudinal models can be fully parametric 
if desired, JLCT can be viewed as a semiparametric approach to the problem.}

JLCT has two major advantages over JLCM.  First,
JLCT is computationally more favorable, for two reasons: (a) it is very efficient 
to fit a tree to uncover complex relationships between covariates, 
and (b) JLCT separates fitting the survival and 
longitudinal outcomes from uncovering the latent classes,
whereas JLCM fits longitudinal, survival, and latent classes altogether.
The second advantage lies in flexible modeling: JLCT addresses the time-invariant
limitation of JLCM by using time-varying covariates 
as the splitting variables to construct the tree. 
Furthermore, once a tree is constructed,
it is up to the user to decide which type of survival models and which covariates
to use within each terminal node, depending on the analyst's modeling choices and preferences.
Both advantages are verified by our experimental results:
JLCT is orders of magnitude faster, and it achieves better 
prediction performances than JLCM by adopting more flexible models.

The rest of the paper is organized as follows. In
Section~\ref{sec:method} we introduce the setup of the 
modeling problem, and describe our joint latent class tree (JLCT)
method. In Section~\ref{sec:sim_time_var} we use simulations to compare JLCT with
JLCM in terms of prediction performance and running time under
various latent class scenarios. 
Finally,
in Section~\ref{sec:application} we apply JLCT to a real data set,
and demonstrate that JLCT admits competitive (or superior)
prediction performance, while being potentially orders of magnitude
faster than both the JLCM and a popular alternative approach to 
joint modeling, the shared random effects model (SREM).

%--------------------
\section{Constructing a tree to uncover conditional independence}
\label{sec:method}

\subsection{Modeling setup}
\label{subsec:setup}

Assume there are $N$ subjects in the sample. For each subject $i$,
we observe $n_i$ repeated measurements of a longitudinal outcome at times
$\vt_i = ( t_{i1}, \dots, t_{in_i})^\prime$.
We denote the vector of longitudinal outcomes by
$\vY_i = (y_{i t_{i1}}, \dots, y_{i t_{in_i}})^\prime$.
In addition, for each subject $i$, we observe a vector of covariates
$\vX_{it}$ at each measurement time $t\in \vt_i$.
These covariates can be either time-invariant or time-varying.
Each subject is also associated with a time-to-event tuple $(T_i, \delta_i)$,
where $T_i$ is the time of the event, and $\delta_i$ is the censor indicator with
$\delta_i=0$ if subject $i$ is censored at $T_i$, and $\delta_i=1$ otherwise.

We assume there exist $G$ latent classes, and let $g_{it} \in \{1,\dots,G\}$
denote the latent class membership of subject $i$ at time $t$.
Let $\vg_i=(\vg_{it_{i1}}, \cdots, \vg_{it_{in_i}})$ denote the vector
of latent class memberships of subject $i$ at each measurement time $t\in\vt_i$.
We assume the latent class membership $g_{it}$ is determined by a subset 
of covariates $\vX_{it}$, denoted by $\vXg_{it}$. 
\red{The latent classes are assumed to be homogeneous, in the sense that 
the joint distribution of the longitudinal and time-to-event variables is 
the same for all observations within a latent class.}

The joint latent class modeling problem proposed here makes the same key assumption as JLCM:
a patient's time-to-event $(T_i,\delta_i)$ and longitudinal outcomes $(\vY_i)$
are independent conditioning on her latent class membership $(\vg_{i})$.
Unlike JLCM where the membership of a subject does not vary over time, 
JLCT allows the elements in $\vg_i$ to change.
Thus we need a slightly more involved definition of ``conditional independence.'' 
For subject $i$ and the time interval
between any two consecutive observations, $I_{ik} = [t_{ik}, t_{i(k+1)})$,
conditioning on the event that subject $i$'s membership equals 
$g\in G$ on this interval $\big(\{g_{it} =g \mid t\in I_{ik}\}\big)$, 
the longitudinal process on the time interval
$(y_{it} \mid t\in I_{ik})$ and the 
time-to-event outcome $(T_i, \delta_i)$ are independent. More precisely,
for every subject $i$, every corresponding time interval $I_{ik}$,
and every $g\in G$,
\begin{align*}
    (T_i, \delta_i) \perp (y_{it} \mid t \in I_{ik}) 
    \text{, conditioning on the event } \{g_{it} = g \mid t \in I_{ik}\}.
\end{align*}
\red{The motivation for the conditional independence assumption in joint latent 
class modeling comes from the hypothesis that observed associations between the 
longitudinal and time-to-event variables over the entire population are spurious, 
being driven by heterogeneity in levels of the variables between latent classes.}
Without controlling the latent class membership $\vg_{i}$, time-to-event and longitudinal
outcomes may appear to be correlated because each is related to the latent class,
but given $\vg_{i}$ the two are independent of each other, and therefore the longitudinal
outcomes have no prognostic value for time-to-event given the latent class.

Construction of the tree is based on working assumptions for the longitudinal 
and time-to-event processes (recall, however, that once tree-based latent 
classes are determined, the analyst is free to fit any models they wish to the 
observations in each estimated class).
Conditioning on latent class membership, the working assumption is that the time-to-event tuple
$(T_i,\delta_i)$ depends on a subset of covariates $\vXs_{it}$ observed at time $t\in \vt_i$,
through the extended Cox model for time-varying covariates \cite{cox1972regression}:
\begin{equation}\label{jlct:survival}
    h_i(t \vert g_{it}=g) = h_{0g}(t) e^{\vXs_{it} \vb_g},
\end{equation}
where $\vb_g$ is the vector of class-specific slope coefficients,
and $h_{0g}$ the class-specific baseline hazard function.
In Section~\ref{sec:jlct}, we show how to use this extended Cox 
model~\eqref{jlct:survival} to design the tree splitting criterion.
In order to fit the extended Cox model~\eqref{jlct:survival} using the longitudinal
outcome variable as a predictor, and perhaps other time-varying covariates,
we need to convert the original data into
left-truncated right-censored (LTRC) data, which is often referred to 
as a  ``counting process approach'' \cite{andersen1982cox,SurvivalTreeLTRC}.
The conversion is described in detail in Appendix~\ref{app:ltrc}.

The working assumption for the longitudinal outcomes is that they 
come from a linear mixed-effects model \cite{laird1982random}:
\begin{align}\label{eq:y}
    y_{it} \vert_{g_{it}=g} = \vXf_{it} \vbeta + \vXr_{it} \vu_{g}
    + v_i + \varepsilon_{it},\quad
    \vu_g \sim \mathcal{N}(0, \vSigma_r), \quad
    v_i \sim \mathcal{N}(0, \sigma_1^2 ), \quad
    \varepsilon_{it} \sim \mathcal{N}(0, \sigma_2^2 ) .
\end{align}
where $\vXf_{it}$ and $\vXr_{it}$ are the subsets of covariates associated with 
a fixed effect vector $\vbeta$ and a latent
class-specific random effect vector $\vu_{g}$, respectively.  
The random effect vector
$\vu_g$ is independent across latent classes $g=1,\dots,G$,
the subject-specific random intercept $v_i$ is
independent across subjects and independent of the random
effects $\vu_g$, and the error $\varepsilon_{it}$ is
independent and normally distributed with mean $0$ 
and variance $\sigma_2^2$, and independent of all of the random 
effects. 
\red{
To reduce model complexity, we only use a random intercept for each subject $i$ here,
but in principle we can include any random effects on the subject level.
Model selection for predicting longitudinal outcomes does not affect the construction 
of JLCT tree and is beyond the scope of this paper.}

We have introduced four subsets of covariates so far: $\vXg_{it}$ for the latent classes,
$\vXf_{it}$ for the fixed effects, $\vXr_{it}$ for the random effects,
and $\vXs_{it}$ for the time-to-event.
Each of the four subsets can contain time-varying covariates,
and the four subsets can be either identical, or share common covariates,
or share no covariates at all.

\subsection{Joint Latent Class Tree (JLCT) methodology} \label{sec:jlct}
In this section we formally introduce JLCT, the joint latent class tree approach.
As discussed in the introduction, JLCT consists of two steps:
the first and primary step is to construct a tree-based partitioning
of the population; in the second step, JLCT independently fits survival and 
longitudinal models, using the proposed latent class memberships from the first step.

\paragraph{Step 1: Construct a tree-based partitioning.} 
\red{Since the joint latent class model assumes conditional independence 
within each latent class},
JLCT looks for a tree-based partitioning such that within each estimated class 
defined by a terminal node, the time-to-event and longitudinal outcomes display a lack of association.
JLCT recursively splits a node into 
two children nodes according to a splitting criterion
until some stopping criteria are met.
The splitting criterion ensures that
the two children nodes are more ``homogeneous'' than their parent node,
where the notion of ``homogeneous'' is measured by 
how apparently independent the two variables are
conditioning on the node.

\red{Specifically}, the splitting criterion repeatedly uses the test statistic
for the hypothesis test
\begin{align}\label{eq:h0h1}
    H_0 : b_y = 0, \quad \text{vs.} \quad H_1 : b_y \ne 0,
\end{align}
under the extended Cox model
\begin{equation}\label{eq:cox_split_cri}
    h(t,\vXs_i,\vY_i) = h_0(t) e^{y_{it}b_y  + \vXs_{it}\vb_x },
\end{equation}
which is fitted using all the data within the node to split.
Thus, the null hypothesis ($b_y=0$)
corresponds to the longitudinal outcome having no relationship with
the time-to-event in the node. 
We use the log-likelihood ratio test (LRT) for the hypothesis test~\eqref{eq:h0h1}
in all experiments; simulations indicate that the Wald test gives similar results.
Note that the time-to-event formulation~\eqref{eq:cox_split_cri} is {\bf only} 
being used as a splitting criterion, {\bf not} as a representation of the true relationship 
between $\vY_i$ and $(T_i, \delta_i)$.

We will denote the test statistic of the hypothesis test~\eqref{eq:h0h1} as \TS\@.
The smaller the value of \TS\ is, the less related longitudinal outcomes
$\vY_{i}$ are to the time-to-event data $(L_{it},R_{it}, \delta_{it})$
within the current node.
JLCT seeks to partition observations using covariates in $\vXg$,
such that \TS\ is small within each leaf node,
until \TS\ is less than a specified stopping parameter $s$.
More formally, the tree splitting procedure works as follows:
\begin{enumerate}[label*=\arabic*.]
    \item At the current node, compute the test statistic \TS$_{\text{parent}}$.
    \item If \TS$_{\text{parent}} < s$, stop splitting. Otherwise proceed.
    \item For every split variable $X_j\in\vXg$, and possible split point $C$,
        \begin{enumerate}[label*=\arabic*.]
            \item Define two children nodes
                \begin{align*}
                    R_{\text{left}}(j,C)  =\{(i,t): x_{itj} \leq C\}, \quad
                    R_{\text{right}}(j,C)  =\{ (i,t): x_{itj} > C\}.
                \end{align*}
            %\item Ignore this split if the number of events in either node
            %    is less than a threshold. Otherwise proceed.
            \item Ignore this split if either node violates the restrictions
                specified in the control parameters.
                (Details are given near the end of this section.)
                Otherwise proceed.

            \item At each child node,  determine the test statistics
                \TS$_{\text{left}}(j,C)$, \TS$_{\text{right}}(j,C)$ respectively.
            \item Compute the score
                $S(j,C)=\TS_{\text{parent}}-
                \TS_{\text{left}}(j,C) -\TS_{\text{right}}(j,C)$.
        \end{enumerate}
    \item Scan through all pairs of $(j,C)$ to find
        $(j^*, C^*)=\argmax_{j,C} S(j,C)$,
        and split the current node on variable $j^*$ at split point $C^*$.
\end{enumerate}
JLCT recursively splits nodes according to the above procedure, until
none of the terminal nodes can be further split.
Under the null model $H_0$ in~\eqref{eq:h0h1},
the distribution of \TS\ is approximately a $\chi^2_1$ distribution.
By default, we adopt the stopping criterion $\TS_{\text{parent}} < 3.84$,
which corresponds to the 5\% tail of the $\chi^2_1$ distribution.
\red{We have also explored alternative stopping criteria $\TS_{\text{parent}} < 2.71$ and 
$\TS_{\text{parent}} < 6.63$, which corresponds to the 10\% and 1\% tail of 
the $\chi^2_1$ distribution, respectively, and find that these three stopping 
criteria have very similar performance in both simulations (Section~\ref{subsec:morestops}) and the
PAQUID dataset (Section~\ref{subsec:paquidresults}).
Note that this hypothesis test-based stopping criterion is similar in 
spirit (although different in purpose) to the approach used in conditional 
inference trees \cite{hothorn2006unbiased}, as opposed to the greedy overfit-and-prune-back 
strategy of CART-type trees.}

Standard control parameters for trees, such as the minimal number of observations
in any terminal node, the minimal improvement in the ``purity'' by a split, etc.,
also apply to the JLCT method. In addition, we introduce two control parameters
that are specific to JLCT:
\begin{itemize}
\item Minimum number of events in any terminal node.
    This parameter ensures that the survival model in each terminal node
    is fit on meaningful survival data with enough occurrences of events
    (that is, uncensored observations).
    By default, this parameter is set to the number of covariates used for
    the Cox PH model.
\item Upper bound on the variance of the estimated coefficients in all survival
    models at tree nodes. This parameter ensures that the fitted coefficients
    of the survival models are numerically stable. By default this parameter 
    is set to $10^5$. 
\end{itemize}
These two new control parameters enforce reliability of the fitted survival models
at all nodes, which further produces reliable test statistics to
reflect the relationships between the time-to-event and the longitudinal outcomes.

\paragraph{Step 2: Fit survival and longitudinal models.}
Once the tree is constructed, we view each terminal
node as an estimated latent class, and obtain the ``predicted''
latent class membership $\hat g_{it}$
for each subject $i$ at time $t$. We then fit the following 
two models separately:
\begin{itemize}
\item Fit a survival model to time-to-event data $(T_i, \delta_i)$, 
    using input covariates $\vXs_{it}$ and the predicted latent class membership $\h g_{it}$.
\item Fit a longitudinal model to longitudinal outcomes $y_{it}$, 
    using input covariates $\vXf_{it}$, $\vXr_{it}$, 
    and the predicted latent class membership $\h g_{it}$.
\end{itemize}
By default, we use the extended Cox model~\eqref{jlct:survival} for the 
survival data, and the linear mixed-effects model~\eqref{eq:y} for the 
longitudinal data,  but the user is free to make any 
modeling choices they wish.

\section{Simulation results}\label{sec:sim_time_var}
In this section, we use simulations to study the behavior of JLCT,
and compare JLCT with JLCM\@. 
We mainly focus on the case where all of the covariates are time-varying,
and provide more simulation results for the time-invariant case 
in Appendix~\ref{app:sim_time_inv}.
\red{
We demonstrate via these simulations that JLCT is a competitive alternative 
to JLCM for the following reasons:
\begin{enumerate}
    \item It provides more accurate predictions. JLCT significantly outperforms JLCM in terms 
        of survival and longitudinal prediction accuracies. We further
        show that the advantage comes primarily from using time-varying predictors.
    \item It is more effective at uncovering true latent classes. 
        In the simulations JLCT is highly effective at identifying the true 
        latent classes when the latent structure is consistent with the form of a tree. 
        Further, under the null scenario where there are no
        latent classes, JLCT almost always makes the correct decision of making 
        no splits. 
    \item It is orders of magnitude faster. 
        JLCT provides effective prediction and latent class identification while also being
        orders of magnitude faster to fit, due to its 
        use of a fast tree algorithm to identify latent classes.
\end{enumerate}
}

\subsection{Data} 
We simulate five randomly distributed, time-varying covariates $X_1,\cdots, X_5$, 
and use them to generate latent class memberships, time-to-event, and longitudinal 
outcomes. The data are generated such that the time-to-event is correlated with the
longitudinal outcomes, but the two are independent conditioning on the latent classes.
Thus, the key assumption of conditional independence holds on the simulated data.
We consider various data generating schemes along two directions: 
\begin{enumerate}
\item The structure of latent classes. 
    We consider four ways of generating latent classes (as functions of covariates 
    $X_1$ and $X_2$): tree partition, linear separation,  non-linear separation, 
    and asymmetric tree partition, which are shown in Figure~\ref{fig:struct}.
    Class membership for a subject at any time $t$ is randomly drawn with probability $p_0$ being
    the class determined by $X_{1t}$ and $X_{2t}$, and $1-p_0$ being any of the remaining three classes.
    When $p_0=1$, the latent class membership corresponds to a deterministic partitioning based on $X_{1t},X_{2t}$;
    on the other hand, when $p_0=0.25$, the latent class membership is 
    randomly determined and independent of $X_{1t},X_{2t}$. 
    \red{In addition, we include the null scenario where there are no latent classes. 
        The value of $p_0$ does not matter here, and we set $p_0$ to 1 only as a placeholder. }

\item The distribution of time-to-event data. 
    We consider three distributions for baseline hazards of time-to-event data:
    exponential, Weibull with decreasing hazards with time (Weibull-D),
    and Weibull with increasing hazards with time (Weibull-I). We also consider 
    various censoring rates: no censoring, light censoring, and heavy censoring,
    where approximately 0\%, 20\%, and 50\% observations are censored, respectively.
\end{enumerate}
We give a full description of the simulation setup
in Appendix~\ref{app:sim_time_var}, as well as a simpler
setup with time-invariant covariates in Appendix~\ref{app:sim_time_inv}.

{
% -------- Structure graph -----------
\begin{figure}[t]
    \centering
    \begin{subfigure}[b]{0.24\textwidth}
        \centering
        \includegraphics[width=\textwidth]{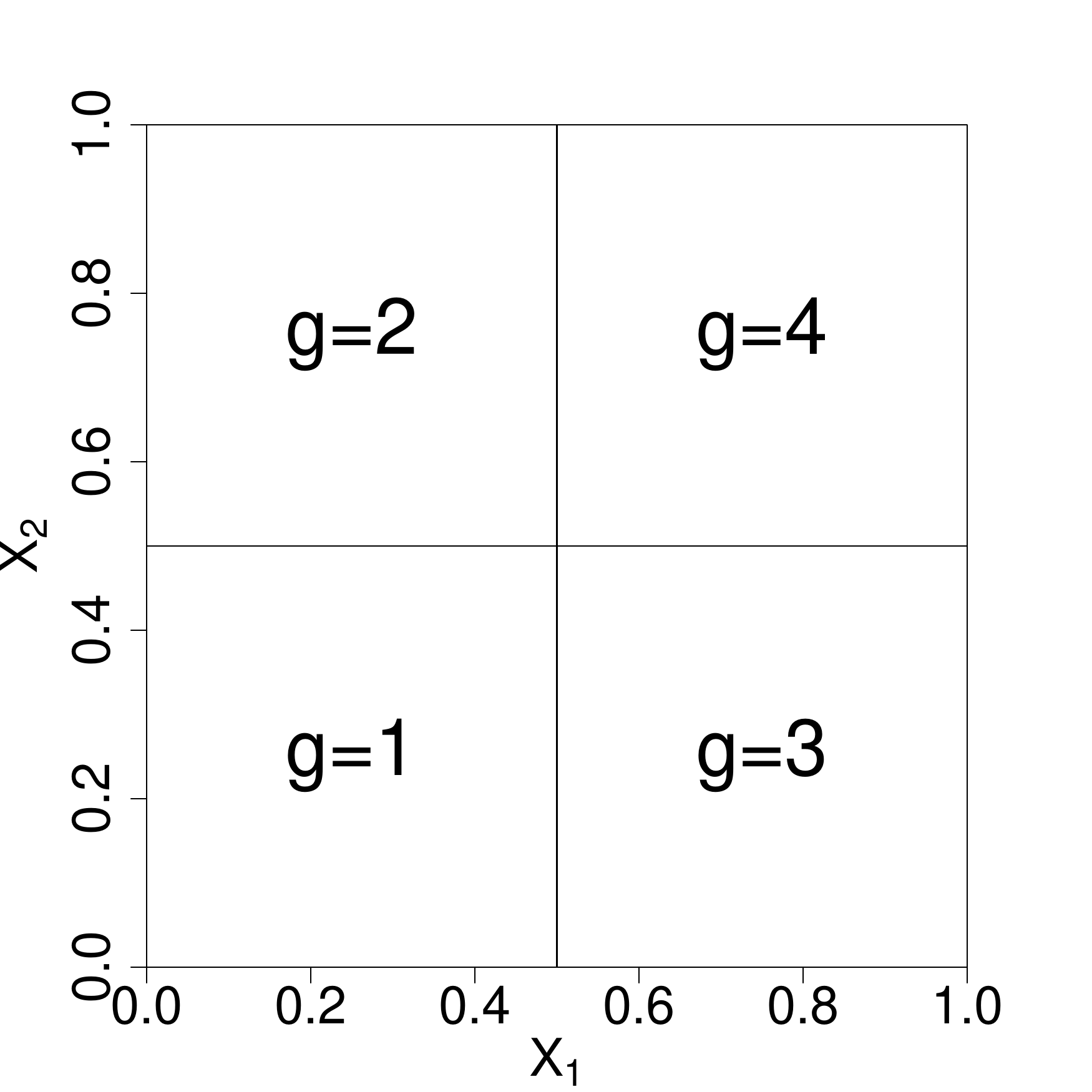}
        \caption[]%
        {{\small Tree}}
        \label{fig:tree}
    \end{subfigure}
    \begin{subfigure}[b]{0.24\textwidth}
        \centering
        \includegraphics[width=\textwidth]{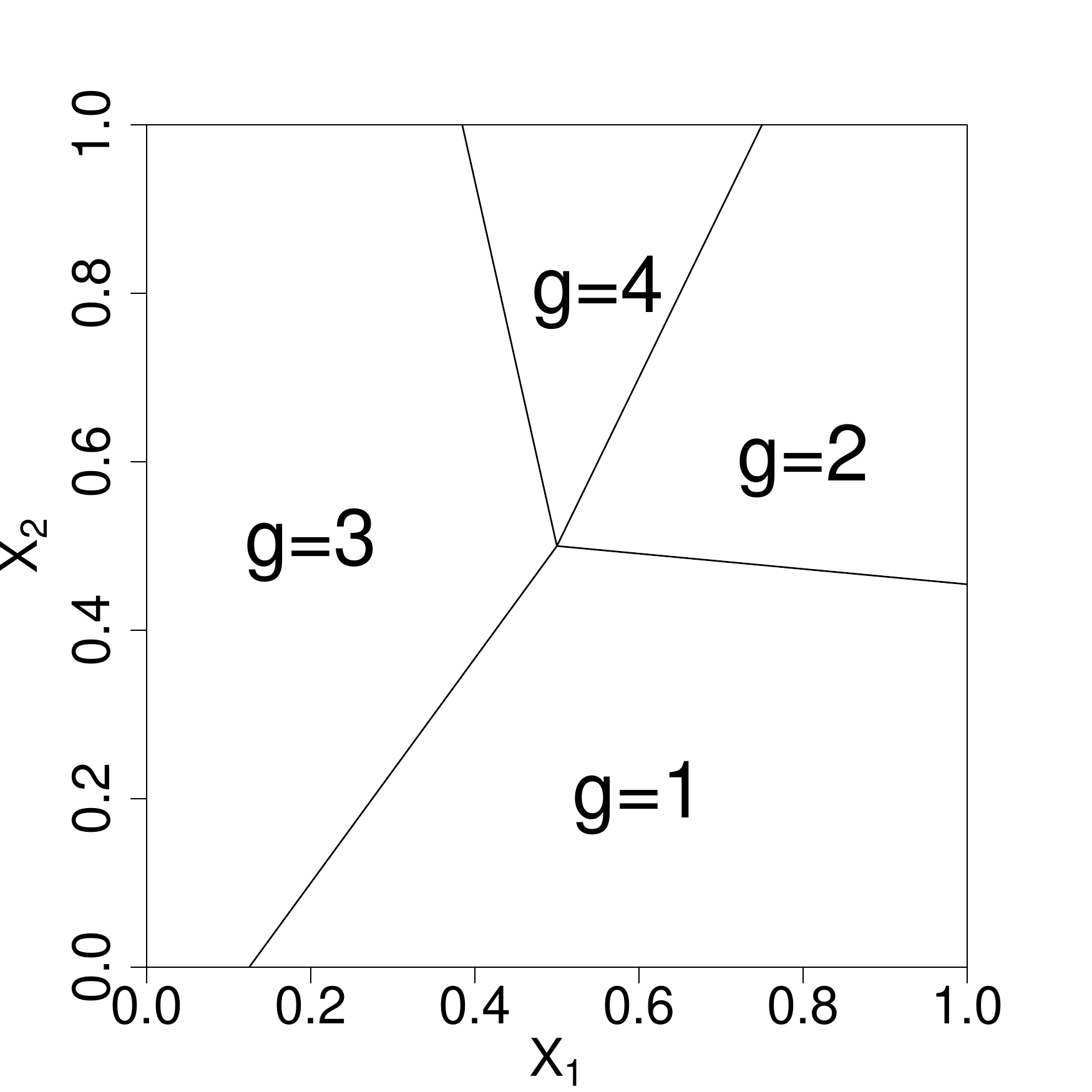}
        \caption[]%
        {{\small Linear}}
        \label{fig:linear}
    \end{subfigure}
    \begin{subfigure}[b]{0.24\textwidth}
        \centering
        \includegraphics[width=\textwidth]{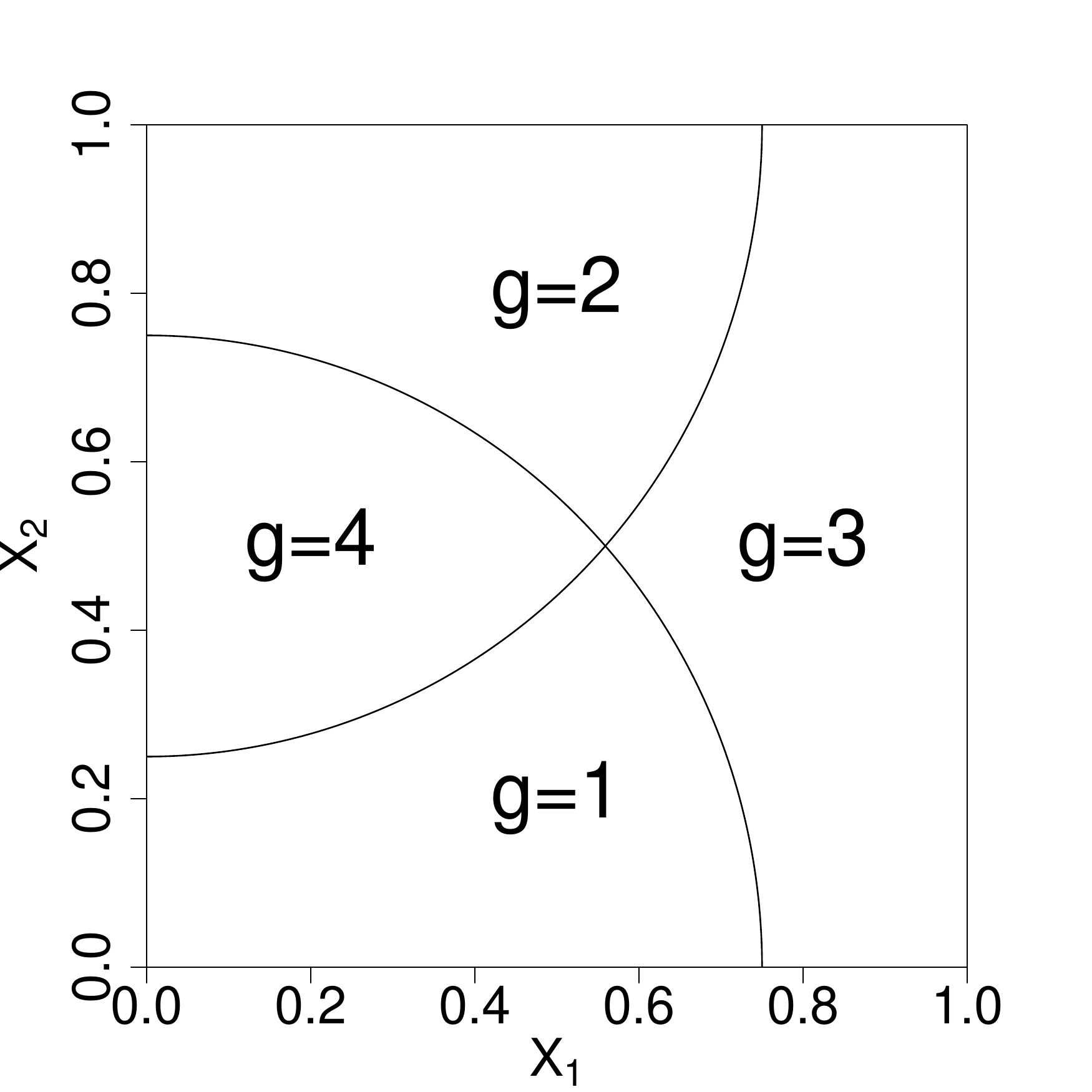}
        \caption[]%
        {{\small Nonlinear}}
        \label{fig:nonlinear}
    \end{subfigure}
    \begin{subfigure}[b]{0.24\textwidth}
        \centering
        \includegraphics[width=\textwidth]{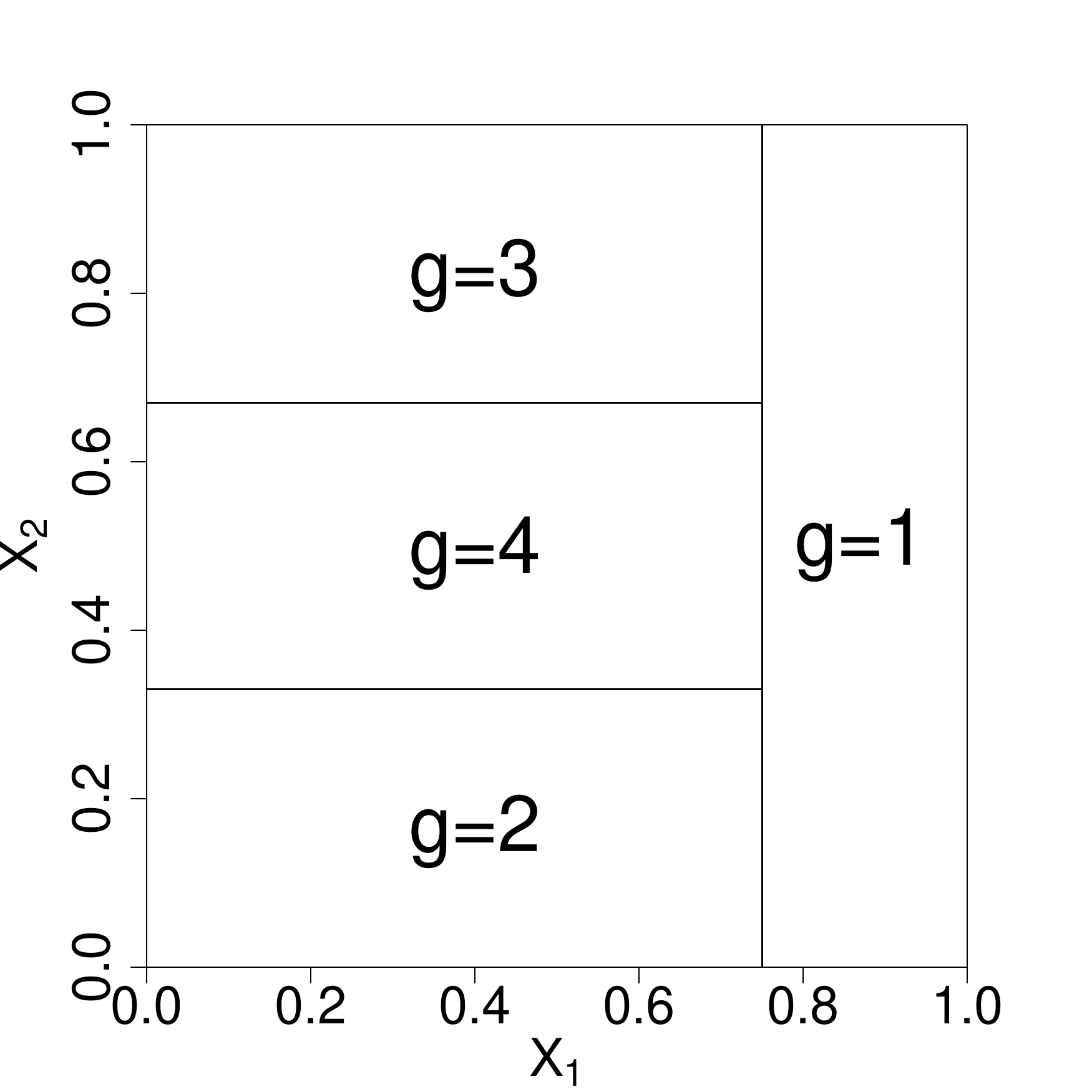}
        \caption[]%
        {{\small Asymmetric}}
        \label{fig:asym}
    \end{subfigure}
    \caption{Four structures of latent class membership based on $X_1$ and $X_2$:
    (a) Tree partition, (b) Linear partition,
    (c) Nonlinear partition, (d) Asymmetric tree partition.}
    \label{fig:struct}
\end{figure}
}

\subsection{Models}
Given the simulated data, JLCT uses 
$\vXs=\{X_3,X_{4},X_{5}\}$ to model time-to-event data,
$\vXf=\vXr=\{X_1,\dots,X_5\}$ to model fixed and random effects for
longitudinal outcomes, and ${\vXg}^{(\text{JLCT})} = \{X_1,\dots,X_5\}$ to
model latent class memberships. In order to match the 
maximum number of latent classes allowed for JLCM
\footnote{The running time of JLCM grows exponentially as a function
of number of latent classes \cite{zhangsimonoff}, thus we cap it at $6$.}, 
all JLCT models are pruned to have no more than
6 terminal nodes.  We fit the JLCT models using the 
\texttt{jlctree} function of our package \texttt{jlctree}
(see Appendix~\ref{app:package} for more details).
\red{Note that this allows both methods to potentially uncover the true latent classes; 
clearly if the maximum number of classes allowed is smaller than the true number neither 
method can be very effective. Unlike for JLCM, the computing time for JLCT is not 
sensitive to the choice of the maximum number of terminal nodes, so if desired this 
number can be varied to provide a sensitivity analysis in the analysis of real data.}

JLCM uses almost the same subsets of covariates as JLCT, except for
latent classes. Since tree-based approaches automatically use
interactions between covariates to partition the population,
for fair comparison we include an additional interaction term
in JLCM's latent classes modeling:
${\vXg}^{(\text{JLCM})}=\{X_1,\dots,X_5, X_1X_2\}$.
We allow class-specific coefficients 
and class-specific Weibull baseline risk functions for its survival model.
The optimal number of latent classes is chosen from $\{2,3,4,5,6\}$ using BIC\@.
We fit the JLCM models using the \texttt{Jointlcmm} function of the
\texttt{lcmm} package \cite{lcmm-package}.

We can directly fit JLCT to the simulated data where all covariates are 
time-varying. However, JLCM does not support time-varying covariates
in either the latent class or the survival model.
In the implementation of JLCM, i.e.\ the \texttt{Jointlcmm} function of the 
\texttt{lcmm} package, if a time-varying covariate is used in either model,
by default \texttt{Jointlcmm} will take the first encountered value per subject
as a baseline value and only use that. Thus, we can still fit JLCM with 
the simulated time-varying covariates, but JLCM automatically replaces 
$\vXs,\vXg$ with $\toinv \vXs, \toinv \vXg$, respectively,
where $\toinv X$ denotes the first encountered value 
of a time-varying covariate $X$. 

Since JLCM only uses the first encountered values per subject for covariates
$\vXs,\vXg$, it is not clear whether differences in performance of JLCM and JLCT
are due to the additional information contained in later values of time-varying
covariates, or because of the differences in the methodology itself.
For the purpose of decomposing the difference,
we also examine versions of JLCT that are restricted to using
$\toinv \vXs$ and $\toinv \vXg$.
Table~\ref{tb:simulate_models} summarizes the various combinations of covariates
we used to fit JLCT and JLCM models. 
Note that JLCM uses the same amount of information
as JLCT$_2$, while JLCT$_4$ is the default JLCT method.

\begin{table}
    \caption{Simulation: covariates used to fit JLCT and JLCM models}
    \label{tb:simulate_models}
    \centering
    \begin{tabular}{ccccc} \toprule
        Model      &Time-to-event &Latent class                   &Longitudinal \\ \midrule
        JLCT$_1$   &$\vXs$        & no splitting                  &$\vXf,\vXr$ \\
        JLCT$_2$   &$\toinv \vXs$ & $\toinv {\vXg}^{(\text{JLCT})}$     &$\vXf,\vXr$ \\
        JLCT$_3$   &$\toinv \vXs$ & ${\vXg}^{(\text{JLCT})}$            &$\vXf,\vXr$ \\
        JLCT$_4$   &$\vXs$        & ${\vXg}^{(\text{JLCT})}$            &$\vXf,\vXr$ \\
        JLCM       &$\toinv \vXs$ & $\toinv {\vXg}^{(\text{JLCM})}$     &$\vXf,\vXr$ \\ \bottomrule
    \end{tabular}
\end{table}

\subsection{Evaluation metrics}\label{sec:evaluation}
We use the following metrics to measure and compare model performances.
To compute the out-of-sample measures, at each simulation run
we generate a new random sample of $N$ subjects,
using the same data generating process as for the in-sample data.

\paragraph{Time-to-event prediction} We use the integrated squared error (ISE)
    to measure the divergence between the true and the predicted survival curves.
    The ISE for a set of $N$ subjects is defined as
    \begin{equation*}
        \text{ISE} = \frac{1}{N} \sum_{i=1}^N
        \frac{1}{\max_i T_i} \int_0^{\max_j T_j}
        \big(\h S_i(t) - S_i(t)\big)^2 dt,
    \end{equation*}
    where $\h S_i(t)$ and $S_i(t)$ are the predicted and the true
    survival probability for subject $i$ at time $t$, respectively.
    We compute ISE on in-sample subjects (ISE$_{\text{in}}$)
    and on out-of-sample subjects (ISE$_{\text{out}}$).
    In order to predict survival probabilities beyond subject $i$'s 
    event time $T_i$, we assume that  covariates for subject $i$
     remain unchanged  since the last observed time.

\paragraph{Longitudinal prediction} 
    We use the mean squared error (MSE$_y$) 
    between the true and the predicted longitudinal outcomes to measure prediction performance.
    The MSE$_y$ for a set of $N$ subjects is defined as
    \begin{equation*}
        \text{MSE}_y = \frac{1}{\sum_{i=1}^N n_i}
        \sum_{i=1}^N \sum_{j=1}^{n_i} (\h y_{ij} - y_{ij})^2,
    \end{equation*}
    where $n_i$ is the number of observations for subject $i$,
    and we denote by $\h y_{ij}$ and $y_{ij}$ the predicted and the true
    $j$-th longitudinal outcome of subject $i$, respectively.
    We compute MSE$_y$ on in-sample subjects (MSE$_{y\text{in}}$)
    and on out-of-sample subjects (MSE$_{y\text{out}}$).

    \paragraph{Parameter estimation} We use the mean squared error (MSE$_b$) 
    to measures the difference between the true and the estimated Cox PH slope coefficients.
    MSE$_b$ on a set of $N$ subjects is defined as:
    \begin{equation*}
        \text{MSE}_b = \frac{1}{\sum_{i=1}^N n_i} \sum_{i=1}^N \sum_{j=1}^{n_i}
        \norm{\hat{\boldsymbol{b}}_{\h d_{ij}} - \boldsymbol{b}_{g_{ij}}}_2^2,
    \end{equation*}
    where $\hat{\boldsymbol{b}}_{k}$ and $\boldsymbol{b}_{k}$
    are the estimated and the true Cox PH slopes for latent class $k$,
    and where $\h d_{ij}$ and $g_{ij}$ are the predicted and the true latent
    class memberships for the $j$-th observation of subject $i$, respectively.

    \paragraph{Latent class membership recovery} We use classification accuracy 
    ($\text{Acc}_g$) to measure how well JLCT recovers the true latent class membership. 
    Since JLCT constructs a tree, $\text{Acc}_g$ is only computed for the setups where the true latent 
    classes are determined by trees, thus only for ``Tree'' and ``Asymmetric'' setups.  
    Given a constructed JLCT tree, all observations falling into the same terminal node
    are classified into the majority of actual classes of these observations. 
    $\text{Acc}_g$ is then defined as the classification accuracy on the out-of-sample data,
    \[
        \text{Acc}_g = \frac{1}{\sum_{i=1}^N n_i} \sum_{i=1}^N \sum_{j=1}^{n_i}  
        1_{\{ g_{ij} = \h d_{ij}\}},
    \]
    where $g_{ij}$ and $\h d_{ij}$ denote the true and the predicted latent 
    class membership for the $j$-th observation of subject $i$, respectively.

It is worth emphasizing that JLCM uses extra information, such as
the longitudinal outcomes and time-to-event,
to predict latent class membership for in-sample subjects.
The quality of latent class membership prediction directly affects that of
time-to-event and longitudinal outcomes predictions,
and thus JLCM is advantaged compared to JLCT for in-sample performance.
A comparison between the two is only fair on out-of-sample subjects, 
since the longitudinal
outcomes and time-to-event are no longer available to JLCM at prediction time,
and therefore the two methods use the same amount of information 
for prediction. In view of this, we focus on comparing the out-of-sample 
measures in this section, and present the in-sample prediction results 
in Appendix~\ref{app:moreresults_time_inv}.

\subsection{Results}
Figures~\ref{fig:survvar_N500_ISEout} to~\ref{fig:survvar_N500_purity_var}
show the boxplots of ISE$_{\text{out}}$, MSE$_{y_\text{out}}$, and MSE$_b$ 
on $\log_{10}$ scale, and Acc$_g$, for $N=500$, light censoring, and Weibull-I distributions,
for combinations of latent class structure and concentration level:
$\{$Tree, Linear, Nonlinear, Asymmetric, \red{Null} $\} \times
\{ p_0=0.5,\, p_0=0.7,\, p_0=0.85, \,p_0=1\}$, for the five models listed 
in Table~\ref{tb:simulate_models}.
The experiments are repeated 100 times for JLCT and JLCM under each setting.
The results for other baseline hazards distributions and in-sample performance measures
are given in Appendix~\ref{app:moreresults_time_var}.

\red{In the discussion below, we first focus on the cases 
    where latent classes are generated as in Figure~\ref{fig:struct}, 
    and then comment on the null scenario where there are no latent classes.}

\paragraph{ISE$_{\text{out}}$} 
Figure~\ref{fig:survvar_N500_ISEout} demonstrates that
ISE$_{\text{out}}$ keeps improving as JLCT uses time-varying data
in more modeling components,
and eventually JLCT$_4$ significantly outperforms JLCM\@.
JLCT$_2$ uses the converted ``time-invariant''(baseline) data,
and it has slightly larger ISE values than JLCM (which uses the same converted data).
Once we allow JLCT$_3$ to use the original time-varying covariates in the survival model,
the ISE becomes smaller than that of JLCM\@.
When we further allow JLCT$_4$ to use the original time-varying
covariates in the class membership model, the ISE
decreases even more and becomes significantly better than that of JLCM\@.
That is, if each subject is allowed to switch between latent classes throughout 
the time of study, and the estimated membership is also allowed to switch,
we can achieve considerable improvement in survival predictions.

The performances of the default JLCT (i.e. JLCT$_4$) improve when the
latent class membership becomes less noisy, i.e. $p_0$ increases,
and JLCT is a considerably better performer than JLCM when the latent classes are generated
by a nearly deterministic partitioning ($p_0\geq 0.85$).
In the extreme case where the partitioning is deterministic ($p_0=1$)
and the underlying structure is a tree (``Tree'', and ``Asymmetric Tree''),
JLCT is expected to perform well since data are generated according
to its underlying model, and it does indeed outperform JLCM by a
much larger margin in deterministic tree setups.

\paragraph{MSE$_{y\text{out}}$ and MSE$_b$} 
Similar patterns to those for ISE$_{\text{out}}$ occur for MSE$_{y\text{out}}$ (Figure~\ref{fig:survvar_N500_MSEyout}),
and MSE$_b$ (Figure~\ref{fig:survvar_N500_MSEb}) as well,
that using time-varying covariates yields stronger performance 
when time-varying variables are predictive. 
These two measures have less variability relative to overall levels
than does ISE.

\paragraph{Acc$_g$}
Figure~\ref{fig:survvar_N500_purity_var} shows the boxplots of Acc$_g$ 
for JLCT$_{\{2,3,4\}}$.
When $p_0=1$ and thus the actual latent classes are generated by a 
deterministic partitioning,
Acc$_g$ of JLCT$_3$ and JLCT$_4$ concentrate near 100\% with very small variations for 
both ``Tree'' and ``Asymmetric'' structures, indicating that JLCT trees are 
robust and almost always identical to the true underlying latent class trees.
In contrast, since JLCT$_2$ only uses time-invariant covariates to model the
time-varying memberships, it is not surprising that it achieves a significantly
lower accuracy.  
When $p_0<1$, the actual latent class memberships contain random noise, and thus on average
no prediction method can achieve an accuracy higher than $p_0$. Indeed, 
when $p_0$ decreases, Acc$_g$ decreases as well, 
yet JLCT$_3$ and JLCT$_4$ manage to remain concentrated
near the theoretical cap $p_0$. \ignore{Furthermore, the variation of Acc$_g$
increases as $p_0$ decreases, suggesting that it becomes more difficult to recover
the latent class structure when the actual latent class memberships become more noisy.}

When the proposed latent classes closely align with the true latent classes,
our experience is that inferences 
from the fitted survival and longitudinal models are reasonably accurate, 
with confidence intervals close to nominal coverage.
On the other hand, if the proposed latent classes do not align with the
true latent classes, then accurate predictions are unlikely.

\paragraph{Running time}
We report average running times of the default JLCT 
(JLCT$_4$)\footnote{The running
    times of all three JLCT methods (JLCT$_{\{2,3,4\}}$) are comparable,
thus we only discuss the default JLCT$_4$ in our comparison.}
and JLCM over 100 runs in Table~\ref{tb:simtime_time_var}. 
The running time of JLCT includes constructing the tree and
fitting the survival and linear mixed-effects model. 
The running time of JLCM includes
fitting with all numbers of latent classes $g\in \{2,3,4,5,6\}$.
Clearly, JLCT is orders of magnitude faster than JLCM across all settings.
The running time for other censoring levels and baseline hazard distributions
are similar.
All of the simulations are performed
on high performance computing nodes with 3.0GHz CPU and 62 GB of memory.
\ignore{Compared to the time-invariant %TODO: revisit this once time-inv data is in appendix.
scenario (Figure~\ref{tb:simtime_time_inv}), the change in JLCT's
running time is negligible, while JLCM takes much longer to run.
This observation is surprising, since internally JLCM still fits to a
time-invariant data set. The longer running time of JLCM suggests
that the data generated by time-varying covariates contain more
complex signals, which results in longer running times in fitting
the (incorrect) JLCM model.}

\newgeometry{margin=1in}
{
\begin{figure}[h!]
    \centering
    \begin{subfigure}[b]{0.5\textwidth}
        \centering
        \includegraphics[width=\textwidth]{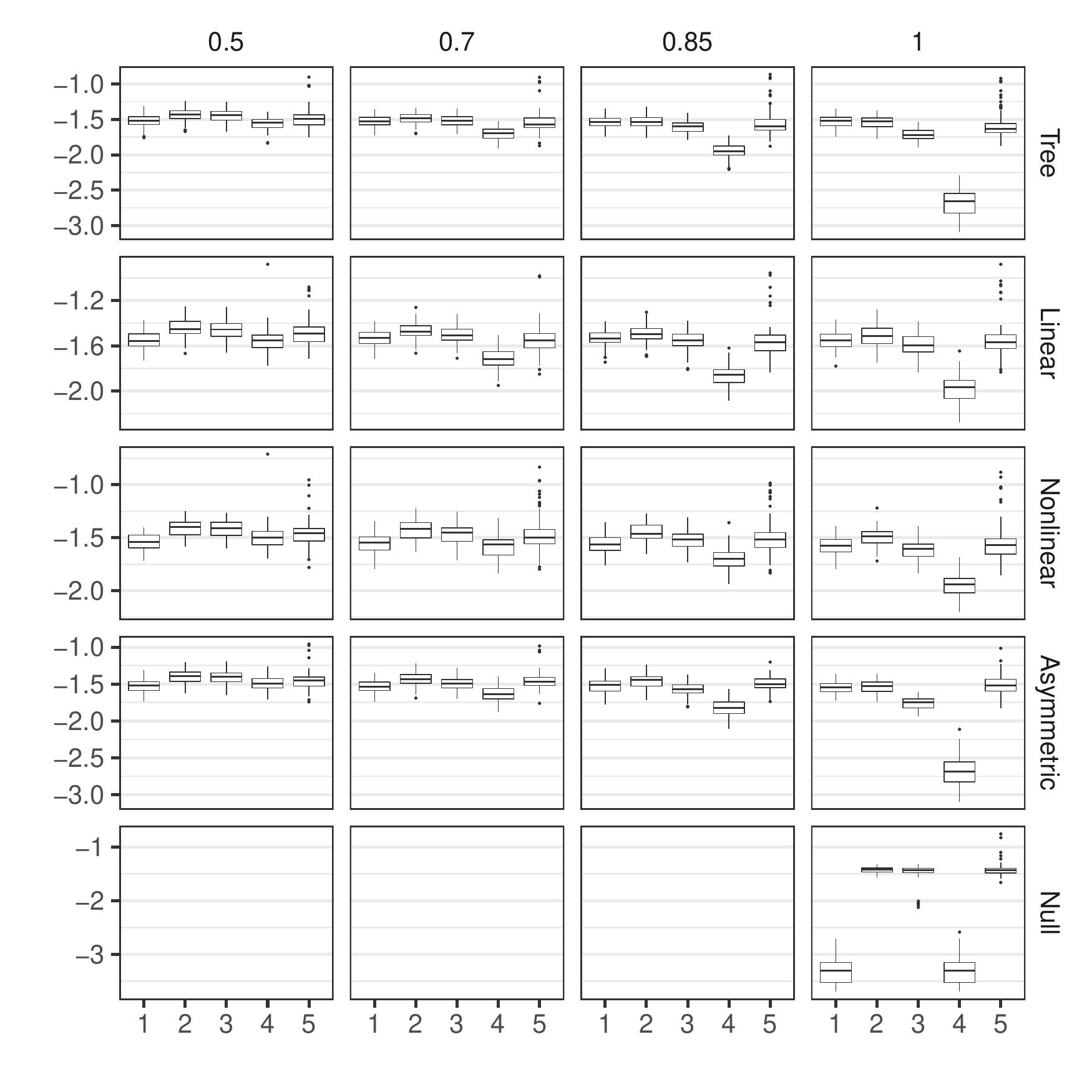}
        \caption[]{{\small $\log_{10}$ ISE$_{\text{out}}$}}%: integrated squared error of time-to-event predictions }}
        \label{fig:survvar_N500_ISEout}
    \end{subfigure}\hfill
    \begin{subfigure}[b]{0.5\textwidth}  
        \centering 
        \includegraphics[width=\textwidth]{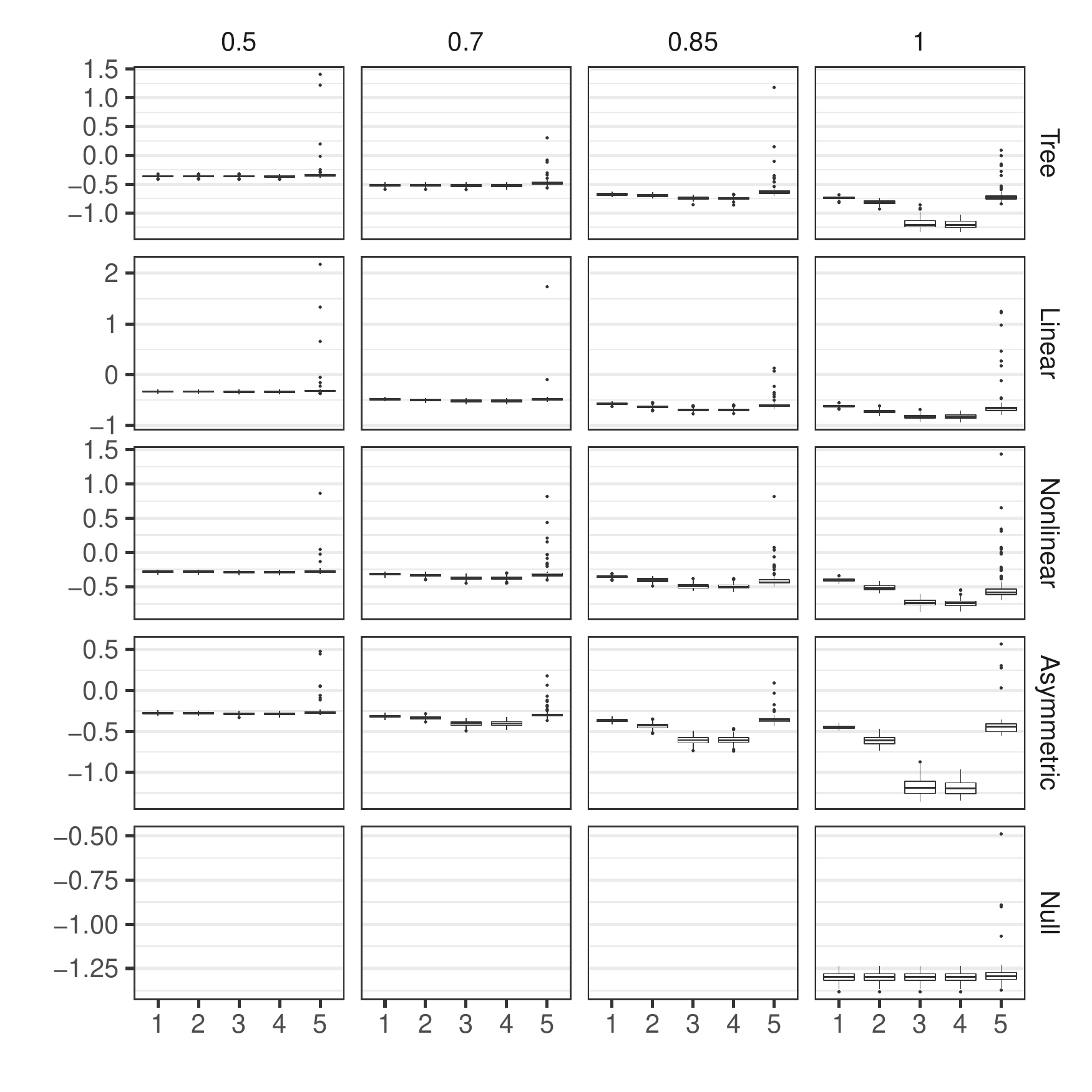}
        \caption[]{{\small $\log_{10}$MSE$_{y\text{out}}$}}%: mean squared error of longitudinal outcome predictions }}
        \label{fig:survvar_N500_MSEyout}
    \end{subfigure}
    \vspace\baselineskip
    \vskip 0.2cm
    \begin{subfigure}[b]{0.5\textwidth}  
        \centering 
        \includegraphics[width=\textwidth]{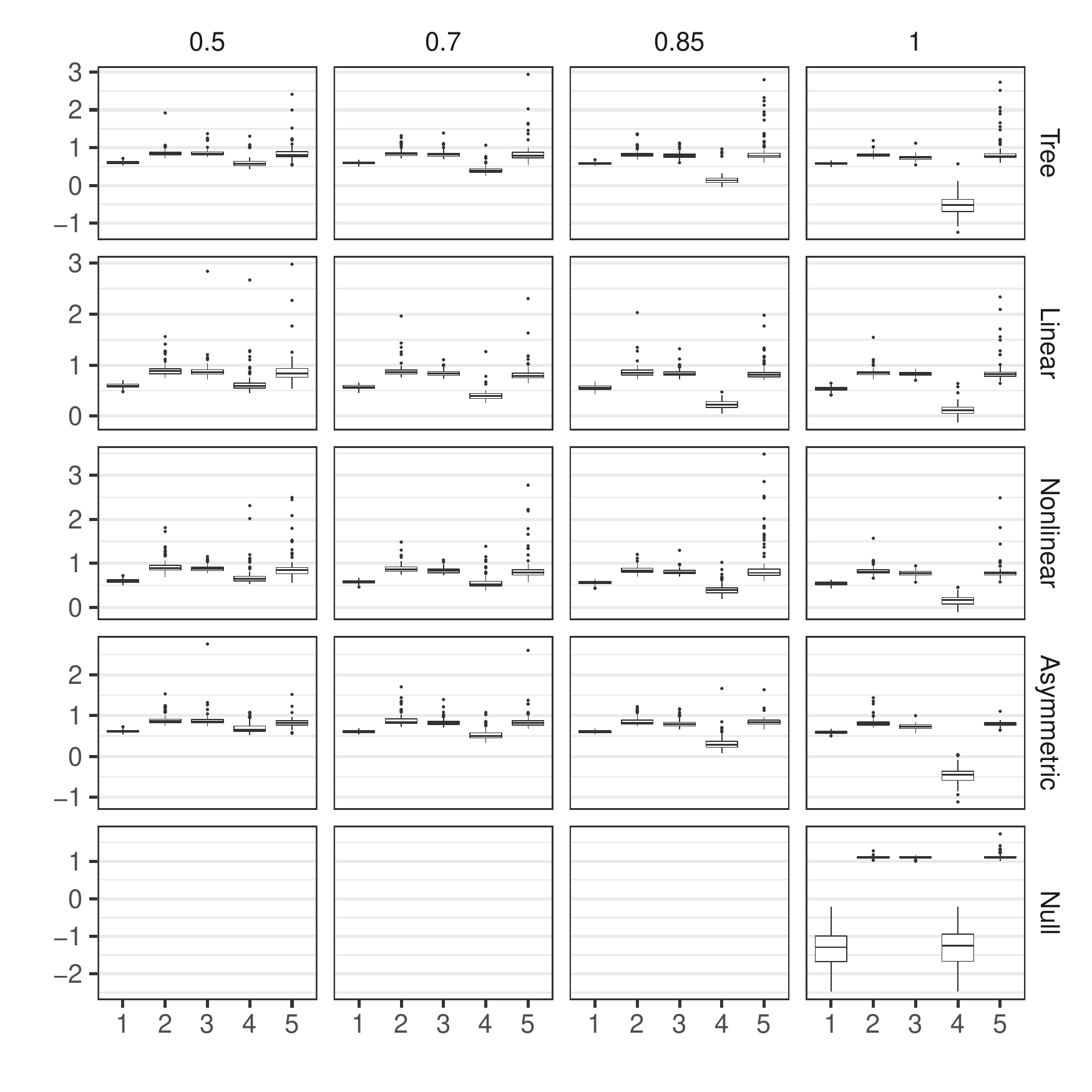}
        \caption[]{{\small $\log_{10}$ MSE$_{b}$}}%: mean squared error of estimated Cox PH slope coefficients }}
        \label{fig:survvar_N500_MSEb}
    \end{subfigure}
    \begin{subfigure}[b]{0.49\textwidth}
        \centering
        \includegraphics[width=\textwidth]{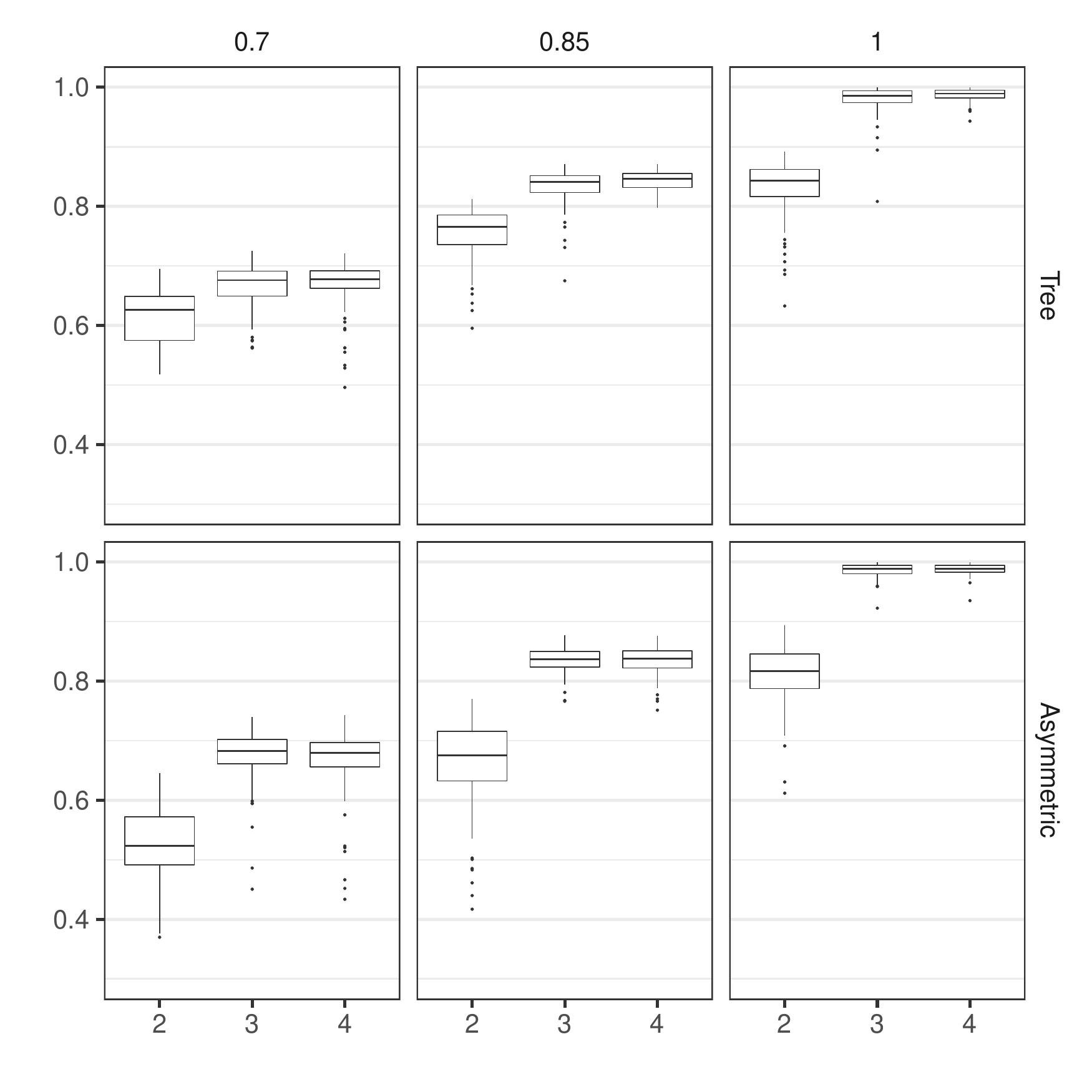}
        \caption[]%
        {{\small Acc$_g$ }}
        \label{fig:survvar_N500_purity_var}
    \end{subfigure}
    \vskip 0.3cm
    \caption{Boxplots of model performance: ISE$_{\text{out}}$, MSE$_{y_\text{out}}$, and MSE$_b$ 
        on $\log_{10}$ scale, and Acc$_g$, for $N=500$, light censoring, and Weibull-I distributions.
        The five models are given in Table~\ref{tb:simulate_models}. }
        \label{fig:survvar_N500}
\end{figure}
}
\restoregeometry
\advance\linewidth -\rightmargin
\advance\linewidth -\leftmargin

\begin{table}
    \caption{The average running time in seconds (standard deviation in parentheses)
        on a data set with time-varying covariates, $N=500$,
    light censoring and Weibull-I distribution. }
    \label{tb:simtime_time_var}
    \centering
    \begin{tabular}{llll}\toprule
        Structure   &$p_0$  & JLCT  & JLCM \\\midrule
        \multirow{4}{*}{Tree}
        &0.5    &13.97\ (0.41)  &1855.64\ (484.12)\\
        &0.7    &12.76\ (0.36)  &1777.32\ (464.07)\\
        &0.85   &10.84\ (0.49)  &1897.43\ (461.45)\\
        &1      &7.79\ (0.54)   &2008.06\ (461.2)\\\midrule
        \multirow{4}{*}{Linear}
        &0.5    &23.51\ (0.92)  &1963.11\ (437.16)\\
        &0.7    &23.27\ (1.21)  &1916.92\ (423.97)\\
        &0.85   &21.3\ (1.27)   &2015.88\ (366.85)\\
        &1      &20.8\ (1.3)    &2130.13\ (519.01)\\\midrule
        \multirow{4}{*}{Nonlinear}
        &0.5    &13.7\ (0.82)   &2052.86\ (342.14)\\
        &0.7    &13.6\ (0.44)   &2040.33\ (309.27)\\
        &0.85   &12.69\ (0.51)  &2095.46\ (368.12)\\
        &1      &11.93\ (0.91)  &2093.55\ (393.01)\\\midrule
        \multirow{4}{*}{Asymmetric}
        &0.5    &13.25\ (0.62)  &1979.74\ (198.41)\\
        &0.7    &13.37\ (1.73)  &1992.53\ (203.28)\\
        &0.85   &13.4\ (1.1)    &2061.09\ (204.94)\\
        &1      &8.94\ (0.86)   &2068.58\ (165.99)\\\midrule
        \red{Null} &\red{1}      &\red{0.61\ (1.78)}   &\red{1275.76\ (549.94)}\\\bottomrule
    \end{tabular}
\end{table}

\red{
\paragraph{Null scenario}
Under the null scenario where there are no latent classes, 
the average number of terminal nodes chosen by the four 
JLCT models are 1, 1.08,  1.06, 1.08, respectively, indicating that 
JLCT almost always makes the correct decision of not splitting.
JLCT$_1$ and JLCT$_4$ have comparable performances, 
while JLCT$_2$, JLCT$_3$ are significantly worse, yet still comparable with JLCM, 
in survival prediction (ISE$_{\text{out}}$) and survival parameter estimation (MSE$_b$). 
This is because JLCT$_2$, JLCT$_3$ and JLCM use time-invariant survival predictors 
$\toinv \vXs$, whereas JLCT$_1$ and JLCT$_4$ uses the original 
time-varying $\vXs$. This again reflects the prognostic value of 
time-varying covariates.  By the definition of Acc$_g$, all JLCT
models have Acc$_g$=1 under the null scenario, and thus we skip its discussion.
As for running time, since JLCT correctly decides to not split,
its running time is significantly less than when there are actual latent classes.
JLCM also has reduced running time under the null scenario (roughly 35\% less),
but it is still much more computationally intensive than JLCT.
}

\subsection{Additional simulation results}\label{subsec:morestops}

\red{As mentioned in Section~\ref{sec:jlct}, we 
also compare performance using the default stopping criterion $\TS_{\text{parent}} < 3.84$ to two 
alternatives, $\TS_{\text{parent}} < \{2.71, 6.63\}$,
under the above simulation setups for JLCT$_4$ model.
Table~\ref{tb:nnodes_time_var} reports the average number of terminal nodes 
using the three stopping rules, and as can be seen,
JLCT$_4$ splits into very similar numbers of terminal nodes 
regardless of the stopping criterion. As a result, 
the rest of the metrics are also comparable, which are given in 
Appendix~\ref{app:more_stopthre}.
}

\begin{table}
    \caption{\red{The average number of terminal nodes (standard deviation in parentheses)
        using JLCT$_4$ in simulations based on a data set with time-varying covariates, $N=500$,
light censoring and Weibull-I distribution, for three different stopping rules. }}
    \label{tb:nnodes_time_var}
    \centering
    \begin{tabular}{lllll}\toprule
        Structure   &$p_0$  & $\TS_{\text{parent}}< 2.71$ & $\TS_{\text{parent}}< 3.84$ & $\TS_{\text{parent}}< 6.63$\\\midrule
        \multirow{4}{*}{Tree}
        &0.5    &5.96\ (0.2)    &5.96\ (0.2)    &5.96\ (0.2)\\
        &0.7    &5.96\ (0.24)   &5.96\ (0.24)   &5.96\ (0.24)\\
        &0.85   &6\ (0) &6\ (0) &6\ (0)\\
        &1      &4.27\ (0.57)   &4.15\ (0.41)   &4.04\ (0.2)\\\midrule
        \multirow{4}{*}{Linear}
        &0.5    &5.94\ (0.24)   &5.94\ (0.24)   &5.93\ (0.26)\\
        &0.7    &5.99\ (0.1)    &5.99\ (0.1)    &5.99\ (0.1)\\
        &0.85   &5.99\ (0.1)    &5.99\ (0.1)    &5.99\ (0.1)\\
        &1      &6\ (0) &6\ (0) &5.99\ (0.1)\\\midrule
        \multirow{4}{*}{Nonlinear}
        &0.5    &5.76\ (0.51)   &5.76\ (0.51)   &5.77\ (0.51)\\
        &0.7    &5.95\ (0.22)   &5.95\ (0.22)   &5.95\ (0.22)\\
        &0.85   &5.99\ (0.1)    &5.99\ (0.1)    &5.99\ (0.1)\\
        &1      &5.98\ (0.28)   &5.92\ (0.31)   &5.9\ (0.36)\\\midrule
        \multirow{4}{*}{Asymmetric}
        &0.5    &5.86\ (0.47)   &5.86\ (0.47)   &5.84\ (0.49)\\
        &0.7    &5.91\ (0.29)   &5.91\ (0.29)   &5.92\ (0.27)\\
        &0.85   &5.83\ (0.43)   &5.84\ (0.42)   &5.85\ (0.41)\\
        &1      &4.36\ (0.63)   &4.2\ (0.43)    &4.03\ (0.17)\\\midrule
        Null    &1    &1.18\ (0.52)   &1.08\ (0.34)   &1.01(0.10) \\\bottomrule
    \end{tabular}
\end{table}

Simulations were also run based on a mix of time-invariant 
and time-varying covariates (results not reported).
Unsurprisingly, the patterns are similar to those with all 
time-varying covariates. 

Finally, we also examined using the median of the time-varying covariate 
values rather than the first encountered value when running JLCM, 
but the improvements in performance are negligible and the 
results are thus omitted.

%--------------------
\section{Application}\label{sec:application}
In this section, we illustrate the application of JLCT to a real
data set, the PAQUID (Personnes Ag\'ees Quid) data set, which was also examined in
\cite{proust2017estimation}.
The PAQUID dataset consists of 2250 records of 500 subjects from the PAQUID study
\cite{letenneur1994incidence}, which collects five
time-varying values, including three cognitive tests (\texttt{MMSE,
IST, BVRT}), a physical dependency score (\texttt{HIER}), and a
measure of depressive symptomatology (\texttt{CESD}), along with age
at visit (\texttt{age}). The time-to-event is the age at
dementia diagnosis or last visit, which is recorded
in the tuple (\texttt{agedem}, \texttt{dem}).
The PAQUID data set also collects three time-invariant covariates: education
(\texttt{CEP}), gender (\texttt{male}), and age at the entry of the
study (\texttt{age\_init}). As suggested by
\cite{proust2017estimation}, we normalize the highly asymmetric
covariate \texttt{MMSE} and only consider its normalized version
\texttt{normMMSE}; we also construct a new covariate
$\texttt{age65}=(\texttt{age}-65)/10$. Our goal for this data set is
to jointly model the trajectories of \texttt{normMMSE} (longitudinal
outcomes) and the risk of dementia (time-to-event), using the
remaining covariates.

We further compared JLCT's performance to 
that of JLCM and another joint modeling baseline,
the shared random effects model (SREM) \cite{wulfsohn1997joint,henderson2000joint,
tsiatis2004joint,rizopoulos2010jm}.
The name ``shared random effects'' comes from the modeling assumption 
that a set of random effects accounts for the association between 
longitudinal outcomes and time-to-event. The longitudinal outcomes 
are modeled by linear mixed-effects models, with random effects for 
each subject. These random effects affect the hazards of the event 
through a proportional hazard model. SREM is limited in the use of 
covariates in the survival model in the same way JLCM is, 
as it only allows time-invariant baseline covariates. 
\cite{blanche2015quantifying} showed that SREM and JLCM
can be viewed as special cases of a general parametric joint modeling 
of longitudinal and time-to-event outcomes, with the variable 
that ties these two parts together either being continuous (SREM) or 
discrete (JLCM). Thus, while SREM can represent the underlying distributions
of the time-to-event and longitudinal variable, respectively, while accounting
for their joint association, it does not produce the desired latent classes 
described in the introduction section.

\subsection{Models}

We consider two JLCM, one SREM, and four JLCT models.
\begin{enumerate} 
    \item (JLCM$_1$) We adopt the time-invariant JLCM model
        in \cite{proust2017estimation}: The trajectories of \texttt{normMMSE}
        depend on fixed effects $\vXf=\{$\texttt{age65}, $\texttt{age65}^2$,
        \texttt{CEP}, \texttt{male}$\}$, and random effects
        $\vXr=\{$\texttt{age65}, $\texttt{age65}^2\}$.
        The risk of dementia depends on $\vXs=\{$\texttt{CEP}, \texttt{male}$\}$,
        with class-specific Weibull baseline hazards function.
        The class membership is modeled by $\vXg=\{$\texttt{CEP}, \texttt{male}$\}$.

    \item (JLCM$_2$) We extend JLCM$_1$ to using additional covariates:
        the survival model uses covariates $\vXs=\{$\texttt{CEP},
            \texttt{male}, \texttt{age\_init}, $\toinv {\texttt{BVRT}}$,
            $\toinv {\texttt{IST}}$, $\toinv {\texttt{HIER}}$,
            $\toinv {\texttt{CESD}}$ $\}$;
        the class membership model uses covariates $\vXg=\{$\texttt{CEP},
            \texttt{male}, $\toinv {\texttt{age65}}$, $\toinv {\texttt{BVRT}}$,
            $\toinv {\texttt{IST}}$, $\toinv {\texttt{HIER}}$,
            $\toinv {\texttt{CESD}}$ $\}$.
        The rest of the model remains the same as in the time-invariant JLCM$_1$.
        Note that \texttt{Jointlcmm} automatically uses the first encountered value $\toinv X$
        of any time-varying covariate $X$.

    \item  (SREM) The shared random effects model (SREM).
        Since fitting a SREM model becomes difficult when there are multiple predictors,
        we consider a simple SREM model that uses the same
        sets of time-invariant covariates as JLCM$_1$:
        the trajectories of \texttt{normMMSE}
        depend on fixed effects $\vXf=\{$\texttt{age65}, $\texttt{age65}^2$, \texttt{CEP}, \texttt{male}$\}$
        and the shared random effects
        $\vXr=\{$\texttt{age65}, $\texttt{age65}^2\}$.
        The risk of dementia depends on $\vXs=\{$\texttt{CEP}, \texttt{male}$\}$,
        as well as the shared random effects $\{$\texttt{age65}, $\texttt{age65}^2\}$.
        We use a Weibull baseline hazards function.

    \item (JLCT$_1$) The first JLCT model uses the same sets of covariates as JLCM$_1$ and SREM.

    \item (JCLT$_2$) The second JLCT model uses the same sets of covariates as JLCM$_2$.
        In particular, JLCT$_2$ also uses $\toinv {X}$ for any time-varying
        covariate $X$.

    \item (JCLT$_3$) The third JLCT model uses the same sets of covariates as JLCM$_2$.
        However, JLCT$_3$ uses all values of the time-varying covariates
        for splitting $\vXg$, but it still uses $\toinv \vXs$ to model time-to-event outcome.

    \item (JLCT$_4$) The last JLCT model adopts the same sets of covariates as JLCM$_2$,
        but using all values of any time-varying covariate:
        $\vXs=\{$\texttt{CEP}, \texttt{male}, \texttt{age\_init}, \texttt{BVRT},
        \texttt{IST}, \texttt{HIER}, \texttt{CESD}$\}$;
        and $\vXg=\{$\texttt{CEP}, \texttt{male}, \texttt{age65}, \texttt{BVRT},
        \texttt{IST}, \texttt{HIER}, \texttt{CESD}$\}$.
        JLCT$_4$ is our main model with no comparable competitors.
\end{enumerate}

As in the simulations, we fit the JLCM and JLCT models using the R packages 
\texttt{lcmm} and \texttt{jlctree}, respectively.
For the two JLCM models, the number of latent classes is chosen from 2 to 6
according to the BIC selection criterion.  For the four JLCT models,
we set the stopping threshold to $3.84$ and prune the trees to have no more than
6 terminal nodes. For all JLCT and JLCM models, the survival models 
share the same slope coefficients across latent classes,
a setup adopted by \cite{proust2017estimation}.
We fit the SREM model using the \texttt{jointModel} function of the
\texttt{JM} \cite{rizopoulos2010jm} package, and then uses its associated \texttt{predict.jointModel}
and \texttt{survfitJM} functions to predict longitudinal outcomes and survival curves
\footnote{Note that once we have constructed a JLCT tree, 
we are free to fit any models we wish, including SREM, to the data within each terminal node. 
However, on the PAQUID dataset we have not been able to fit SREM 
models within terminal nodes, due to numerical issues with \texttt{jointModel}.}.

\subsection{Evaluation metrics}
We use the root mean squared error (RMSE) to measure the 
accuracy of the predicted longitudinal outcomes, as was done in the simulations.
However, unlike in the simulations where we know the true survival curve for each subject,
in this application we only have the empirical survival curves. Thus,
to evaluate the accuracy of the time-to-event predictions,
we take the commonly used measure, the Brier score and its integrated version, IBS \cite{graf1999assessment}.
The Brier score (BS) at a fixed time $t$ is defined as
\begin{equation*}
    \text{BS}(t) = \frac{1}{N} \sum_{i=1}^N \Big(I(Y_i > t)
    - \widehat{S}(t \vert X_i)\Big)^2,
\end{equation*}
where $\widehat{S}(t\vert X_i)$ is the predicted probability of survival
at time $t$ conditioning on subject $i$'s predictor vector $X_i$,
and $Y_i$ is the time-to-event of subject $i$.
The Integrated Brier score (IBS) is therefore defined as
$$ \text{IBS} = \frac{1}{\max{Y_i}} \int_{0}^{\max{Y_i}} \text{BS}(Y)dY.$$

We compute the accuracy measure on out-of-sample subjects using
10-fold cross-validation as follows. We first randomly divide the
data set into 10 folds of equal size, where we take care such that
observations of a single subject belong to the same fold. Next, we
hold out one fold of data and run the model on the remaining nine
folds. The performance of the model is then evaluated on the held
out data. The procedure is repeated 10 times, where each of the 10
folds is used for out-of-sample evaluation.

\subsection{Results}\label{subsec:paquidresults}
We report the average prediction measure and running time over the 10 folds in
Table~\ref{tb:paquid}.  
When using only time-invariant covariates, JLCT$_1$ performs
similarly to its counterpart JLCM$_1$ in prediction accuracy (IBS and RMSE),
while SREM outperforms both JLCT$_1$ and JLCM$_1$ in survival prediction (IBS).
By adding four ``time-invariant'' covariates (which are converted from
time-varying ones) to the class membership and survival models,
the performance of JLCT$_2$ remains similar,
but the performance of JLCM$_2$ becomes much worse.
In fact, JLCM$_2$'s IBS is even worse than a simple prediction of
$\widehat{S}=0.5$ for every observation (which gives IBS $=0.25$),
as JLCM fails to converge.
When using the original time-varying covariates in the class membership,
both JLCT$_3$ and JLCT$_4$ improve their time-to-event prediction accuracies and
outperform all other methods by a noticeable margin on that measure.
When further allowing time-varying covariates to model time-to-event outcomes,
JLCT$_4$ gives an additional lift over JLCT$_3$ on time-to-event prediction.
When we look at the running times, JLCT is much faster than JLCM: fitting using
JLCT took no more than 2 minutes even for the most complex model (JLCT$_4$),
while fitting using JLCM took from~40 to~60 minutes. The model fitting is performed
on a desktop with 2.26GHz CPU and 32GB of memory.

The results in Table~\ref{tb:paquid} demonstrate two key advantages of the
tree-based approach JLCT over the parametric JLCM and SREM:
JLCT is capable of providing significantly better prediction performance with the
use of time-varying covariates in all of its modeling components,
and it can be orders of magnitude faster to fit JLCT than to fit JLCM.
Although SREM runs fast with simple models, it can be difficult
to fit SREM to complex joint models, and its prediction performance 
from a practical point of view is therefore limited. 
\red{Further, of course, it does not provide estimated latent classes at all.}

\begin{table}
    \centering
    \caption{Performance of JLCM, SREM, and JLCT methods on the PAQUID data set.}
    \label{tb:paquid}
    \begin{tabular}{ccccccccc} \toprule
        & JLCM$_1$ & JLCM$_2$  &SREM  & JLCT$_1$ & JLCT$_2$ & JLCT$_3$ & JLCT$_4$  \\ \midrule
        IBS         &0.1731     &0.4467     &0.1271    &0.1611     &0.1690      &0.1060  &0.0966         \\
        RMSE        &14.7588    &18.3544    &14.6689   &14.5503    &14.2912    &14.5590 &14.5014        \\
        Time (secs) &2448.70  &4107.62  &63.39   &1.70     &40.91    &59.86 &87.91        \\ \bottomrule
    \end{tabular}
\end{table}

Figures~\ref{fig:paquid_tree_inv}-\ref{fig:paquid_tree_var} give the
four JLCT tree structures, which are fit using the entire PAQUID
data set. The numbers in each box display the test statistics \TS,
and the proportion of observations contained in the current node. We
make the following observations:
\begin{itemize}
\item When fitting with only time-invariant covariates \texttt{CEP} and
    \texttt{male}, JLCT$_1$ first splits into \texttt{CEP}$=0$ and \texttt{CEP}$=1$,
    then splits on gender within the node of $\texttt{CEP}=1$.
    Two of the three terminal nodes have final \TS\ greater than the stopping
    criterion, 3.84, which indicates potential association between longitudinal
    and survival data within these two nodes. However, since \texttt{CEP}
    and \texttt{male} take binary values $\{0,1\}$,
    JLCT$_1$ cannot split further. Thus, it seems unlikely that
    using only the original time-invariant covariates provides adequate fit for these data.
\item JLCT$_2$ uses more splitting covariates,
    $\vXg=\{$\texttt{CEP}, \texttt{male},
        $\toinv {\texttt{age65}}$, $\toinv {\texttt{BVRT}}$,
        $\toinv {\texttt{IST}}$, $\toinv {\texttt{HIER}}$,
    $\toinv {\texttt{CESD}}$ $\}$,
    with some of the covariates converted from time-varying ones.
    JLCT$_2$ makes multiple splits on $\toinv {\texttt{CESD}}$,
    and then makes a final split on $\toinv {\texttt{ISE}}$.
    With additional covariates to split on, JLCT$_2$ ends up with five terminal
    nodes, each having a test statistic less than 3.84, and thus JLCT$_2$ has uncovered
    a good partitioning in the sense that the terminal nodes lack evidence of
    association between the longitudinal and survival outcomes.  However,
    the prediction performance of JLCT$_2$ does not show improvement
    over JLCT$_1$ based on cross-validation,
    suggesting that the JLCT models
    have reached a limit with only time-invariant original and constructed covariates to use.

\item Both JLCT$_3$ and JLCT$_4$ use all of the time-varying values of the $\vXg$ covariates in JLCT$_2$.
    The difference is that JLCT$_3$ only uses time-invariant values in modeling the time-to-event outcome, whereas
    JLCT$_4$ uses the time-varying counterparts.
    JLCT$_3$ first splits based on \texttt{age} (splitting at ages 85),
    and further splits based on \texttt{IST} for the younger group.
    The JLCT$_4$ tree splits into three nodes based only on
    \texttt{age} (splitting at ages 82 and 90), suggesting that people
    transition into different dementia statuses as they get older,
    which are reflected in both cognitive test score (\texttt{normMMSE})
    and time until a dementia diagnosis.

\item The different tree structures of JLCT$_3$ and JLCT$_4$ indicate that
    \texttt{IST} is potentially an important predictor for dementia diagnosis.
    Among the extended Cox PH models fit within each terminal node of JLCT$_4$,
    \texttt{IST} has a $p$-value of $0.02$ and $0.12$ within the groups
    of $\texttt{age} < 82$ and $82 \leq \texttt{age}< 90$, respectively.
    In fact, it is the most statistically significant covariate among all covariates in all
    Cox PH models fit within each JLCT$_4$ terminal node,
    suggesting that \texttt{IST} is the most prognostic covariate for dementia among
    the younger age groups ($\texttt{age}<90$).
    Since JLCT$_3$ can only use the first encountered value of \texttt{IST} to
    model the time-to-dementia, it apparently exploits the prognostic power of \texttt{IST} by
    splitting on \texttt{IST} among the younger age groups.
    The time-to-dementia prediction accuracy of JLCT$_3$ is only slightly worse than that of
    JLCT$_4$, suggesting that both ways of using \texttt{IST} reasonably capture the relationship
    between \texttt{IST} and dementia statuses.

\item All three nodes of JLCT$_3$ and JLCT$_4$ obtain a final test statistic
    less than 3.84, which indicates a good partitioning of the population
    in terms of preserving conditional independence within groups.
    With time-varying latent class memberships,
    both JLCT$_3$ and JLCT$_4$ achieve significant improvement in prediction performance
    of time to dementia diagnosis.
\end{itemize}

Finally, to study the effect of stopping parameter $s$ on the JLCT model,
we consider two additional values: $s=2.71$ and $s=6.63$, which
correspond to the 10\% and 1\% tails of $\chi_1^2$ distribution, respectively.
Recall that the default value $s=3.84$ is the 5\% tail of $\chi_1^2$ distribution.
We run JLCT$_4$ with these three stopping values, and compare the 10-fold cross-validation
results in Table~\ref{tb:paquid_mores}.
With a smaller $s$ value, JLCT$_4$ tends to split into more terminal nodes, which for this data set
improves the cross-validation-estimated survival prediction performance accuracy (IBS), and slightly
improves the longitudinal predictions (RMSE),
\red{
although of course it is impossible to know if these small cross-validation gains 
correspond to real improvements in representing the underlying longitudinal and 
time-to-event processes.
}

\red{The fitted JLCT models point out that any apparent predictive power of the 
cognitive assessment \texttt{MMSE} for time to dementia could easily be spurious, 
since once age is taken into account in the form of three broad age groups, 
any association between \texttt{MMSE} and \texttt{agedem} is gone.  
This suggests that the use of the \texttt{MMSE} test 
to predict time to dementia is problematic in older people. 
Furthermore, \texttt{IST} is a strong predictor 
for age to dementia in the two younger age groups, but not the oldest group, 
suggesting once again that for older people the tests 
have limited use for this purpose. It is known that \texttt{MMSE} tends to decrease 
with age \cite{pradier2014mini}, and since \texttt{MMSE} is 
widely-used, and considered a standard by which other cognitive tests are 
evaluated \cite{tsoi2015cognitive}, our finding points to a possibly 
inappropriate use of this test for determining time to dementia diagnosis, 
which might also apply to other assessment tools as well.}

\begin{table}
    \centering
    \caption{Performance of JLCT$_4$ on the PAQUID data set, with various values for stopping parameter $s$.}
    \label{tb:paquid_mores}
    \begin{tabular}{cccc} \toprule
        & $s=2.71$ & $s=3.84$  & $s=6.63$  \\ \midrule
        IBS         &0.0842     &0.0966     & 0.0988 \\
        RMSE        &14.4698    &14.5014    &14.5817\\
        Time (secs) &87.54    &87.91    &51.46\\ \bottomrule
    \end{tabular}
\end{table}

%---------- JLCT tree structure ----------
\newgeometry{margin=1in}
{
\begin{figure}[h!]
    \centering
    \begin{subfigure}[b]{0.47\textwidth}
        \centering
        \includegraphics[width=\textwidth]{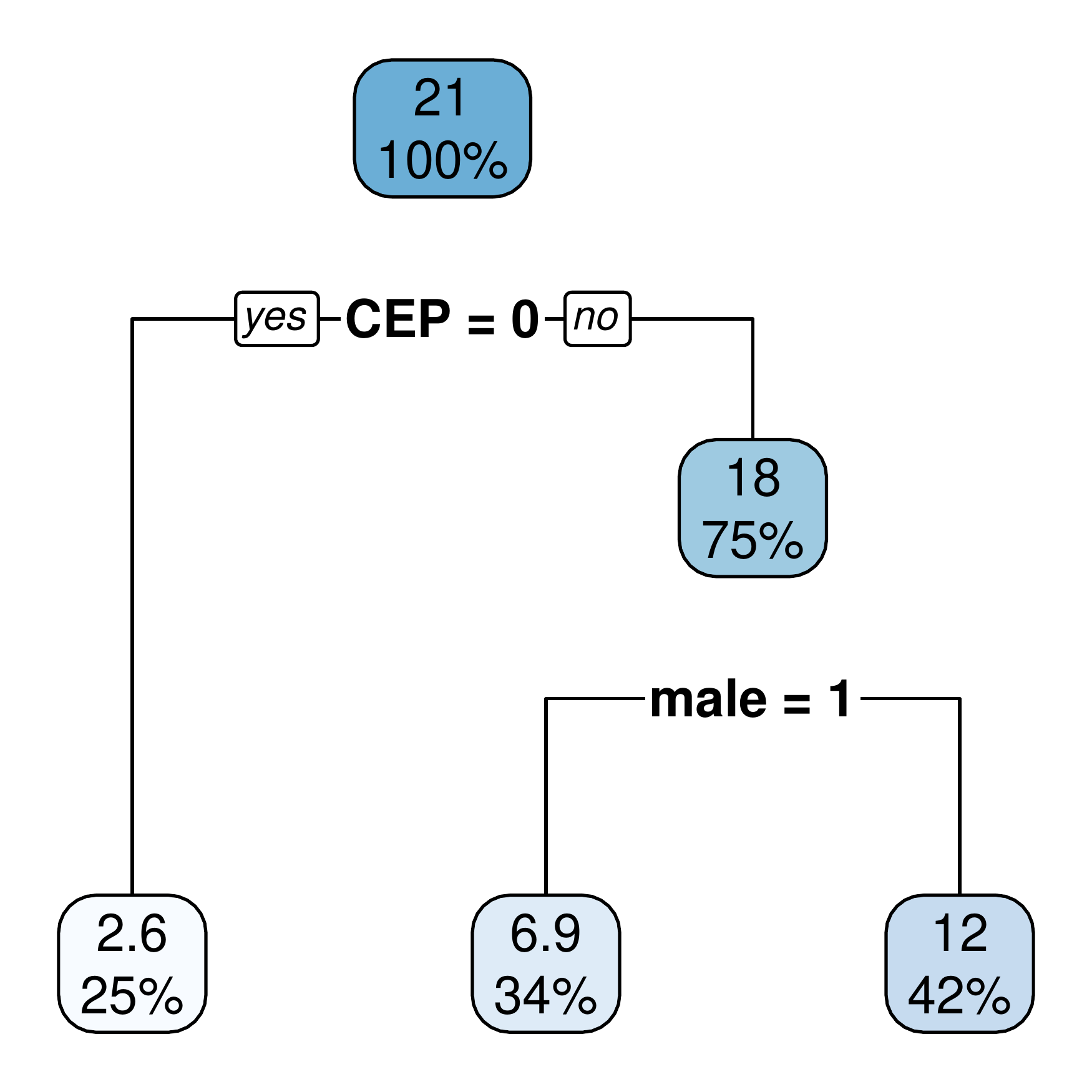}
        \caption[]%
        {{\small JLCT$_1$ tree structure.}}
        %, constructed with $\vXg=\{$\texttt{CEP}, \texttt{male}$\}$.}}
        \label{fig:paquid_tree_inv}
    \end{subfigure}\hfill
    \begin{subfigure}[b]{0.47\textwidth}
        \centering
        \includegraphics[width=\textwidth]{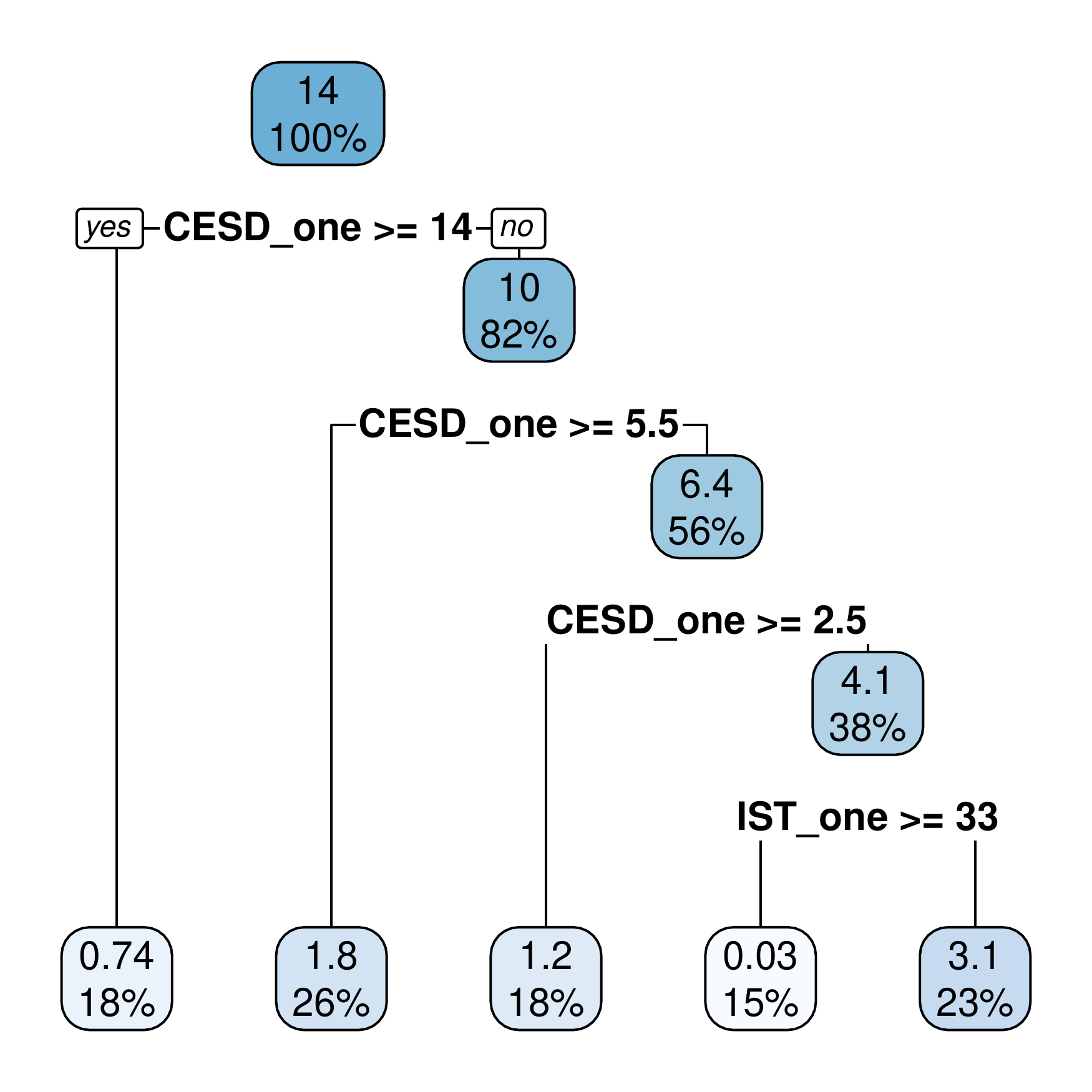}
        \caption[]%
        {{\small JLCT$_2$ tree structure.}}
        %, constructed with
        %$\vXg=\{$\texttt{CEP}, \texttt{male},
        %$\toinv {\texttt{age65}}$, $\toinv {\texttt{BVRT}}$,
        %$\toinv {\texttt{IST}}$, $\toinv {\texttt{HIER}}$,
        %$\toinv {\texttt{CESD}}$ $\}$.}}
        \label{fig:paquid_tree_var1_var1}
    \end{subfigure}
    \vspace\baselineskip
    \begin{subfigure}[b]{0.47\textwidth}
        \centering
        \includegraphics[width=\textwidth]{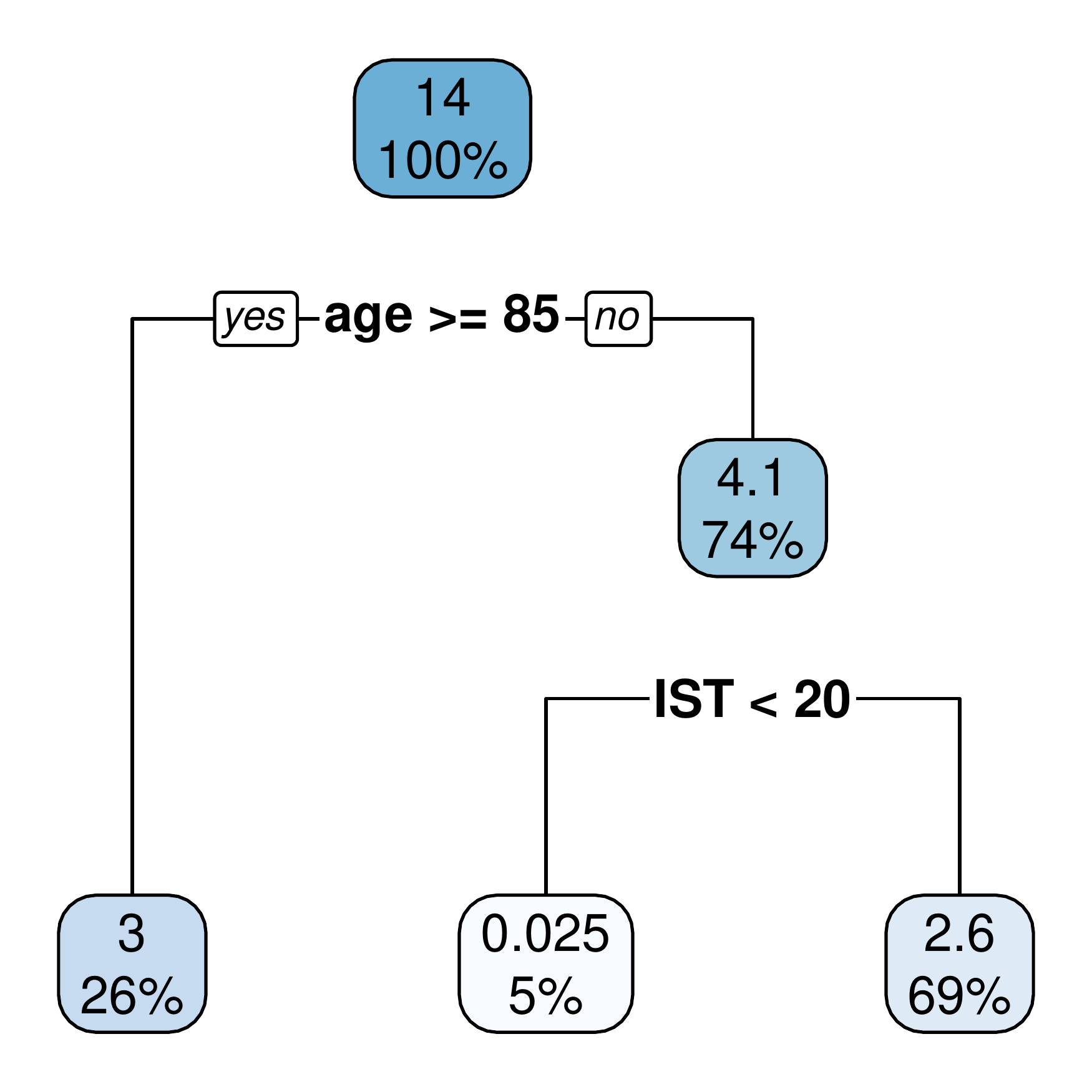}
        \caption[]%
        {{\small JLCT$_3$ tree structure.}}
        %, constructed with
        %$\vXg=\{$\texttt{CEP}, \texttt{male},
        %\texttt{age}, \texttt{BVRT},
        %\texttt{IST}, \texttt{HIER}, \texttt{CESD}
        %$\}$.}}
        \label{fig:paquid_tree_var_var1}
    \end{subfigure}\hfill
    \begin{subfigure}[b]{0.47\textwidth}
        \centering
        \includegraphics[width=\textwidth]{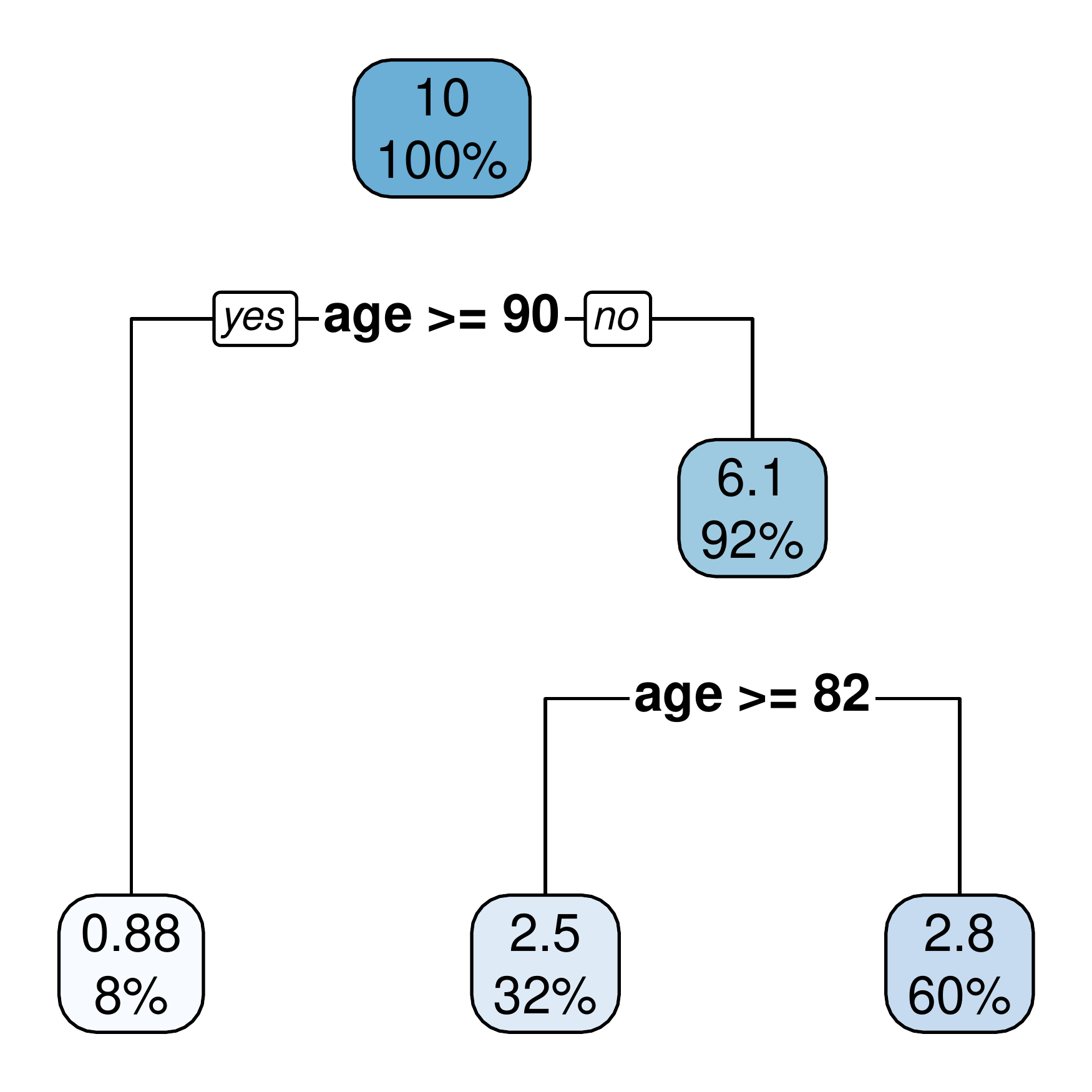}
        \caption[]%
        {{\small JLCT$_4$ tree structure.}}
        %, constructed with
        %$\vXg=\{$\texttt{CEP}, \texttt{male},
        %\texttt{age}, \texttt{BVRT},
        %\texttt{IST}, \texttt{HIER}, \texttt{CESD}
        %$\}$.}}
        \label{fig:paquid_tree_var}
    \end{subfigure}
    \caption{The tree structures returned by JLCT methods.
    JLCT$_1$: constructed with $\vXg=\{$\texttt{CEP}, \texttt{male}$\}$.
    JLCT$_2$: constructed with $\vXg=\{$\texttt{CEP}, \texttt{male},
        $\toinv {\texttt{age65}}$, $\toinv {\texttt{BVRT}}$,
        $\toinv {\texttt{IST}}$, $\toinv {\texttt{HIER}}$,
        $\toinv {\texttt{CESD}}$ $\}$.
    JLCT$_3$ and JLCT$_4$: constructed with $\vXg=\{$\texttt{CEP}, \texttt{male},
        ${\texttt{age65}}$, ${\texttt{BVRT}}$,
        ${\texttt{IST}}$, ${\texttt{HIER}}$,
        ${\texttt{CESD}}$ $\}$.}
\end{figure}
}
\restoregeometry
\advance\linewidth -\rightmargin
\advance\linewidth -\leftmargin

\section{Conclusion}
In this paper, we have proposed a tree-based approach (JLCT) to
model longitudinal outcomes and time-to-event with latent
classes. Simulations and real application on the PAQUID data set show
that JLCT makes full use of the
time-varying information and can demonstrate significant advantages
over JLCM, which can only use a subset of the available time-varying
data. In addition, JLCT is orders of magnitude faster than JLCM
under both scenarios.

Both JLCT and JLCM assume that the time-to-event and the
longitudinal variable are independent given latent class membership. 
It is possible, of course, that subjects come from homogeneous 
latent classes within which the two variables are associated with each other. 
\cite{liu2015joint} proposed such a model, which generalizes JLCM by 
fitting a SREM for the subjects within each latent class. 
This adds more challenges in maximum likelihood fitting 
(some of which are alluded to in the paper), making this model
less likely than JLCM or JLCT to scale to large data sets.

There are several interesting extensions of the JLCT method that could be explored.
The PAQUID data set discussed in \cite{proust2017estimation} and in Section 5
is actually one exhibiting competing risks, since there is a risk of death before
dementia is exhibited, but this was ignored in the analysis.
It would be of interest to generalize JLCT to account for this.
Often survival values are only known to within an interval of time (interval-censoring),
and JLCT could be adapted to that situation as well.
In addition, the analysis here only allows for one longitudinal outcome,
but sometimes several biomarkers are available for a patient,
and it would be useful to generalize JLCT to allow for that.
The PAQUID data is a good example of this: besides the current biomarker \texttt{MMSE},
both \texttt{IST} and \texttt{BVRT} are cognitive tests that can be used as biomarkers as well.
One possible solution for extending JLCT to the scenario of multiple biomarkers
is to perform simultaneous hypothesis tests on all biomarkers
in the tree splitting procedure.

An \texttt{R} package, \texttt{jlctree}, that implements JLCT is available at CRAN.
Appendix~\ref{app:package} provides code illustrating its use.

\newpage
\bibliographystyle{vancouver}
\bibliography{jlctree}

\newpage
\appendix

\section{Joint Latent Class Models (JLCM)}\label{app:jlcm}
In this section,
we give a brief introduction to JLCM, and discuss its strengths and weaknesses.
More details about JLCM can be found in \cite{proust2014joint}.

In JLCM, the latent class membership $g_i \in \{1,\dots,G\}$ for subject $i$
is determined by the set of covariates $\vXg_{i}$,
($\vXg_{it}$ must be time-invariant in JLCM, so we drop the time indicator $t$),
through the following probabilistic model:
\begin{equation*}
    \pi_{ig} = \Pr(g_i=g \vert \vXg_i) =
    \frac{\exp\{\xi_{0g} + \vXg_i \xi_{1g}\}}
    {\sum_{l=1}^G \exp\{\xi_{0l} + \vXg_i \xi_{1l}\}},
\end{equation*}
where $\xi_{0g},\xi_{1g}$ are class-specific intercept and slope parameters for
class $g=1,\cdots,G$.

The longitudinal outcomes in JLCM are assumed to follow a slightly different
linear mixed-effects model than in~\eqref{eq:y}:
\begin{align*}
%    y_{it} \vert_{g_i=g} &= \vXf_{it} \vu_g + \vXr_{it} \vv_{ig} + \e_{it},\\
%    \vv_{ig} & =\vv_i \vert_{g_i=g}  \sim \cN(\boldsymbol{\mu}_g, \boldsymbol{B}_g),
%    \, \varepsilon_{it} \sim \cN(0,\sigma^2),
    y_{it} \vert_{g_i=g} = \vXf_{it} \vbeta_g + \vXr_{it} \vu_{ig} + \e_{it},\quad
    \vu_{ig}  =\vu_i \vert_{g_i=g}  \sim \cN(\boldsymbol{\mu}_g, \boldsymbol{B}_g),
    \quad \varepsilon_{it} \sim \cN(0,\sigma^2),
\end{align*}
where $\vbeta_g$ is the fixed effect vector for class $g$, and $\vu_{ig}$ is the
random effect vector for subject $i$ and class $g$.
The random effect vector $\vu_{ig}$ is independent across latent classes and subjects,
and normally distributed with mean $\boldsymbol{\mu}_g$ and variance-covariance
matrix $\boldsymbol{B}_g$. The errors $\varepsilon_{it}$ are assumed to be
independent and normally distributed with mean $0$ and variance $\sigma^2$,
and independent of all of the random effects as well.
Let $f(\vY_i \vert g_i=g)$ denote the likelihood of longitudinal outcomes $\vY_i$
given that subject $i$ belongs to latent class $g$.

The time-to-event $T_i$ is considered to follow the proportional hazards model
with time-invariant covariates $\vXs$:
\begin{equation}\label{jlcm:survival}
h_i(t \vert g_i=g) = h_{0g}(t;\zeta_g) e^{\vXs_{i} \eta_g},
\end{equation}
where $\zeta_g$ parameterizes the class-specific baseline hazards $h_{0g}$,
and $\eta_g$ is associated with the set of covariates $\vXs_{i}$
(we drop the time indicator $t$ from $\vXs_{it}$ since it must be time-invariant
in JLCM). Let $S_i(t\vert g_i=g)$ denote the survival probability at time $t$
if subject $i$ belongs to latent class $g$.
Note that the extended Cox model~\eqref{jlct:survival} of JLCT extends
the proportional hazards model~\eqref{jlcm:survival} to allow for
time-varying covariates.

%Let $\theta_G=\big(\xi_{0g},\xi_{1g}, \vu_g, \vv_{ig},\boldsymbol{\mu}_g,
%\boldsymbol{B}_g, \sigma,\zeta_g,\eta_g : g={1,\cdots,G},i={1,\cdots,N}\big)$
Let $\theta_G=\big(\xi_{0g}$, $\xi_{1g}$, $\vbeta_g$, $\vu_{ig}$,
$\boldsymbol{\mu}_g$, $\boldsymbol{B}_g$, $\sigma$, $\zeta_g$, $\eta_g$:
$g={1,\cdots,G}$, $i={1,\cdots,N}\big)$
be the entire vector of parameters of JLCM\@.
These parameters are estimated together via maximizing the log-likelihood function
\begin{align*}
     L(\theta_G) =
     \sum_{i=1}^N \log \left(
    \sum_{g=1}^G \pi_{ig} f(\vY_i \vert g_i=g;\theta_G)
    h_i(T_i \vert g_i=g;\theta_G)^{\delta_i}
    S_i(T_i \vert g_i=g;\theta_G) \right).
\end{align*}
The log-likelihood function above uses the assumption that conditioning on the
latent class membership ($g_i$), longitudinal outcomes ($\vY_i$) and time-to-event
($T_i, \delta_i$) are independent.

As mentioned in the introduction, the concept of latent class membership
is of particular interest in clinical studies. JLCM is designed to give
parametric descriptions of subjects' tendency of belonging to each latent class,
and therefore JLCM is a suitable model when the true latent class
is indeed a random outcome with unknown probabilities for each class.
The multinomial logistic regression that JLCM uses is a flexible tool to
model these unknown probabilities.

Despite the usefulness of latent classes, JLCM has several weaknesses.
First of all, the running time of JLCM does not scale well due to its complicated
likelihood function. Simulation results show that the running time of JLCM
increases exponentially fast as a function of the number of observations,
the number of covariates, and the number of assumed latent classes \cite{zhangsimonoff}.
Secondly, the modeling of time-to-event in JLCM is restricted to the use of
time-invariant covariates.
However, time-varying covariates are helpful in modeling the time-to-event,
especially when treatment or important covariates change during the study,
for instance the patient receives a heart transplant \cite{crowley1977covariance},
or the longitudinal CD4 counts change during the study of
AIDS \cite{tsiatis1995modeling}. Research shows that
using time-varying covariates can uncover short-term associations between
time-to-event and covariates \cite{dekker2008survival,kovesdy2007paradoxical},
and ignoring the time-varying nature of the covariates will lead to time-dependent
bias \cite{jongerden2015role,munoz2016handling}.
The other restriction of JLCM is that the latent class membership model only
uses time-invariant covariates, which implies that the latent class membership
of a subject is assumed to be fixed throughout the time of study.
However, the stage of a disease of a patient is very likely to change during the
course of clinical study, for instance the disease would move from its early stages
to its peak, and then move to its resolution.  When the goal of joint modeling is to
uncover meaningful clustering of the population that leads to definitions of
disease stages, it is important to allow time-varying covariates in the latent class
membership model, so that the model reflects this real world situation.

Although JLCM does not allow time-varying covariates in fitting time-to-event
outcomes and latent class memberships, its implementation in the \texttt{R} package \texttt{lcmm}
can take in time-varying covariates. In that case,
\texttt{lcmm} will automatically take the first encountered values of those covariates
of each subject and use them as if these covariates are time-invariant.

\section{Convert data into the left-truncated right-censored (LTRC) format}\label{app:ltrc}
Here we describe how to convert the original survival data
with time-varying covariates into the left-truncated right-censored (LTRC) format.

For each subject $i$ and measurement time $t$, there is a ``pseudo-observation''
with $y_{it}, \vX_{it}$, and a time-to-event triplet $(L_{it}, R_{it}, \delta_{it})$,
where $L_{it} $ is the left-truncated time, $R_{it}$ is the right-censored time,
and $\delta_{it}$ is the censor indicator. Table~\ref{tb:eg_ltrc}
illustrates how to convert the time-to-event data (Time, Death) with a time-varying covariate (CD4)
into the LTRC format. The left chart shows that the subject with $\text{ID}=1$
is observed at times 0, 10, and 20, respectively, with Age and CD4 recorded,
and the death occurred at time 27. The right chart
shows the corresponding LTRC format, where each pseudo-observation
is left-truncated (at time $L$) and right-censored (at time $R$), with
the corresponding covariates (Age, CD4) and censor indicator $\delta$.

\begin{table}
    \caption{Converting the original data (left chart) into the left-truncated
    right-censored (LTRC) format (right chart).}
    \label{tb:eg_ltrc}
    \parbox{.45\linewidth}{
        \centering
        \begin{tabular}{ccccc} \toprule
            ID  &Age    &CD4    &Time   &Death($\delta$)\\ \midrule
            1   &45     &27     &0      &0\\
            1   &45     &31     &10     &0\\
            1   &45     &25     &20     &0\\
            1   &-      &-      &27     &1\\\bottomrule
        \end{tabular}
    }
    $\Rightarrow$
    \parbox{.5\linewidth}{
        \centering
        \begin{tabular}{cccccc} \toprule
            ID  &Age    &CD4    &$L$    &$R$     &Death($\delta$)\\ \midrule
%            ID  &Age    &CD4    &Left-truncated &Right-censored &Death($\delta$)\\
%                &       &       &Time ($L$)       &Time ($R$)       &\\ \hline
            1   &45     &27     &0      &10 &0\\
            1   &45     &31     &10     &20 &0\\
            1   &45     &25     &20     &27 &1\\\bottomrule
        \end{tabular}
    }
\end{table}

\section{Simulation setup: time-invariant covariates}\label{app:sim_time_inv}
In this section we give details of the data generating scheme in the simulation study
of time-invariant covariates only.

\subsection{Data}
At each simulation run, for each subject $i$ we randomly generate five independent,
time-invariant covariates $X_{i1}, \dots, X_{i5}$.
We assume there are four latent classes $g=1,\dots,4$, which are determined by
covariates $X_1, X_2$. Once the latent classes are determined for each subject $i$,
the time-to-event and the longitudinal outcomes are conditionally independent
given the latent classes, and therefore generated separately.
In particular, the survival outcomes (time-to-event) depend on $X_3,X_4,X_5$,
and the longitudinal outcomes depend on the latent classes. We give more details below.

\paragraph{Covariates}
At each simulation run, for each subject $i$ we draw $X_{i1}$, $X_{i2}$, $X_{i3}$, $X_{i4}$, $X_{i5}$
uniformly from $[0,1]$, $[0,1]$, $\{0,1\}$, $[0,1]$, and $\{1,2,3,4,5\}$ respectively.

\paragraph{Latent classes} \label{app:sim_time_inv_latent}
We determine class membership based on a multinomial logistic model of
$X_1, X_2$, with increasing level of concentration on one class.
%We give brief description of latent class constructions here,
%and the full details are give in Appendix~\ref{app:simulation}.
Our latent class membership generation model matches the setup of JLCM,
and it approaches the setup of JLCT as the concentration level approaches $1$.

For subject $i$, we compute the value of a ``score'' function $f(w_g, X_{i1},X_{i2})$,
where $w_g$ denotes the parameters associated with latent class $g$,
and $X_{i1},X_{i2}$ denote the first two covariates of subject $i$.
The latent class membership for subject $i$, $g_i$, is generated according to
two key values: the ``majority'' class $g_i^0$, and the ``concentration'' level $p_0$.
We define the ``majority'' class as the latent class with largest score for sample $i$,
$$g_i^0 = \argmax_{g\in\{1,2,3,4\}} f (w_g, X_{i1},X_{i2}).$$
We consider four types of score functions $f$, which correspond to four
underlying structures of latent classes: tree partition, linear separation,
non-linear separation, and asymmetric tree partition.  The structure is reflected
by the dependency of $g_i^0$ on $X_1,X_2$, which is shown in Figure~\ref{fig:struct}.

The latent class membership of subject $i$ is drawn according to the probabilities
\begin{equation*}
    \Pr (g_i=g\mid X_i,C) = \frac{\exp\big\{C f(w_g, X_{i1},X_{i2})\big\}}
    {\sum_{l=1}^4 \exp \big\{C f(w_l, X_{i1},X_{i2})\big\}},
\end{equation*}
where the parameter $C$ is chosen such that the probability of ``major'' class
is approximately equal to a pre-specified concentration level
$p_0 \in \{0.25, 0.5,0.7, 0.85, 1\}$. That is $\Pr(g_i = g_i^0 \mid X_i, C)\approx p_0$.
In particular, when $C=0$, $\Pr(g_i=g_i^0)=0.25$, and therefore the latent class
membership is randomly determined and independent of $X_{1},X_{2}$.
On the other hand, when $C=\infty$, $\Pr(g_i=g_i^0)=1$, and therefore the latent class
membership corresponds to a deterministic partitioning based on $X_1,X_2$,
which is consistent with the assumptions underlying a tree partitioning.
The choices of parameters $w_g$ for each latent class structure are given below.
\begin{itemize}
    \item Tree.  Consider the following coefficients $(w_{g1},w_{g2})$:
        \begin{align*}
            & w_{11}=-1  ,\ w_{12}=-1 \\
            & w_{21}=-1,\ w_{22}=1, \\
            & w_{31}=1,\ w_{32}=-1, \\
            & w_{41}=1,\ w_{42}=1.
        \end{align*}
        Define the score function
        \begin{align*}
            & f_{\text{tree}}(w_g,X_{i1},X_{i2}) =w_{g1} (2X_{i1}-1) + w_{g2}(2X_{i2}-1),
        \end{align*}
        and let $ g_i^0 =  \argmax_{g} f_{\text{tree}} (w_g, X_{i1},X_{i2})$ denote
        the latent class with largest score for sample $i$.
        See Figure~\ref{fig:tree} for the dependency of $g_i^0$ on $X_1,X_2$.
        The latent classes are drawn according to the probabilities
        \begin{equation*}
        \Pr (g_i=g\mid X_i,C) = \frac{\exp\big\{C f_{\text{tree}}
        (w_g, X_{i1},X_{i2})\big\}}
        {\sum_{l=1}^4 \exp \big\{C f_{\text{tree}}(w_l, X_{i1},X_{i2})\big\}},
        \end{equation*}
        where the parameter $C$ is chosen such that
        $\Pr(g_i = g_i^0 \mid X_i, C)\approx p_0 $,
        and $p_0$ is some pre-specified level $p_0 \in \{0.25, 0.5,0.7, 0.85, 1\}$.
        In particular, when $C=0$, $\Pr(g_i=g_i^0)=0.25$; when $C=\infty$,
        $\Pr(g_i=g_i^0)=1$.
    \item Linear. Consider the following coefficients $(w_{g1},w_{g2})$
        drawn randomly from the unit sphere,
        \begin{align*}
            & w_{11}=0.8  ,\ w_{12}=-0.6 \\
            & w_{21}=0.9,\ w_{22}=0.5, \\
            & w_{31}=-0.8,\ w_{32}=0.6, \\
            & w_{41}=0.5,\ w_{42}=0.9.
        \end{align*}
        Define the score function
        \begin{align*}
            f_{\text{linear}}(w_g,X_{i1},X_{i2})  = w_{g1}(2X_{i1} -1) + w_{g2}(2X_{i2}-1),
        \end{align*}
        and let $ g_i^0 =  \argmax_{g} f_{\text{linear}} (w_g, X_{i1},X_{i2})$.
        See Figure~\ref{fig:linear} for the dependency of $g_i^0$ on $X_1,X_2$.
        The latent classes are drawn according to the following probabilities
        $$\Pr (g_i=g\mid X_i,C) = \frac{
            \exp\big\{Cf_{\text{linear}}(w_g, X_{i1},X_{i2})\big\}}
        {\sum_{l=1}^4 \exp \big\{Cf_{\text{linear}}(w_l, X_{i1},X_{i2})\big\}},
        $$
        with $C$ again chosen to control the value of
        $\Pr(g_i = g_i^0 \mid X_i, C)\approx p_0 \in \{0.25,0.5,0.7,0.85,1\}$.
    \item Nonlinear.  We can skip the step of defining the score function $f$
        and the $C$ value, but directly work with $g_0$ and $p_0$.
        For each observation,  its  ``most likely'' latent class $g^0_i$
        is determined by whether $(X_{i1}, X_{i2})$ belongs to the circles centered
        at $(0,0)$ and $(0,1)$ with radius $0.75$ :
        \begin{align*}
            g_i^0 = \begin{cases}
                1,\ \left\{X_{i1}^2 + X_{i2}^2 \leq 0.75^2\right\} \,\&\,
                \left\{X_{i1}^2 + (1-X_{i2})^2 > 0.75^2\right\} \\
                2, \ \left\{X_{i1}^2+(1-X_{i2})^2  \leq 0.75^2 \right\} \,\&\,
                \left\{X_{i1}^2 + X_{i2}^2 > 0.75^2\right\} \\
                3, \ \left\{X_{i1}^2 + X_{i2}^2 > 0.75^2\right\} \,\&\,
                \left\{X_{i1}^2 + (1-X_{i2})^2 > 0.75^2\right\} \\
                4, \ \left\{X_{i1}^2 + X_{i2}^2 \leq 0.75^2\right\} \,\&\,
                \left\{X_{i1}^2 + (1-X_{i2})^2 \leq 0.75^2\right\} \\
            \end{cases}
        \end{align*}

        See Figure~\ref{fig:nonlinear} for visualization of $g_i^0$.
        The latent classes are drawn according to the following probabilities:
        $$\Pr (g_i=g\mid X_i) = p_0 \{g=g^0_i\} + \frac{1-p_0}{3} \{g\ne g^0_i\}, $$
        where $p_0 \in \{0.25,0.5,0.7,0.85,1\}$.
    \item Asymmetric. We can skip the step of defining the score function $f$
        and the $C$ value, but directly work with $g_0$ and $p_0$.
        For each observation,  its ``most likely''  latent class $g_i^0$
        is determined by the following asymmetric tree:
        \begin{align*}
            g_i^0 = \begin{cases}
                1,\ \{X_{i1} > 0.75\}, \\
                2, \ \{X_{i1} \leq 0.75\}\, \&\, \{X_{i2} \leq 0.33\},\\
                3, \ \{X_{i1} \leq 0.75\}\, \&\, \{0.33 < X_{i2} \leq 0.67\},\\
                4, \ \{X_{i1} \leq 0.75\}\, \&\, \{X_{i2} > 0.67\}.
            \end{cases}
        \end{align*}
        See Figure~\ref{fig:asym}.
        The latent classes are drawn according to the following probabilities:
        $$\Pr (g_i=g\mid X_i) = p_0 \{g=g^0_i\} + \frac{1-p_0}{3} \{g\ne g^0_i\}, $$
        where $p_0 \in \{0.25,0.5,0.7,0.85,1\}$.
\end{itemize}

\paragraph{Time-to-event}
The survival time (time-to-event) $T_i$ of subject $i$ follows the
proportional hazards model
\begin{align*}
    h(t,\vX_i) = h_0(t) e^{b_{g_i3}X_{i3} + b_{g_i4}X_{i4} + b_{g_i5}X_{i5}},
\end{align*}
where the slope coefficients $b_{g_i3},b_{g_i4},b_{g_i5}$ depend on
latent class $g_{i} \in \{1,2,3,4\}$ for subject $i$.

We use three different distributions for baseline hazards $h_0(t)$:
exponential, Weibull with decreasing hazards with time (Weibull-D),
and Weibull with increasing hazards with time (Weibull-I).
We select parameters and slopes such that the mean values of survival time $T$ across
latent classes remain similar across different distributions.
The distributions and corresponding parameters for generating time-to-event data
are listed below.
\begin{itemize}
    \item Exponential with $\lambda=0.1$, and slopes are
        \begin{align*}
            b_{13}&=0, \ b_{14}=0,\ b_{15}=0 \\
            b_{23}&=0.56, \ b_{24}=0.56,\ b_{25}=0.09, \\
            b_{33}&=0.92, \ b_{34}=0.92,\ b_{35}=0.15, \\
            b_{43}&=1.46, \ b_{44}=1.46,\ b_{45}=0.24.
        \end{align*}
    \item Weibull distribution with shape parameter $\alpha=0.9$,
        which corresponds to decreasing hazards with time (Weibull-D).
        The scale parameter is $\beta=1$, and slopes are
        \begin{align*}
            b_{13}&=-1.17, \ b_{14}=-1.17,\ b_{15}=-0.19 \\
            b_{23}&=-0.66, \ b_{24}=-0.66,\ b_{25}=-0.11, \\
            b_{33}&=-0.55, \  b_{34}=-0.55,\ b_{35}=-0.09, \\
            b_{43}&=0, \ b_{44}=0,\ b_{45}=0.
        \end{align*}
    \item Weibull distribution with shape parameter $\alpha=3$,
        which corresponds to increasing hazards with time (Weibull-I).
        The scale parameter is $\beta=2$, and slopes are
        \begin{align*}
            b_{13}&=-3.22, \ b_{14}=-3.22,\ b_{15}=-0.54 \\
            b_{23}&=-2.26, \ b_{24}=-2.26,\ b_{25}=-0.38, \\
            b_{33}&=-1.53, \ b_{34}=-1.53,\ b_{35}=-0.26, \\
            b_{43}&=0, \ b_{44}=0,\ b_{45}=0.
        \end{align*}
    %\item Log-normal distribution. The slopes are all $0$, and the location parameter
    %    $\mu$ and scale parameter $\sigma$ for each latent class are:
    %    \begin{align*}
    %        \mu_{1}&=2,  \sigma_{1} =0.3, \\
    %        \mu_{2}&=1,7,  \sigma_{2} =0.2,\\
    %        \mu_{3}&=1.3,  \sigma_{3} =0.3,\\
    %        \mu_{4}&=0.5,  \sigma_{4} =0.5.
    %    \end{align*}
\end{itemize}

Left truncation times are generated independently from uniform $U[0,1]$.
Right censoring times are generated independently from an exponential distribution,
with parameters chosen to reflect light censoring
(approximately 20\% observations are censored),
and heavy censoring  (approximately 50\% observations are censored).

\paragraph{Longitudinal outcomes}
The longitudinal outcome $y$ comes from the following linear mixed-effects model:
for subject $i$ at time $t$, let $g_{i}$ denote the latent class membership, and
\begin{align*}
    y_{it} &=  u_{g_{i}}   + v_i +  \e_{it} , \quad
    v_i \sim \cN(0, \sigma_v^2), \quad
    \e_{it} \sim \cN(0,\sigma_e^2),
\end{align*}
where $\sigma_v=0.2,\sigma_e=0.1$, and $u_1=0,\ u_2=1,\ u_3=1,\ u_4=2$
are class-specific random intercepts.
We assume each subject $i$ is measured at its entry (left truncation) time,
together with multiple intermediate measurement times $t_{i1},t_{i2}$, etc.
The number of intermediate measurements for subject $i$ is generated independently
from $1+\text{Poisson}(1)$, thus each subject has at least 2 measurements,
and has 3 measurements on average.  The intermediate measurement time $t_{ij}$
is then sampled independently and uniformly between subject $i$'s left-truncated
and right-censored time, $t_{ij} \sim U[L_i,T_i]$.
Finally, the data are converted to the LTRC format according to Appendix~\ref{app:ltrc}.

Observe that covariates $X_1$ and $X_2$ determine the latent classes,
and thus affect the time-to-event and longitudinal outcomes $y$.
Therefore, time-to-event is correlated with $y$ if $X_1$ and $X_2$ are unknown.
On the other hand, conditioning on one of the four latent classes $g=1,2,3,4$,
time-to-event and longitudinal outcomes are independent:
the former follows a class-specific proportional hazards model and only depends
on $X_3, X_4, X_5$,  while the latter is a constant plus random noise.
Therefore, the simulated data satisfy the conditional independence assumption
made by both JLCM and JLCT.

\subsection{Methods} \label{app:methods_time_inv}
JLCT and JLCM use the same subsets of covariates as in the time-varying case:
$\vXs=\{X_3,X_{4},X_{5}\}$,
$\vXf=\vXr=\{X_1,\dots,X_5\}$,
${\vXg}^{(\text{JLCT})} = \{X_1,\dots,X_5\}$,
and ${\vXg}^{(\text{JLCM})}=\{X_1,\dots,X_5, X_1X_2\}$.

We compare four methods on the simulated data set.
\begin{enumerate}
    \item A baseline JLCT with no splitting. 
    \item A full JLCT model with default splitting and stopping criterion. In the second step, we fit a Cox proportional hazards (PH) model
        with the same baseline hazard function but different slopes across terminal nodes.
    \item A full JLCT model with default splitting and stopping criterion. In the second step, we fit a Cox PH model where
        both baseline hazard functions and Cox PH slopes differ across terminal nodes.
    \item A JLCM model. Since all the simulated covariates are time-invariant, we can fit JLCM using 
        the original data.
\end{enumerate}

\subsection{Predictions}
\paragraph{JLCT prediction}
The prediction procedure for JLCT is as follows.  
Once JLCT returns a tree,
each subject $i$ is assigned to a tree leaf node $d_i$.
We fit the proportional hazards model~\eqref{jlct:survival}
to the time-to-event data $(T_i,\delta_i)$,
with time-invariant covariates $\vXs_i=\{X_{i3},X_{i4},X_{i5}\}$
and slopes $\eta_{d_i} =(b_{d_i 3},b_{d_i4},b_{d_i5})$ for $d_i \in \{1,2,3,4,5,6\}$
(since there are no more than six terminal nodes).
Method 2 assumes a shared baseline hazard function $h_0(t)$,
meanwhile Method 3 assumes class-specific baseline hazard functions $h_{d_i0}(t)$.
%We fit a Cox PH model with class-specific slope coefficients $b_{d3},b_{d4},b_{d5}$,
%and a shared baseline hazard function $h_0(t)$ (Method 2), or
%class-specific baseline hazard functions $h_{d0}(t)$ (Method 3):
%\begin{align*}
%    h(t,\vX) &= h_{0}(t) e^{b_{d3} X_3  + b_{d4} X_4  + b_{d5} X_5 }, \tag{Method 2}\\
%    h(t,\vX) &= h_{d0}(t) e^{b_{d3} X_3  + b_{d4} X_4  + b_{d5} X_5 }. \tag{Method 3}
%\end{align*}
We use the \texttt{R} function \texttt{coxph} from the \texttt{survival} package \cite{survival-package,survival-book}
to get fitted slopes
$\h b_{d3}, \h b_{d4}, \h b_{d5}$ for all $d\in\{1,2,3,4,5,6\}$.
Given the fitted model, we compute the predicted survival
probability for subject $i$ at time $t$, denoted $\h S_i(t)$,
using the \texttt{R} function \texttt{survfit.coxph}.
For longitudinal outcomes, we fit the linear mixed-effects model~\eqref{eq:y}
using the \texttt{R} function \texttt{lmer} from the \texttt{lme4} package \cite{bates2015fitting},
and compute the predicted longitudinal outcomes $\h y_{it}$.
%Finally, we fit the following linear mixed-effects model on the longitudinal outcomes,
%using the \texttt{R} function \texttt{lme4}:
%\begin{align*}
%    y_{it}  & = \beta_1 X_{i1} + \beta_2 X_{i2} +
%    \beta_3 X_{i3}+ \beta_4 X_{i4} + \beta_5 X_{i5}
%     + u_{d_i} + v_{i} + \e_{it},\\
%    u_{d_i} & \sim \cN(0,\sigma_u^2),\ v_i \sim \cN(0,\sigma_v^2), \,
%    \e_{it} \sim \cN(0,\sigma_\e^2),
% \end{align*}
% where $\beta_1,\cdots,\beta_5$ are the fixed effects coefficients,
% $u_d$ and $v_i$ are class-specific and subject-specific random effects coefficients.
% Given the fitted model, we can compute the predicted longitudinal outcomes $\h y_{it}$.
 For any out-of-sample subject $k$, we first determine its leaf node assignment
 $d_{k}$ according to its covariates $\vXg_{k}$ and the constructed
 tree of JLCT\@. Then we proceed to compute predictions $\h S_k(t)$ and $\h y_{kt}$
 as we did for the in-sample subjects.

\paragraph{JLCM prediction}
 The prediction procedure for JLCM is very similar to that of JLCT\@.
 Let $D^*$ be the BIC optimal number of latent classes.
 For each latent class $d\in \{1,\dots,D^*\}$,
 JLCM returns estimated Cox PH coefficients  $\h b_{d3}, \h b_{d4}, \h b_{d5}$
 as well as the baseline survival curves $\h S_{d0}(t)$.  In addition,
 JLCM returns a fitted linear mixed-effects model for longitudinal outcomes.
 For in-sample subjects, JLCM also returns a predicted latent class membership $d_i$
 for each subject $i$, conditioning on all available information:
 covariates $\vX_i$, time-to-event $(T_i,\delta_i)$, and longitudinal outcomes $\vY_i$.
 We can therefore use the estimated parameters for class $d_i$ to compute
 $\h S_{i}(t) \vert_{d_i}$ and $\h y_{it}\vert_{d_i}$.
 Note that JLCM uses much more information than JLCT at the prediction time
 to determine the latent class memberships for in-sample subjects,
 and thus JLCM is expected to perform better than JLCT on in-sample data.
 For any new subject $k$, however, time-to-event and longitudinal outcomes
 are no longer available at prediction time. Instead, we use JLCM's
 fitted multinomial logistic model to get $\pi_{k}  = (p_{k1}, \dots, p_{kD^*})$,
 a vector of predicted probabilities of subject $k$ belonging to each latent class.
 Finally we take weighted averages over all classes as final predictions:
 \begin{align*}
     \h S_k(t) & = \sum_{d=1}^{D^*} p_{kd} \ \h S_{k}(t)\vert_{d},\quad
     \h y_k(t) = \sum_{d=1}^{D^*} p_{kd} \ \h y_{kt}\vert_{d}.
\end{align*}

\section{Simulation setup: time-varying covariates}\label{app:sim_time_var}
In this section we give details of the data generating scheme in the simulation study
of time-varying covariates.

\paragraph{Covariates}
At each simulation run, for each subject $i$ we randomly generate five independent,
time-varying covariates $X_{it1},\dots,X_{it5}$, which are piecewise constant, and change values at time $t_i$.
To be more concrete, for each subject $i$,
we first randomly generate a time point $t_i \sim U[1,3]$.
Next, draw $X_{i1}^\prime$, $X_{i2}^\prime$, $X_{i3}^\prime$, $X_{i4}^\prime$,
$X_{i5}^\prime$ uniformly from $[0,1]$, $[0,1]$, $\{0,1\}$, $[0,1]$,
and $\{1,2,3,4,5\}$ respectively.
Generate $X_{i1}^{\prime\prime}$, $X_{i2}^{\prime\prime}$, $X_{i3}^{\prime\prime}$,
$X_{i4}^{\prime\prime}$, $X_{i5}^{\prime\prime}$ by
\begin{align*}
    X_{i1}^{\prime\prime} &= \Proj_{[0,1]} \Big(X_{i1}^\prime + U[-0.3, 0.3]\Big), \\
    X_{i2}^{\prime\prime} &= \Proj_{[0,1]} \Big(X_{i2}^\prime + U[-0.3, 0.3]\Big),\\
    X_{i3}^{\prime\prime} &= U\{0,1\},\\
    X_{i4}^{\prime\prime} &= \Proj_{[0,1]} \Big(X_{i4}^\prime + U[-0.3, 0.3]\Big), \\
    X_{i5}^{\prime\prime} &= \Proj_{[1,5]} \Big(X_{i5}^\prime + U\{-1,0,1\}\Big),
\end{align*}
where the projection operator \Proj$_{[a,b]}(x)$ projects any real value $x$ onto the
interval $[a,b]$. For example, \Proj$_{[0,1]}(1.3)=1$.
For $k\in \{1,2,3,4,5\}$, the time-varying covariate
$X_{k}$ of object $i$ at time $t$ is defined as
\begin{align*}
    X_{itk} &=1\{t\leq t_i\} X_{ik}^\prime+1\{t>t_i\}X_{ik}^{\prime\prime}.
\end{align*}
Therefore, for each subject $i$, the five time-varying covariates
$X_1,\cdots,X_5$ are piecewise constant, and change values at time point $t_i$.

\paragraph{Latent classes}
We use the same procedure as is described in
Appendix~\ref{app:sim_time_inv} to generate latent classes,
the only difference being that the covariates $X_1, X_2$ become time-varying,
as does the latent class membership.
We denote by $g_{it}$ the latent class membership of subject $i$ at time $t$.

\paragraph{Time-to-event}
The time-to-event data follow the same hazard models as in
Appendix~\ref{app:sim_time_inv}, except that the covariates
and slope coefficients are time-varying.
The hazard of subject $i$ at time $t$ becomes
\begin{align*}
    h(t,\vX_i) = h_0(t) e^{b_{g_{it}3}X_{it3} + b_{g_{it}4}X_{it4}
    + b_{g_{it}5}X_{it5} },
\end{align*}
where the slope coefficients $b_{g_{it}3},b_{g_{it}4},b_{g_{it}5}$ depend on
latent class membership $g_{it} \in \{1,2,3,4\}$ at time $t$.

\paragraph{Longitudinal outcomes}
The longitudinal outcome $y$ follows the same
linear mixed-effects model as in Appendix~\ref{app:sim_time_inv},
the only difference being now the fixed effect $u_{g_{it}}$,
which depends on the latent class membership $g_{it}$, becomes time-varying as well.

\section{Simulation results: time-invariant covariates}
\label{app:moreresults_time_inv}
In this section, we present the complete simulation results:
for three time-to-event distributions (Weibull-I, Weibull-D, and Exponential),
and light censoring.
The results of other censoring levels (no censoring, heavy censoring)
are very similar to those of light centering, and are omitted.
We report results for the four methods discussed in Appendix~\ref{app:methods_time_inv}:
\begin{enumerate}
\item JLCT with no split,
\item JLCT with same baseline hazard function but different Cox PH slopes across terminal nodes,
\item JLCT with different baseline hazard function and different Cox PH slopes across terminal nodes, 
\item JLCM with different baseline hazard function and different Cox PH slopes across terminal nodes,
\end{enumerate}
We use the performance measures described in Section~\ref{sec:evaluation}:
ISE$_{\text{in}}$, ISE$_{\text{out}}$, MSE$_{y\text{in}}$,
MSE$_{y\text{out}}$, and MSE$_b$. The results are given below.

\begin{table}
    \caption{The average running time in seconds (standard deviation in parentheses)
        on a data set with time-invariant covariates, $N=500$,
    light censoring, and Weibull-I distribution. }
    \label{tb:simtime_time_inv}
    \centering
    \begin{tabular}{llll}\toprule
        Structure   &$p_0$  & JLCT   & JLCM \\\midrule
        \multirow{5}{*}{Tree}
        &0.5    &14.79\ (0.41)  &1759.23\ (354.93)\\
        &0.7    &14.12\ (0.27)  &2146.55\ (423.37)\\
        &0.85   &13.91\ (0.55)  &1783.7\ (432.72)\\
        &1      &9.25\ (0.5)    &1580.72\ (333.41)\\\midrule
        \multirow{5}{*}{Linear}
        &0.5    &23.34\ (0.59)  &1609.96\ (297.82)\\
        &0.7    &22.36\ (0.75)  &2045.23\ (402.44)\\
        &0.85   &21.51\ (0.99)  &1752.89\ (476.53)\\
        &1      &20.91\ (1.41)  &1694.08\ (389.78)\\\midrule
        \multirow{5}{*}{Nonlinear}
        &0.5    &14.74\ (0.71)  &1489.01\ (323.39)\\
        &0.7    &14.18\ (0.34)  &1522.28\ (293.78)\\
        &0.85   &13.74\ (0.39)  &1597.38\ (285.09)\\
        &1      &12.52\ (0.87)  &2985.69\ (572.36)\\\midrule
        \multirow{5}{*}{Asymmetric}
        &0.5    &15.07\ (3.44)  &1494.52\ (384.03)\\
        &0.7    &13.96\ (0.39)  &1550.19\ (421.94)\\
        &0.85   &13.35\ (0.72)  &1719.99\ (418.76)\\
        &1      &9.98\ (0.59)   &2559.32\ (696.39)\\\bottomrule
    \end{tabular}
\end{table}
%--------------------
%\subsection{Integrated squared error of time-to-event predictions.}
\begin{figure}[h!]
    \vspace*{-0.5cm}
    \makebox[\linewidth][c]{%
    \begin{subfigure}[b]{0.5\textwidth}
        \centering
        \includegraphics[width=\textwidth,height=0.3\textheight,trim=10 8 5 5]{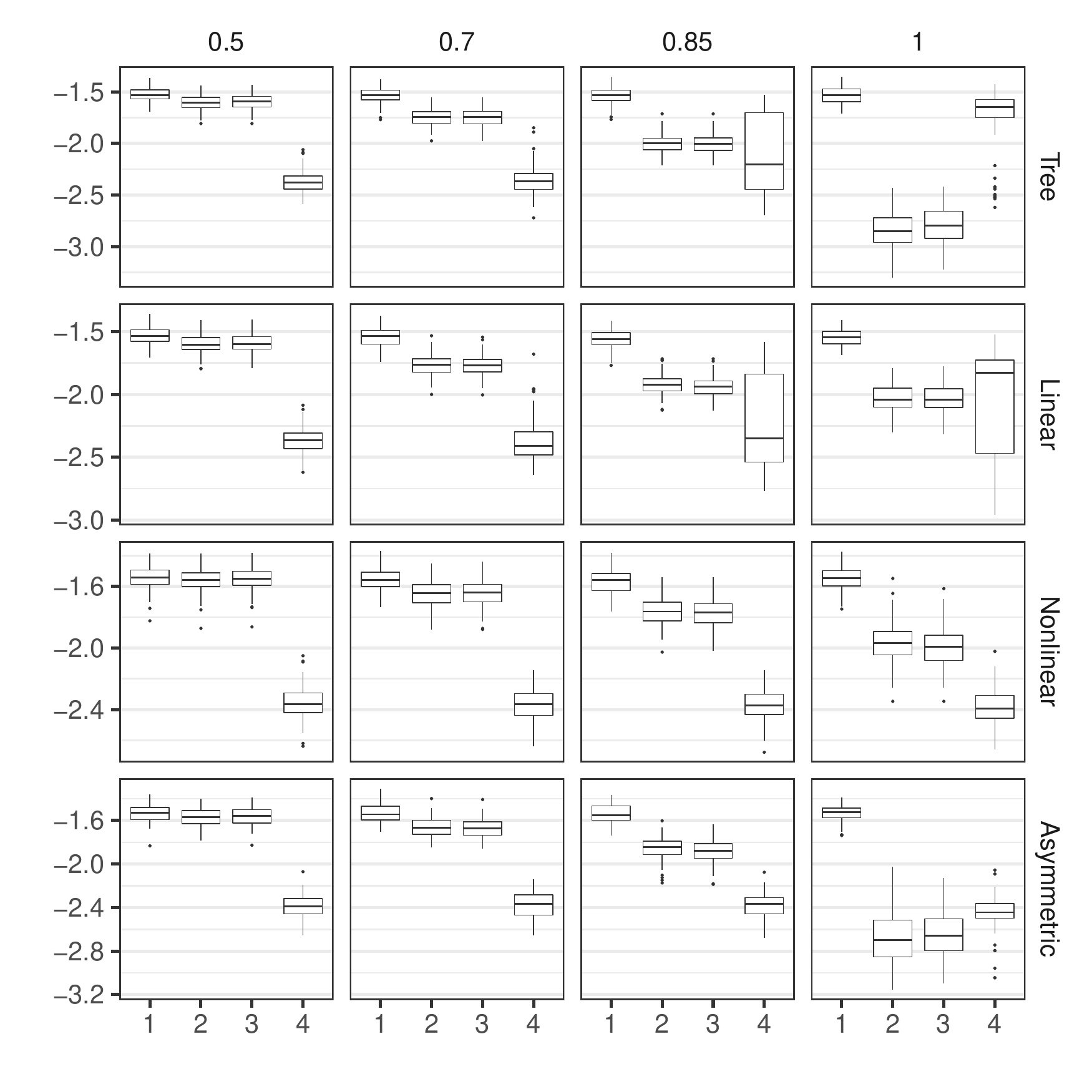}
        \caption[]{{\small Weibull-I, ISE$_{\text{in}}$}}
    \end{subfigure}\hfill
    \begin{subfigure}[b]{0.5\textwidth}  
        \centering 
        \includegraphics[width=\textwidth,height=0.3\textheight,trim=10 8 5 5]{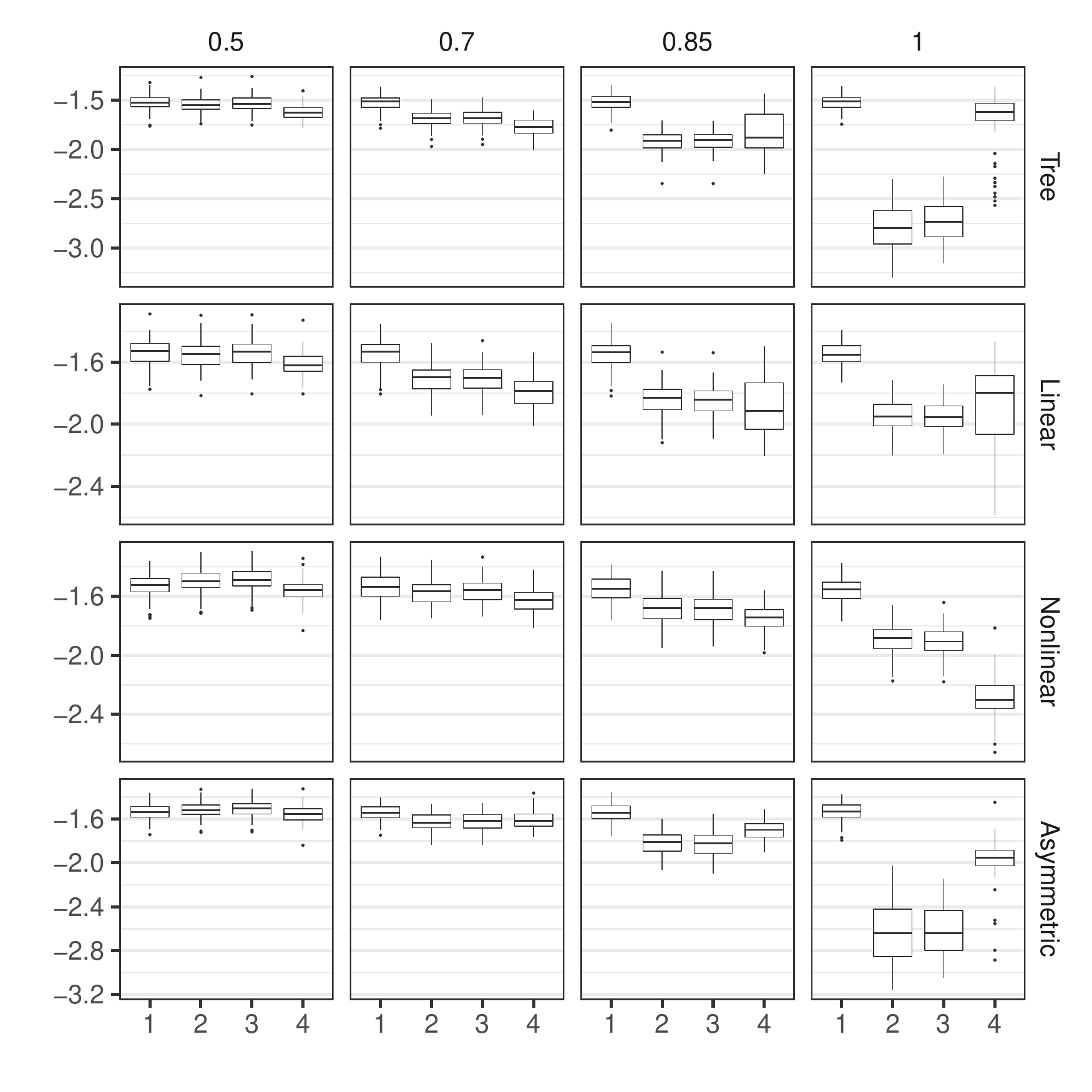}
        \caption[]{{\small Weibull-I, ISE$_{\text{out}}$}}
    \end{subfigure}
}
%    \vspace\baselineskip
    \makebox[\linewidth][c]{%
    \begin{subfigure}[b]{0.5\textwidth}  
        \centering 
        \includegraphics[width=\textwidth,height=0.3\textheight,trim=10 8 5 5]{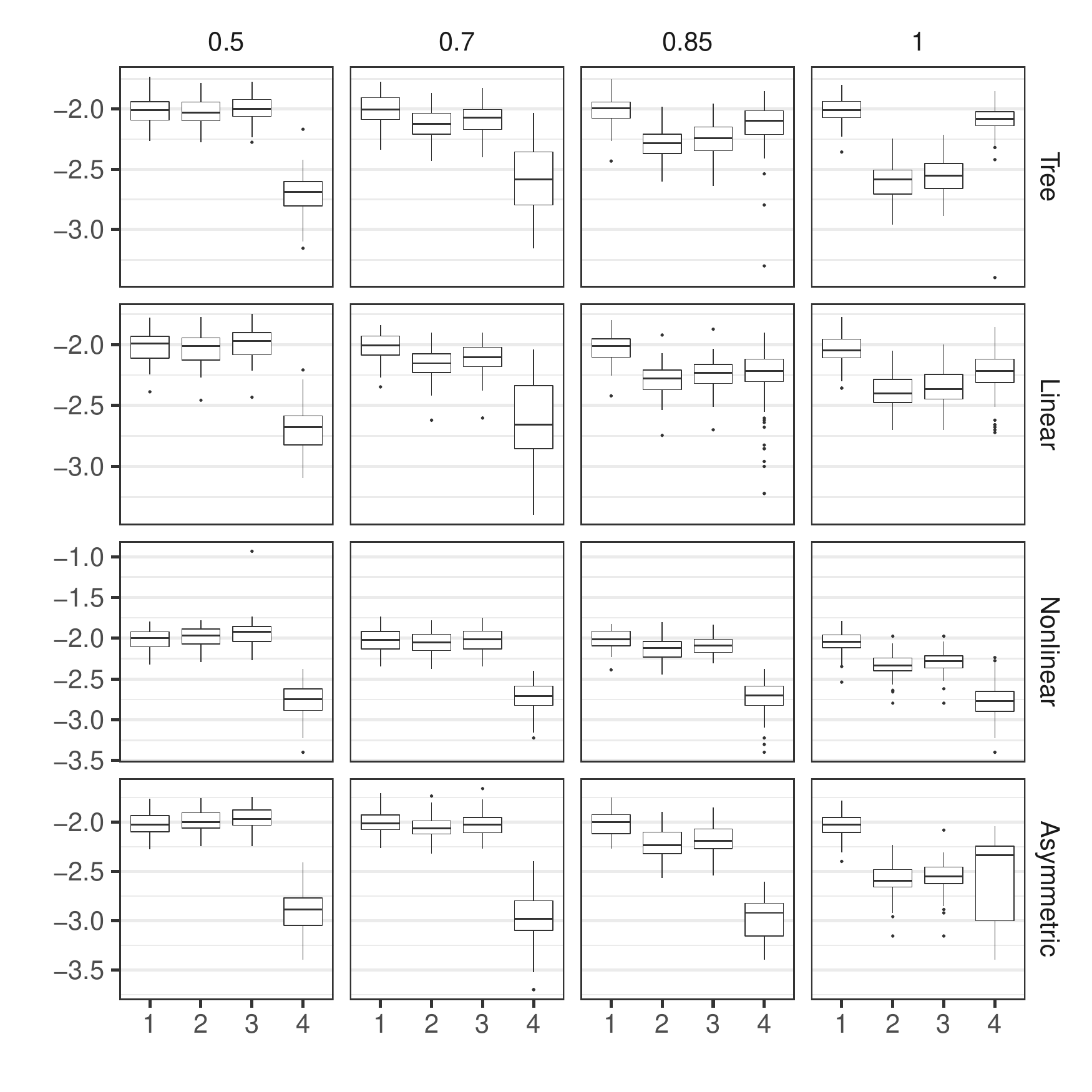}
        \caption[]{{\small Weibull-D, ISE$_{\text{in}}$}}
    \end{subfigure}\hfill
    \begin{subfigure}[b]{0.5\textwidth}  
        \centering 
        \includegraphics[width=\textwidth,height=0.3\textheight,trim=10 8 5 5 ]{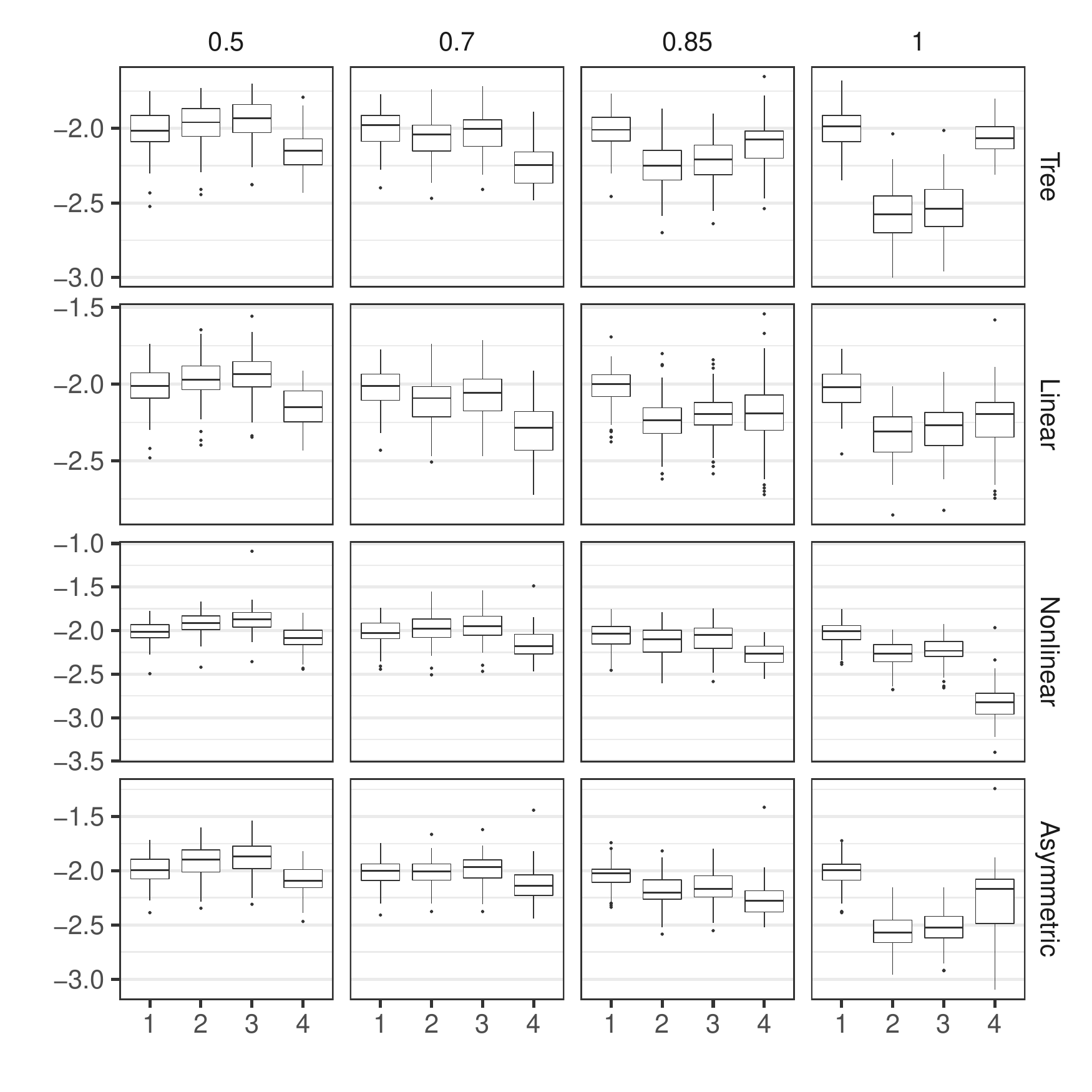}
        \caption[]{{\small Weibull-D, ISE$_{\text{out}}$}}
    \end{subfigure}
}
\makebox[\linewidth][c]{%
    \begin{subfigure}[b]{0.5\textwidth}  
        \centering 
        \includegraphics[width=\textwidth,height=0.3\textheight,trim=10 8 5 5]{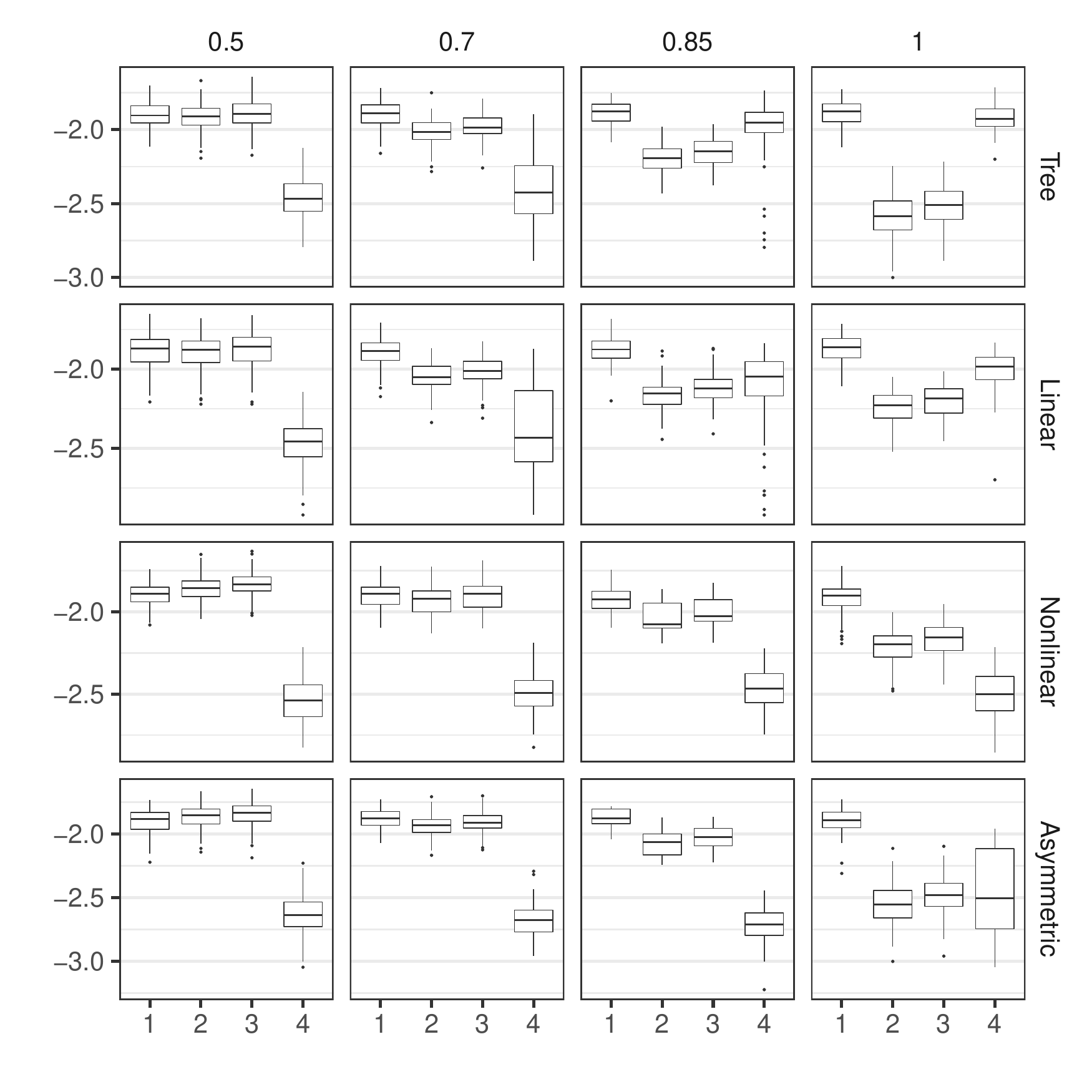}
        \caption[]{{\small Exponential, ISE$_{\text{in}}$}}
    \end{subfigure}\hfill
    \begin{subfigure}[b]{0.5\textwidth}  
        \centering 
        \includegraphics[width=\textwidth,height=0.3\textheight,trim=10 8 5 5]{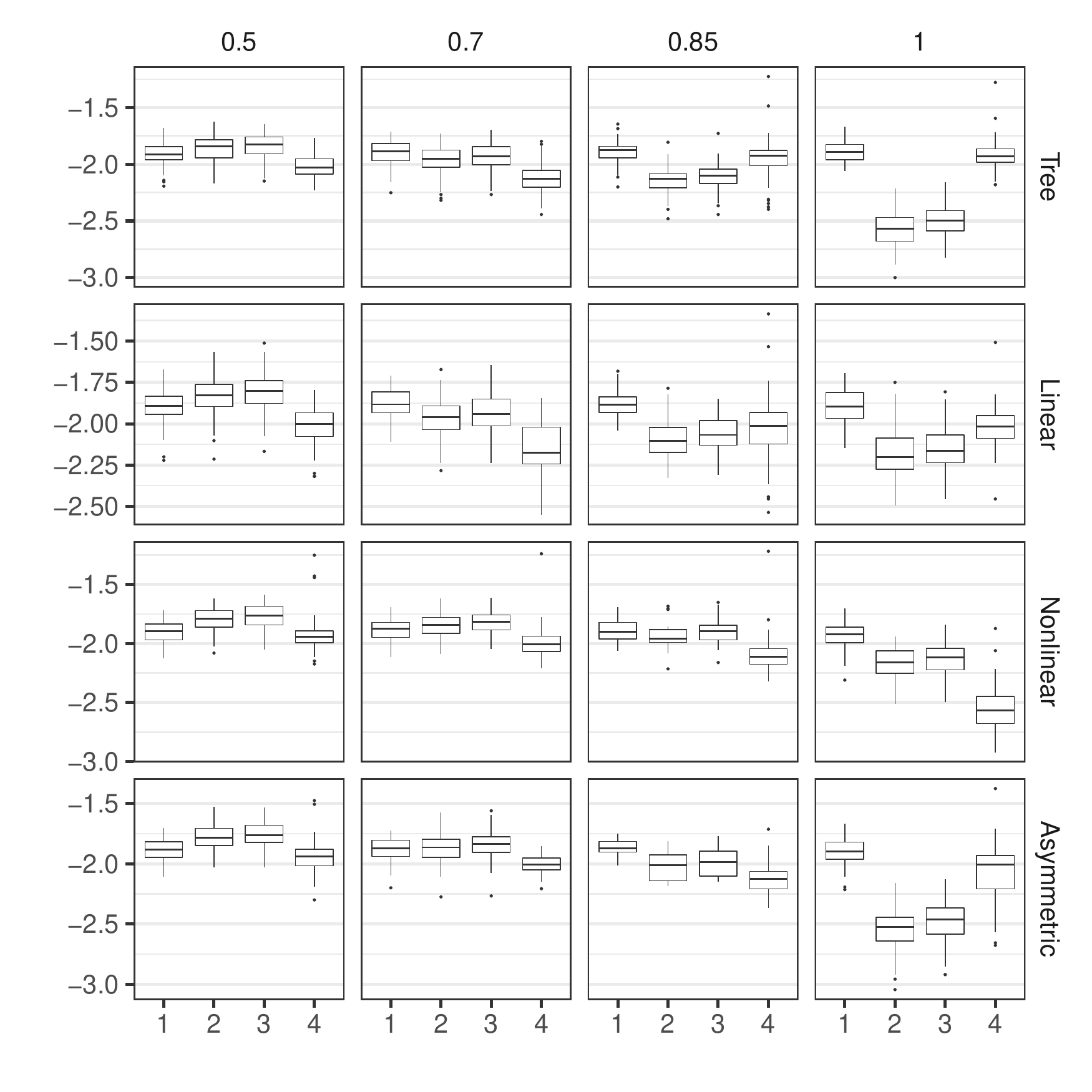}
        \caption[]{{\small Exponential, ISE$_{\text{out}}$}}
    \end{subfigure}
}
\caption{Boxplots of integrated squared error (ISE, both in-sample and out-of-sample) on $\log_{10}$ scale, 
time-invariant covariates, N=500, light censoring, Weibull-I, Weibull-D, and Exponential distribution. }
\end{figure}

%--------------------
\newpage
%\subsection{Mean squared error of longitudinal outcome predictions.}

\begin{figure}[h!]
    \vspace*{-0.5cm}
    \makebox[\linewidth][c]{%
    \begin{subfigure}[b]{0.5\textwidth}
        \centering
        \includegraphics[width=\textwidth,height=0.3\textheight,trim=10 8 5 5]{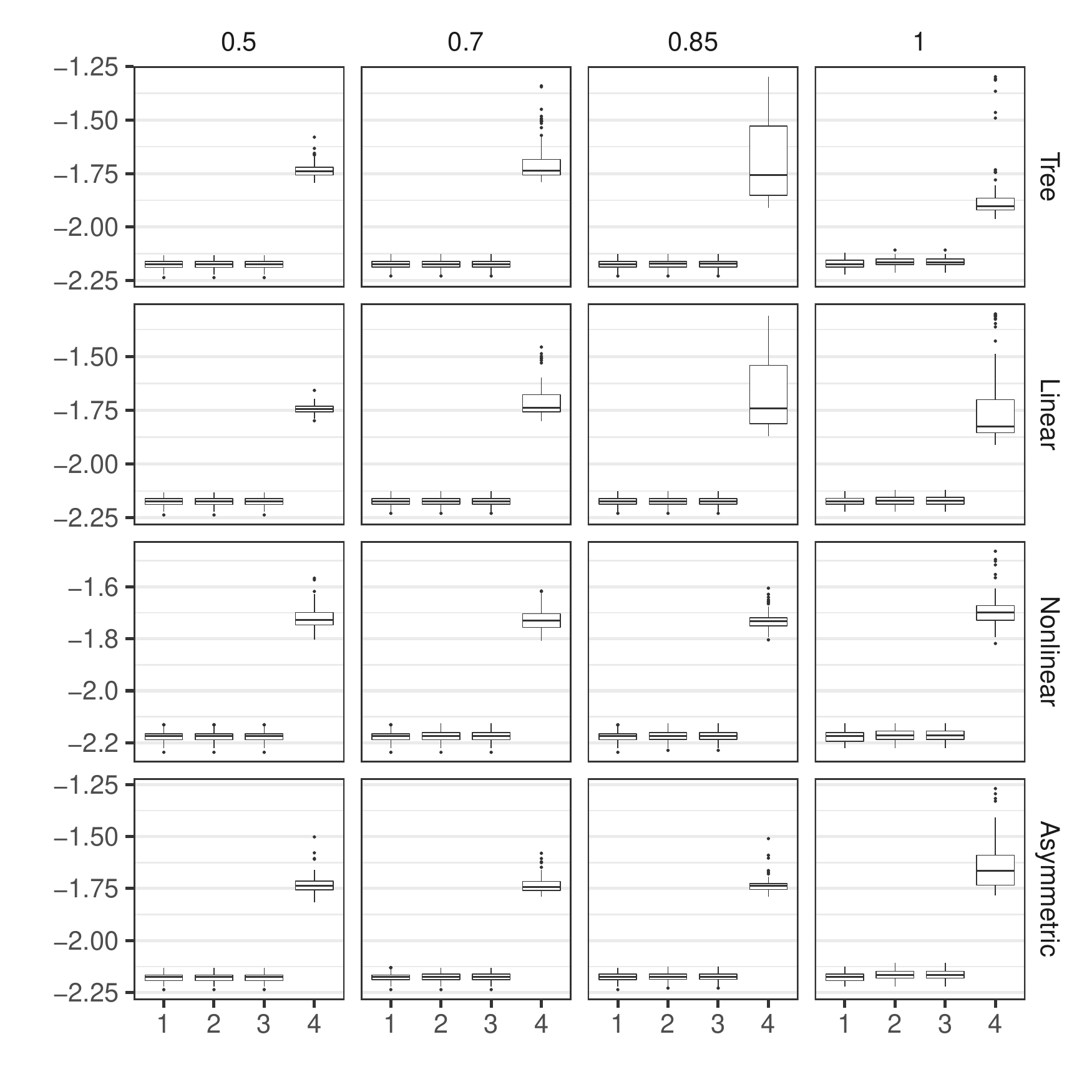}
        \caption[]{{\small Weibull-I, MSE$_{y\text{in}}$}}
    \end{subfigure}\hfill
    \begin{subfigure}[b]{0.5\textwidth}  
        \centering 
        \includegraphics[width=\textwidth,height=0.3\textheight,trim=10 8 5 5]{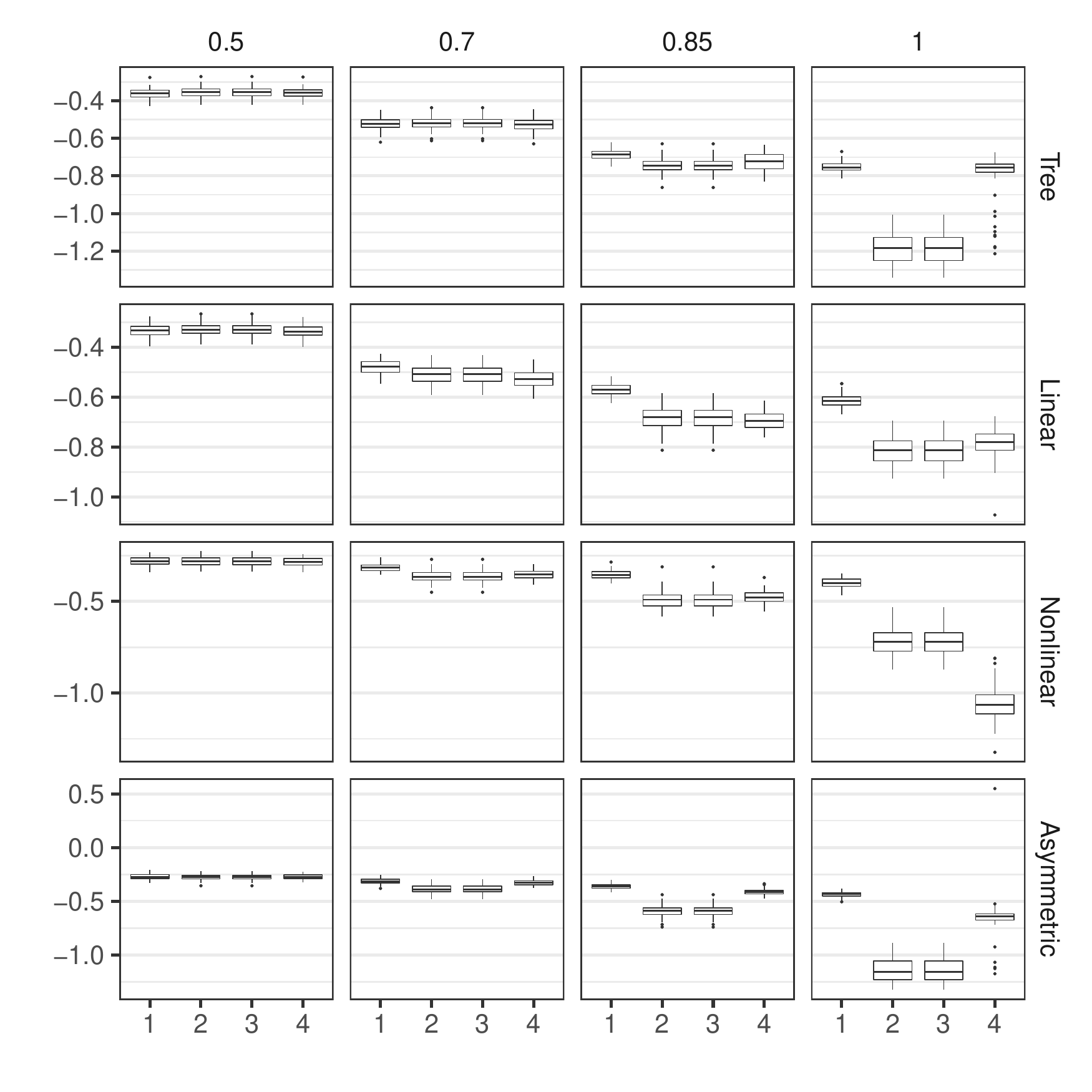}
        \caption[]{{\small Weibull-I, MSE$_{y\text{out}}$}}
    \end{subfigure}
}
%    \vspace\baselineskip
    \makebox[\linewidth][c]{%
    \begin{subfigure}[b]{0.5\textwidth}  
        \centering 
        \includegraphics[width=\textwidth,height=0.3\textheight,trim=10 8 5 5]{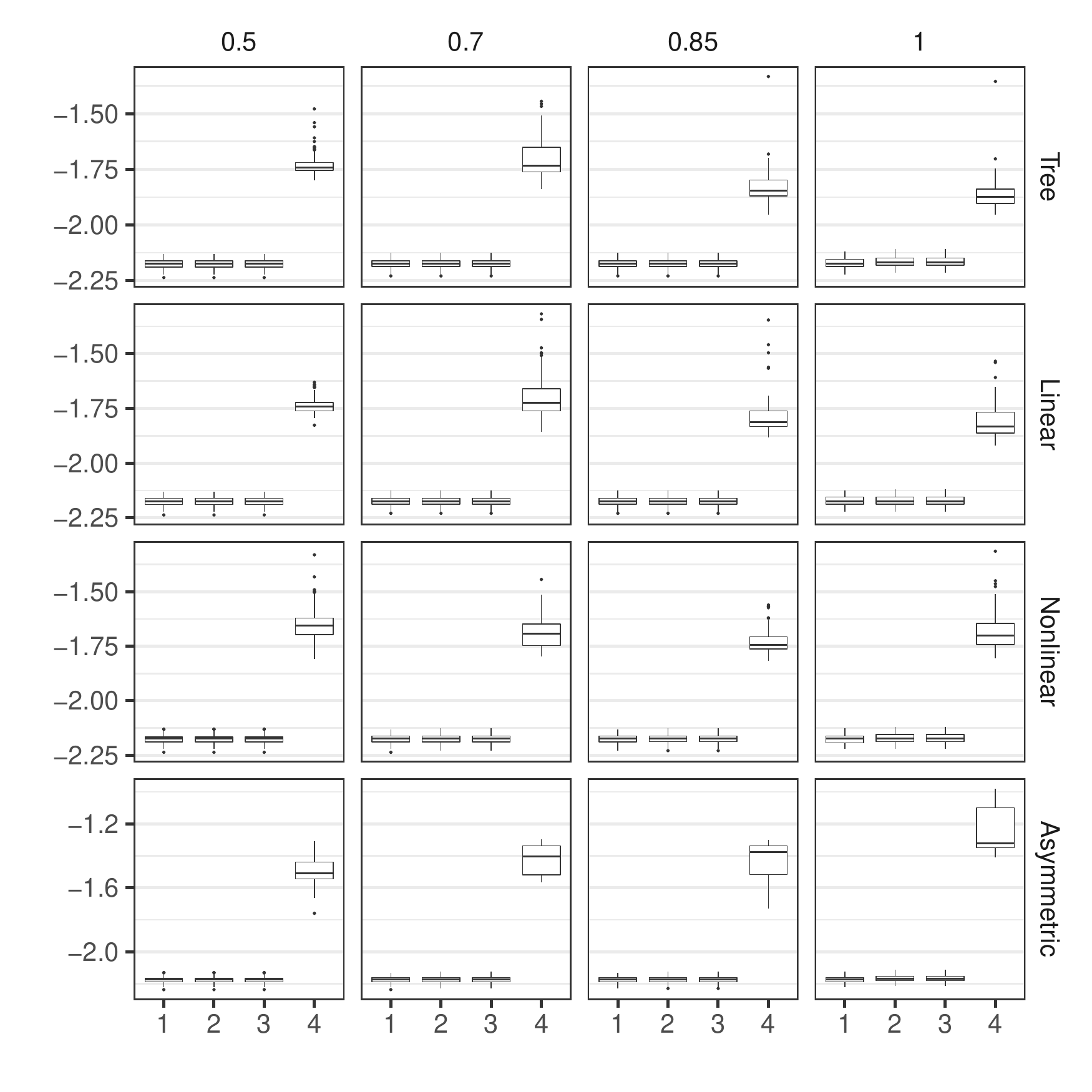}
        \caption[]{{\small Weibull-D, MSE$_{y\text{in}}$}}
    \end{subfigure}\hfill
    \begin{subfigure}[b]{0.5\textwidth}  
        \centering 
        \includegraphics[width=\textwidth,height=0.3\textheight,trim=10 8 5 5 ]{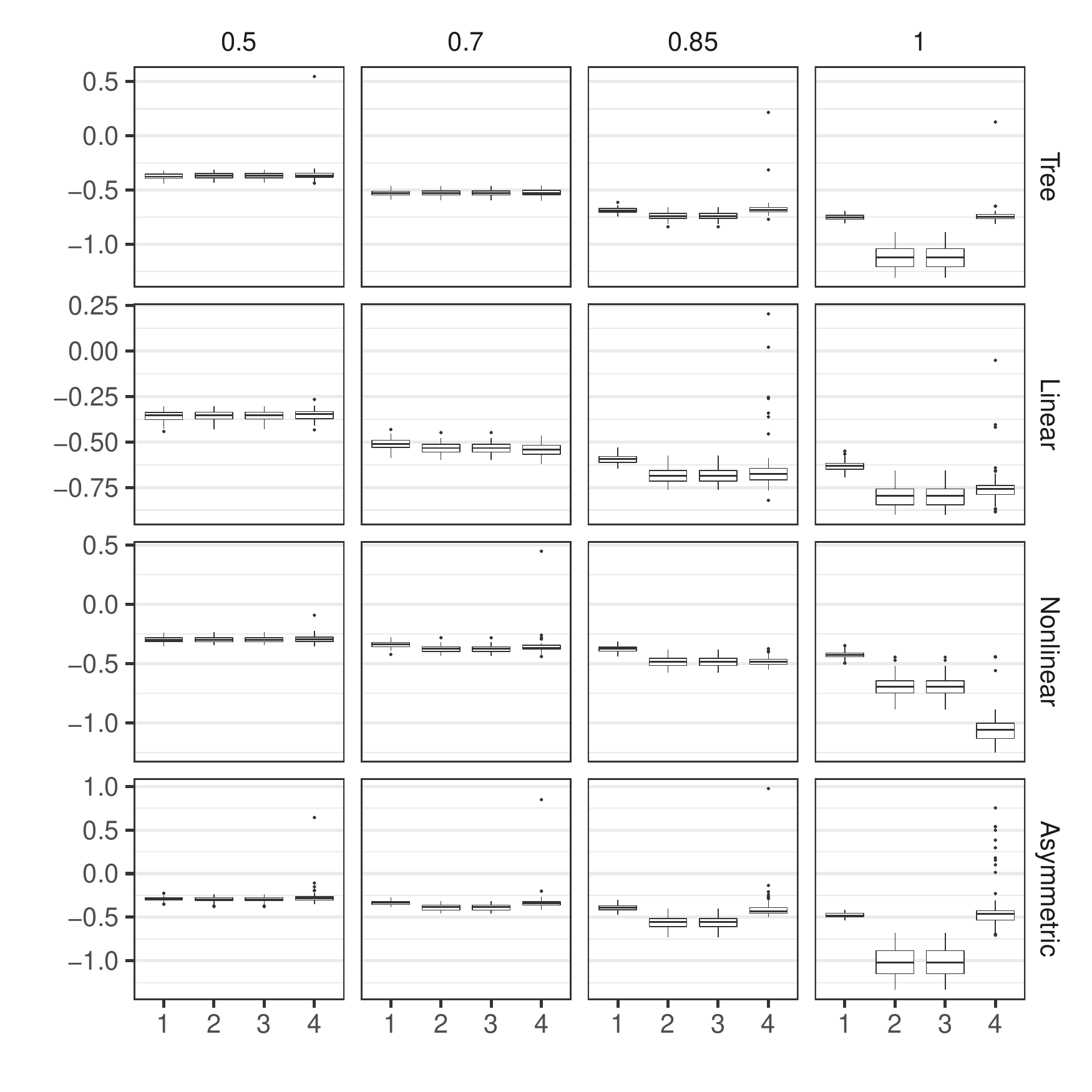}
        \caption[]{{\small Weibull-D, MSE$_{y\text{out}}$}}
    \end{subfigure}
}
\makebox[\linewidth][c]{%
    \begin{subfigure}[b]{0.5\textwidth}  
        \centering 
        \includegraphics[width=\textwidth,height=0.3\textheight,trim=10 8 5 5]{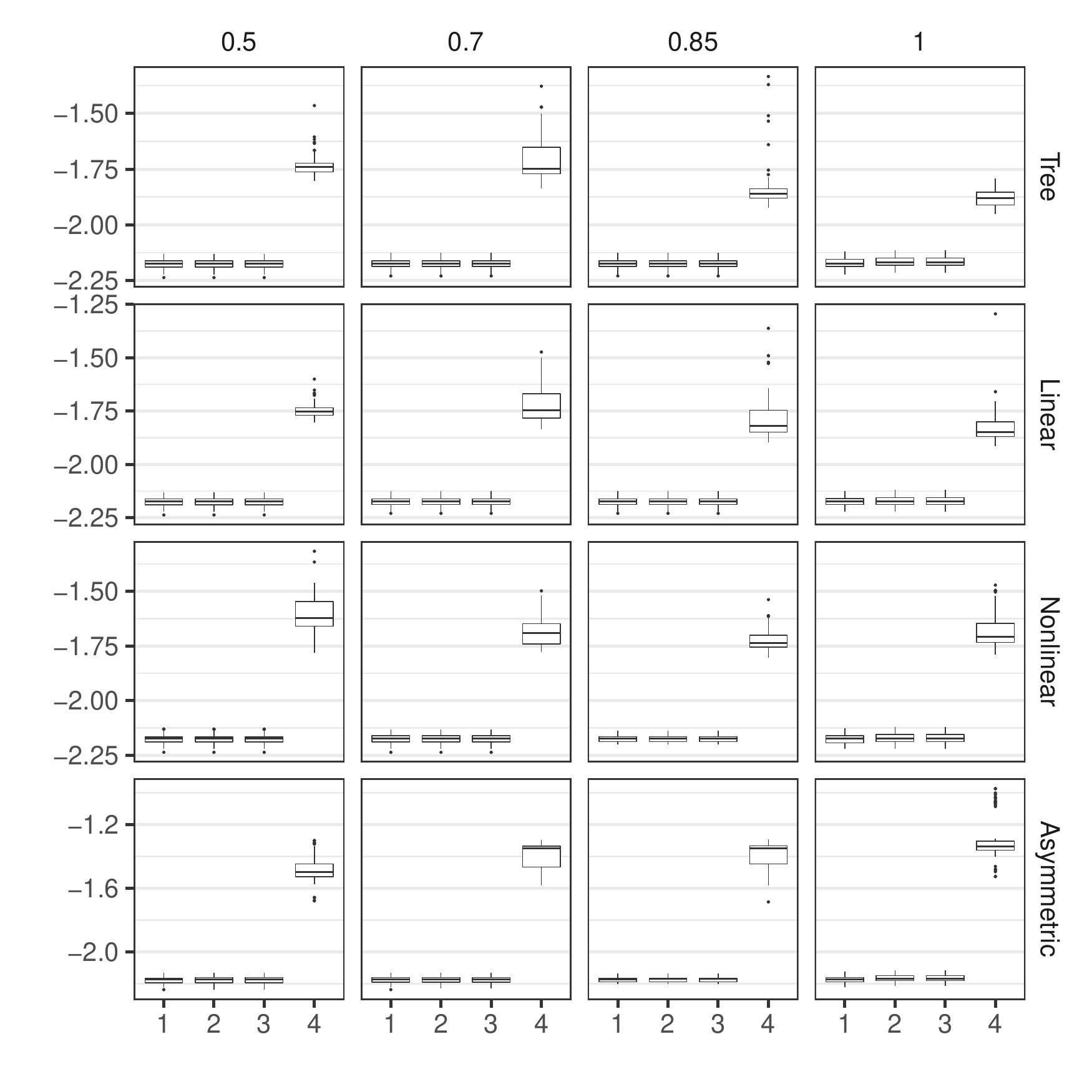}
        \caption[]{{\small Exponential, MSE$_{y\text{in}}$}}
    \end{subfigure}\hfill
    \begin{subfigure}[b]{0.5\textwidth}  
        \centering 
        \includegraphics[width=\textwidth,height=0.3\textheight,trim=10 8 5 5]{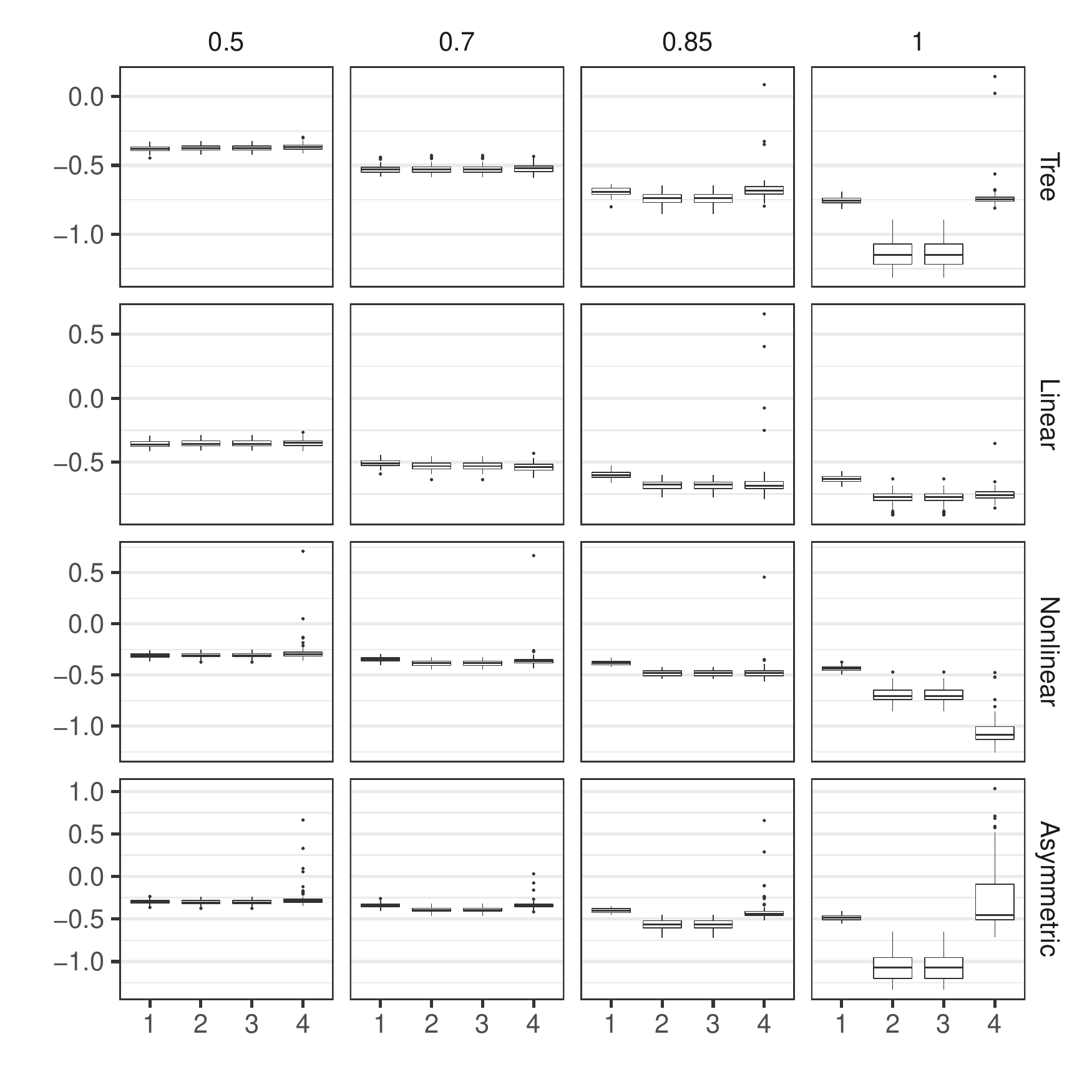}
        \caption[]{{\small Exponential, MSE$_{y\text{out}}$}}
    \end{subfigure}
}
\caption{Boxplots of mean squared error (MSE$_y$, both in-sample and out-of-sample) on $\log_{10}$ scale, 
time-invariant covariates, N=500, light censoring, Weibull-I, Weibull-D, and Exponential distribution. }
\end{figure}

%--------------------
\newpage
%\subsection{Mean squared error of estimated Cox PH slope coefficients.}
\begin{figure}[h!]
%    \vspace*{-0.5cm}
    \makebox[\linewidth][c]{%
    \begin{subfigure}[b]{0.5\textwidth}
        \centering
        \includegraphics[width=\textwidth,height=0.3\textheight,trim=10 8 5 5]{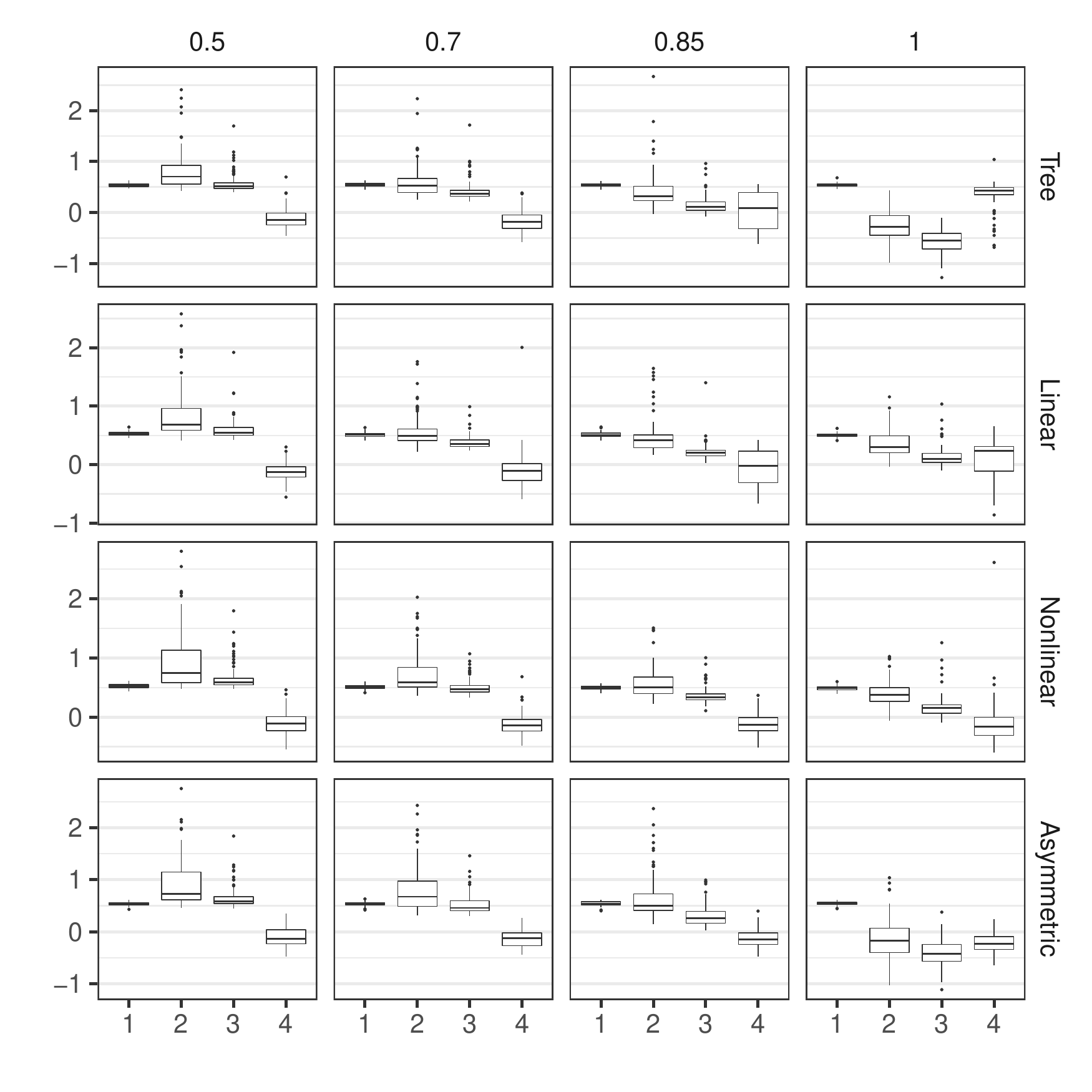}
        \caption[]{{\small Weibull-I, MSE$_{b}$}}
    \end{subfigure}%\hfill
    %\begin{subfigure}[b]{0.5\textwidth}
    %    \centering
    %    \includegraphics[width=\textwidth,height=0.3\textheight,trim=10 8 5 5]{figures/varmaj_Nsub_500_attr_cover_Weibull-I_Light.pdf}
    %    \caption[]{{\small Weibull-I, Cover$_{b}$}}
    %\end{subfigure}
}
%    \vspace\baselineskip
    \makebox[\linewidth][c]{%
    \begin{subfigure}[b]{0.5\textwidth}  
        \centering 
        \includegraphics[width=\textwidth,height=0.3\textheight,trim=10 8 5 5]{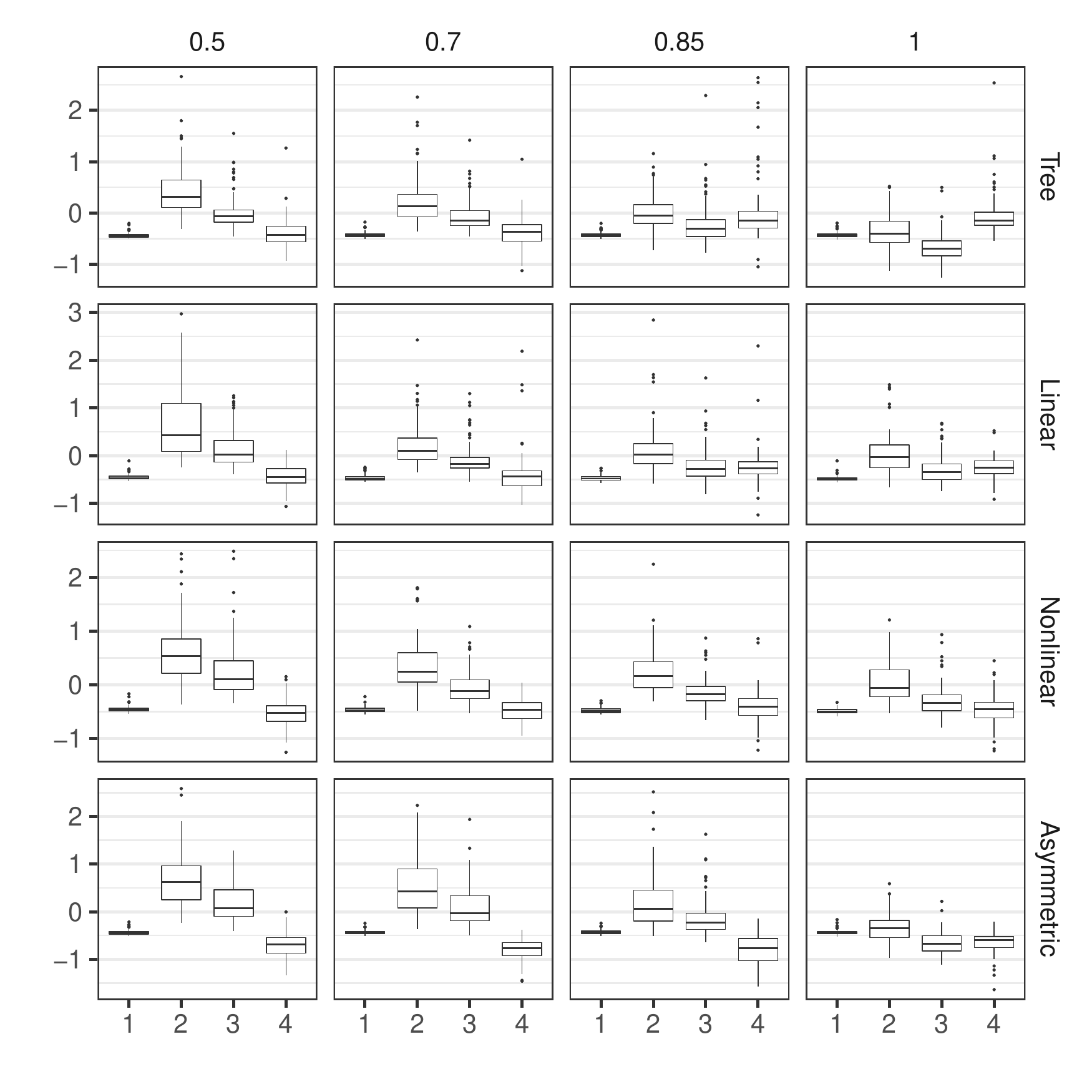}
        \caption[]{{\small Weibull-D, MSE$_{b}$}}
    \end{subfigure}%\hfill
%    \begin{subfigure}[b]{0.5\textwidth}  
%        \centering 
%        \includegraphics[width=\textwidth,height=0.3\textheight,trim=10 8 5 5]{figures/varmaj_Nsub_500_attr_cover_Weibull-D_Light.pdf}
%        \caption[]{{\small Weibull-D, Cover$_{b}$}}
%    \end{subfigure}
}
\makebox[\linewidth][c]{%
    \begin{subfigure}[b]{0.5\textwidth}  
        \centering 
        \includegraphics[width=\textwidth,height=0.3\textheight,trim=10 8 5 5]{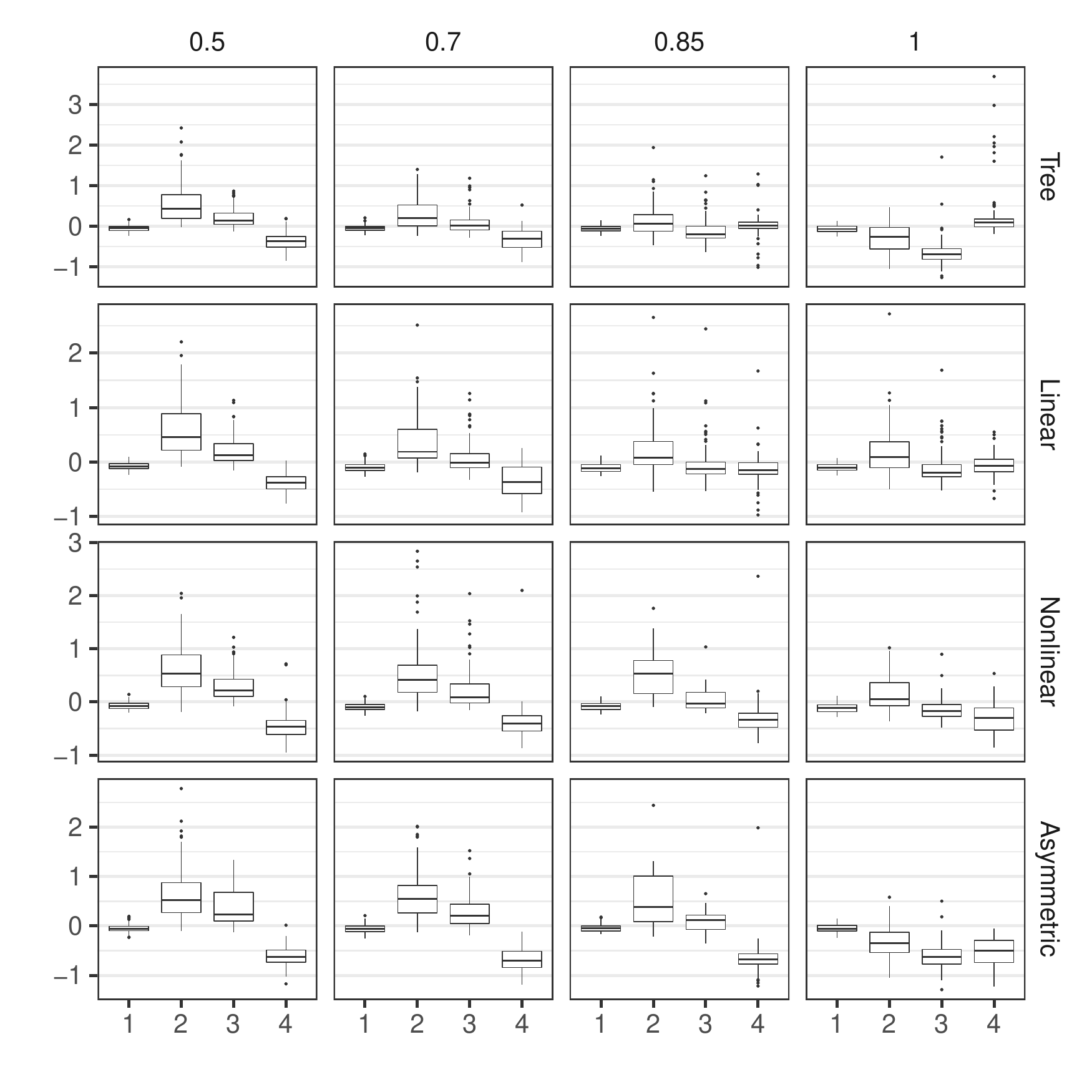}
        \caption[]{{\small Exponential, MSE$_{b}$}}
    \end{subfigure}%\hfill
%    \begin{subfigure}[b]{0.5\textwidth}  
%        \centering 
%        \includegraphics[width=\textwidth,height=0.3\textheight,trim=10 8 5 5]{figures/varmaj_Nsub_500_attr_cover_Exponential_Light.pdf}
%        \caption[]{{\small Exponential, Cover$_{b}$}}
%    \end{subfigure}
}
\caption{Boxplots of mean squared error (MSE$_b$) of estimated Cox PH slopes on $\log_{10}$ scale, %and Cover$_b$, 
time-invariant covariates, N=500, light censoring, Weibull-I, Weibull-D, and Exponential distribution. }
\end{figure}

Boxplots for ISE$_{\text{out}}$ show that the accuracies of
survival predictions of JLCT (Methods 2 \& 3) and JLCM (Method 4)
are comparable, with JLCT being slightly worse, under the settings
where latent classes are generated from probabilistic models
($p_0<1$). In particular, when the latent class structure comes from
a linear separation (Linear) and therefore the simulated data
matches exactly the model of JLCM, JLCT still performs comparably to
JLCM (e.g. $p_0=0.85$, ``Linear''). The gap between JLCT and JLCM
decreases as the concentration level increases from $p_0 = 0.5$ to
$1$.  In the case where $p_0=1$ and thus latent classes are
generated by a deterministic partitioning, JLCT is more effective
than JLCM in all settings except nonlinear partitioning. In
particular, when the latent class model follows the deterministic
partitioning by a tree ($p_0=1$,``Tree'') and thus matches the model
of JLCT, JLCT outperforms JLCM by a significant margin. Furthermore,
JLCM tends to have higher variance in ISE values, due to
occasionally unconverged JLCM runs.
A similar pattern appears for MSE$_{y\text{out}}$,
with JLCT and JLCM comparable for most settings,
and JLCT significantly outperforming JLCM
when the simulated data come from the underlying model of JLCT\@.
Regarding estimation accuracy of Cox PH slopes (MSE$_{b}$)
the same pattern shows up here as well, although there exist more outliers
in all methods.

Table~\ref{tb:simtime_time_inv} shows the average running time of JLCT and JLCM 
(over 100 runs) under the same settings as the plots.
The running time of JLCT includes constructing the tree,
fitting the two survival models of Methods 2 and 3,
and fitting the linear mixed model. The running time of JLCM includes
fitting with all numbers of latent classes $g\in \{2,3,4,5,6\}$.
Clearly, JLCT is orders of magnitude faster than JLCM across all settings:
JLCT completes one run within a minute, while JLCM typically takes~25 to~45 minutes.
The running time for other censoring levels and baseline hazard distributions
are similar.

Finally, note that Methods 2 and 3 give very similar results, as they only differ in the 
survival models that are fitted after tree construction.
Method 2 assumes a single baseline hazard function shared across terminal nodes,
which agrees with the true data generating scheme.
Nevertheless, Methods 2 and 3 have almost identical out-of-sample performances,
which suggests that introducing additional parameters for node-specific baseline hazard
functions does not hurt the prediction performance for the simulated data.
In practice, it is often unclear whether the latent classes share a single
baseline hazard function.
Our results suggest that in such a case, we can assume a separate baseline hazard
function for each terminal node, without worrying about over-fitting the data.
Thus, in the time-varying simulations, we only consider the
Cox PH model with separate baseline hazard functions across terminal nodes.

\section{Additional simulation results: time-varying covariates}
\label{app:moreresults_time_var}
In this section, we present the complete simulation results for time-varying covariates:
for three time-to-event distributions (Weibull-I, Weibull-D, and Exponential),
and light censoring. No censoring and heavy censoring are omitted because of
their similar performances to light censoring.
We report results for the five methods discussed in Table~\ref{tb:simulate_models}:
\begin{enumerate}
\item JLCT with no split, and using time-varying survival covariates,
\item JLCT with ``time-invariant'' latent class and survival covariates,
\item JLCT with ``time-invariant'' latent class covariates and time-varying survival covariates,
\item JLCT with time-varying latent class and survival covariates, 
\item JLCM with ``time-invariant'' latent class and survival covariates.
\end{enumerate}
We use the performance measures described in Section~\ref{sec:evaluation}:
ISE$_{\text{in}}$, ISE$_{\text{out}}$, MSE$_{y\text{in}}$,
MSE$_{y\text{out}}$, and MSE$_b$.
%--------------------
%\subsection{Integrated squared error of time-to-event predictions.}
\begin{figure}[h!]
    \vspace*{-0.5cm}
    \makebox[\linewidth][c]{%
    \begin{subfigure}[b]{0.5\textwidth}
        \centering
        \includegraphics[width=\textwidth,height=0.3\textheight,trim=10 8 5 5]{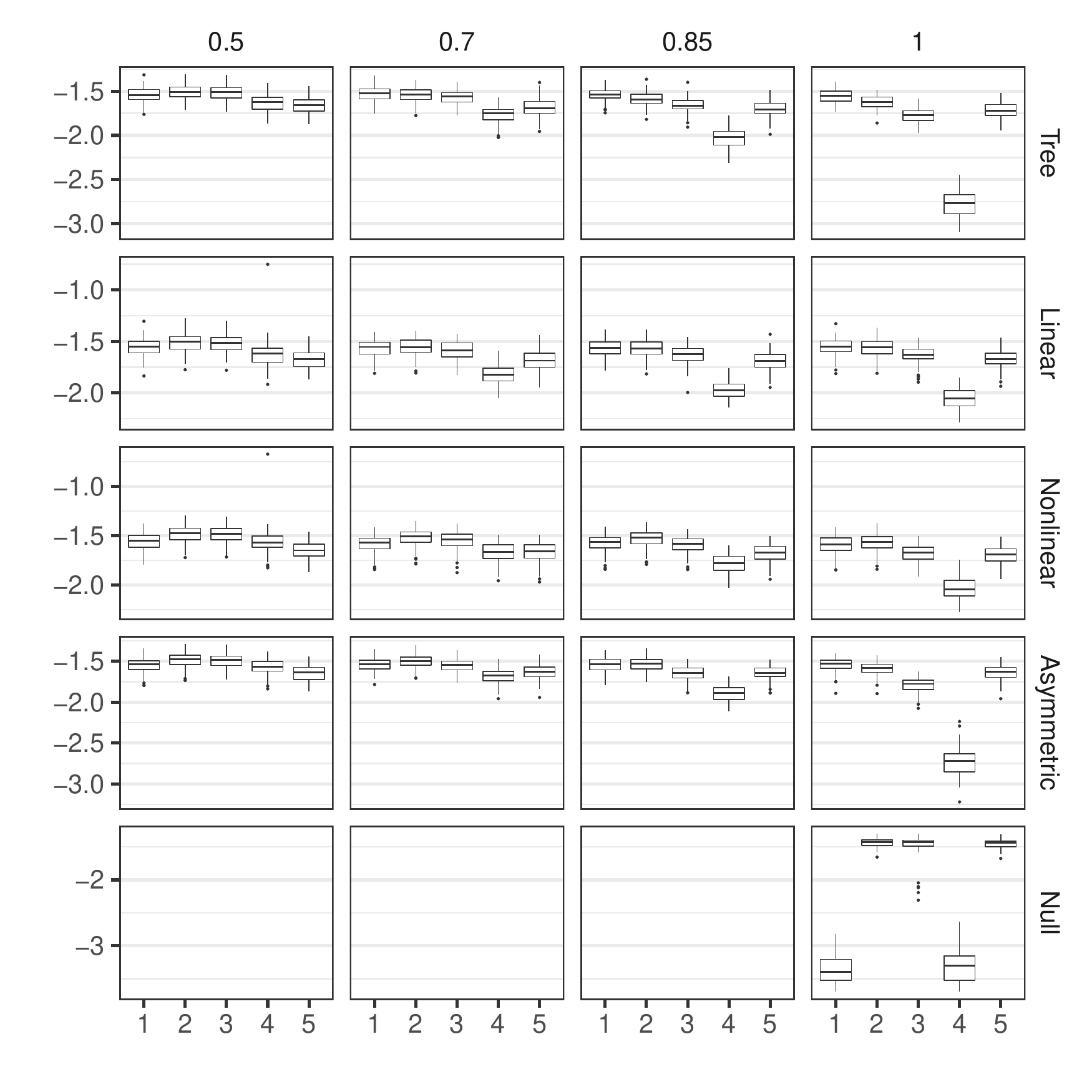}
        \caption[]{{\small Weibull-I, ISE$_{\text{in}}$}}
    \end{subfigure}\hfill
    \begin{subfigure}[b]{0.5\textwidth}  
        \centering 
        \includegraphics[width=\textwidth,height=0.3\textheight,trim=10 8 5 5]{survvar_Nsub_500_attr_ISE_test_Weibull-I_Light.pdf}
        \caption[]{{\small Weibull-I, ISE$_{\text{out}}$}}
    \end{subfigure}
}
%    \vspace\baselineskip
    \makebox[\linewidth][c]{%
    \begin{subfigure}[b]{0.5\textwidth}  
        \centering 
        \includegraphics[width=\textwidth,height=0.3\textheight,trim=10 8 5 5]{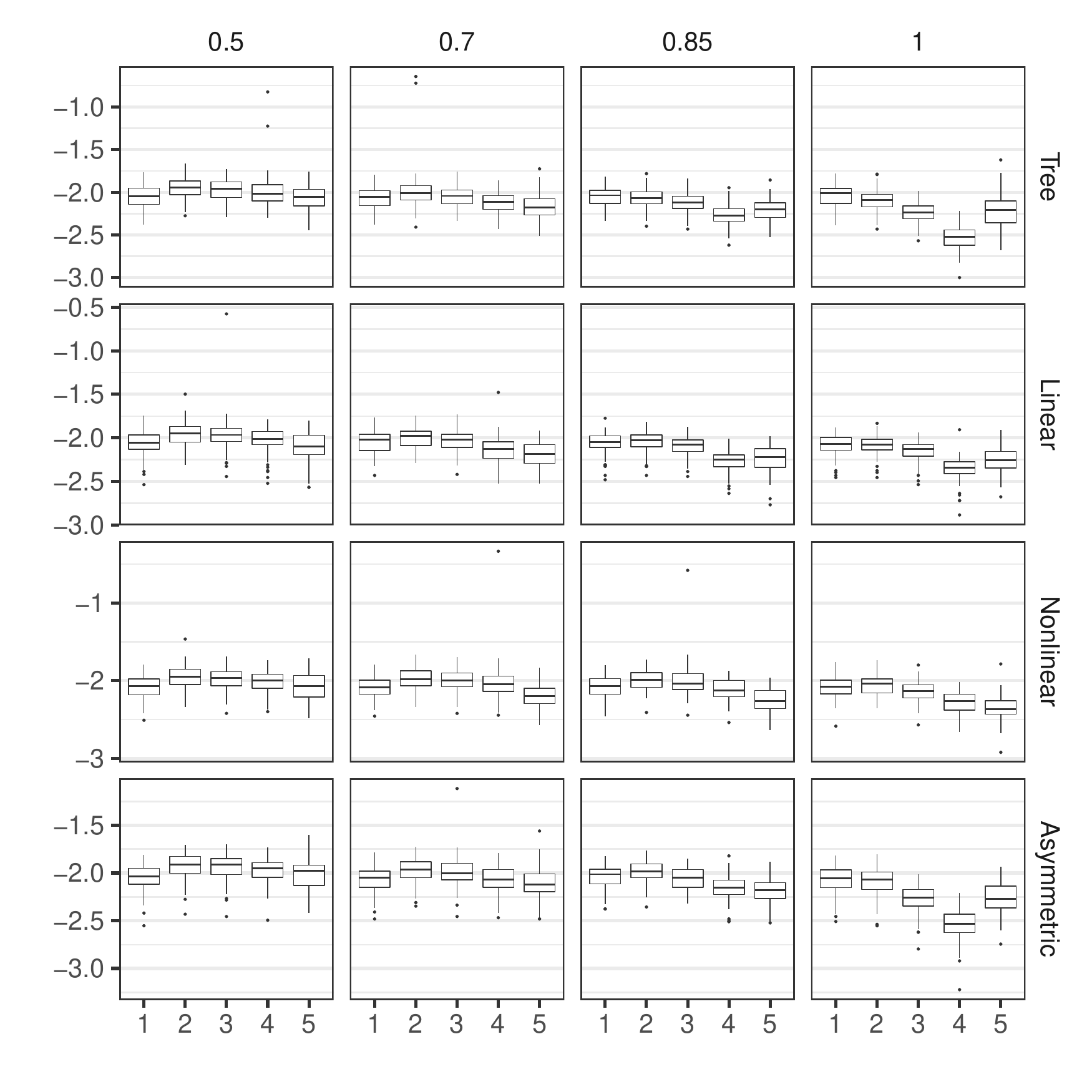}
        \caption[]{{\small Weibull-D, ISE$_{\text{in}}$}}
    \end{subfigure}\hfill
    \begin{subfigure}[b]{0.5\textwidth}  
        \centering 
        \includegraphics[width=\textwidth,height=0.3\textheight,trim=10 8 5 5 ]{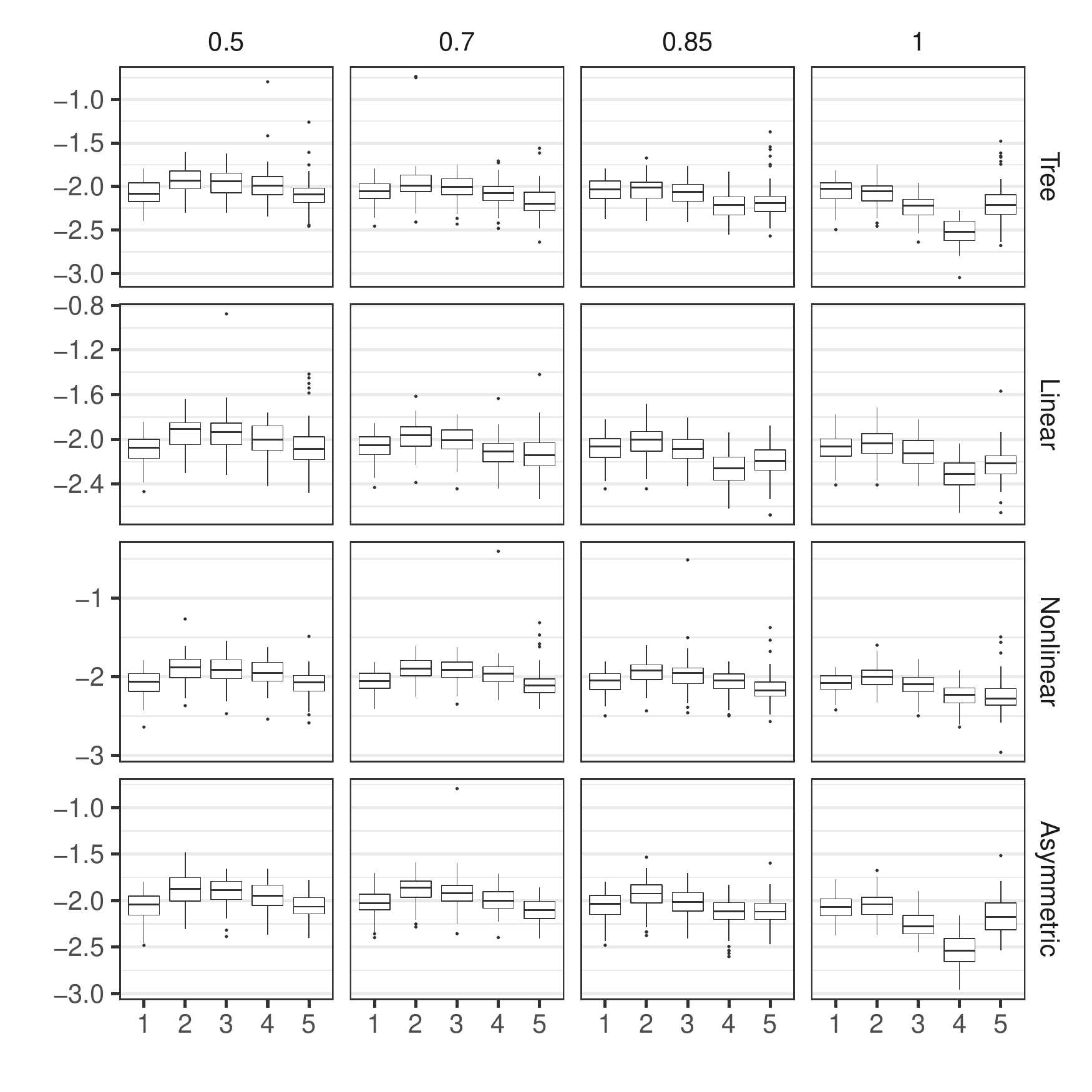}
        \caption[]{{\small Weibull-D, ISE$_{\text{out}}$}}
    \end{subfigure}
}
\makebox[\linewidth][c]{%
    \begin{subfigure}[b]{0.5\textwidth}  
        \centering 
        \includegraphics[width=\textwidth,height=0.3\textheight,trim=10 8 5 5]{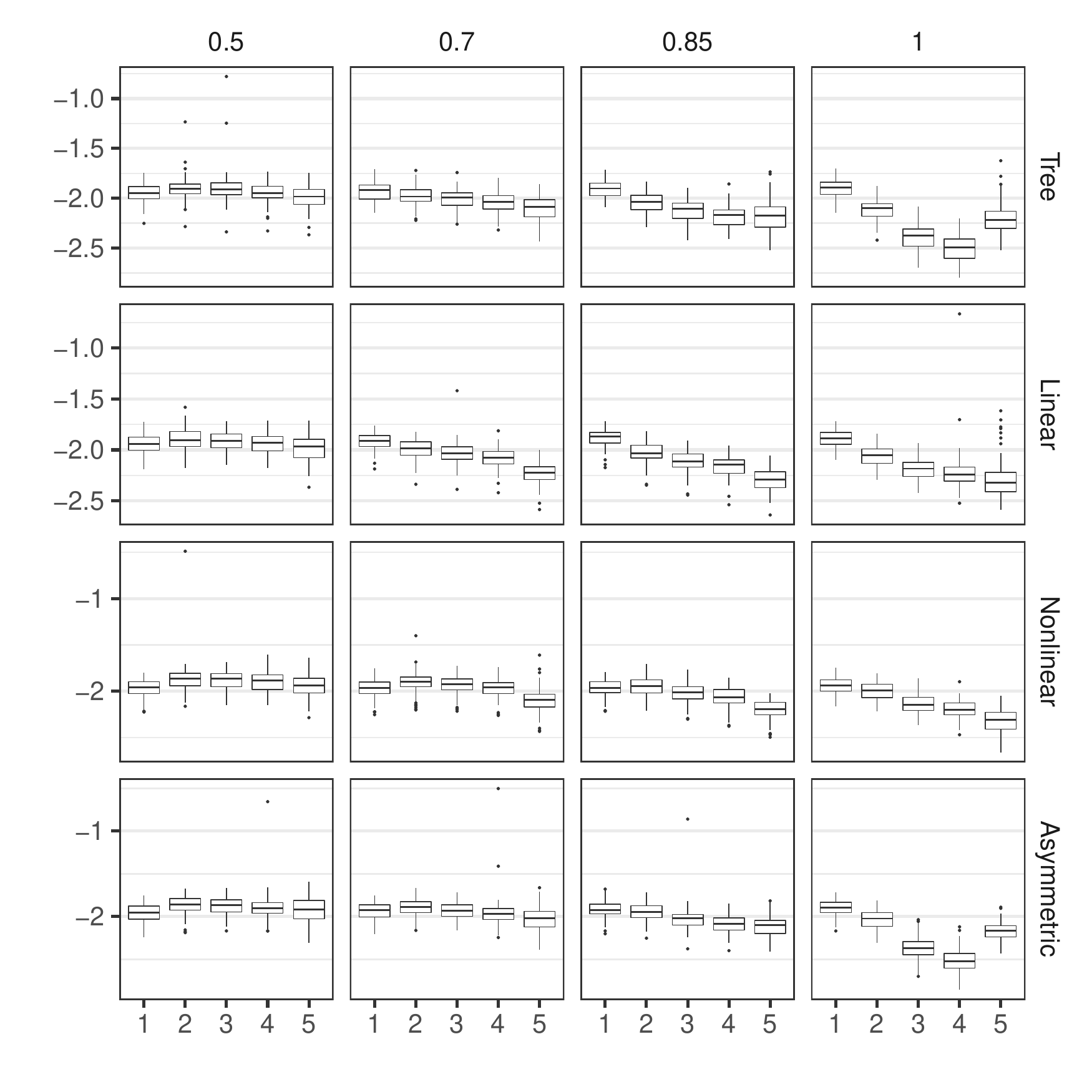}
        \caption[]{{\small Exponential, ISE$_{\text{in}}$}}
    \end{subfigure}\hfill
    \begin{subfigure}[b]{0.5\textwidth}  
        \centering 
        \includegraphics[width=\textwidth,height=0.3\textheight,trim=10 8 5 5]{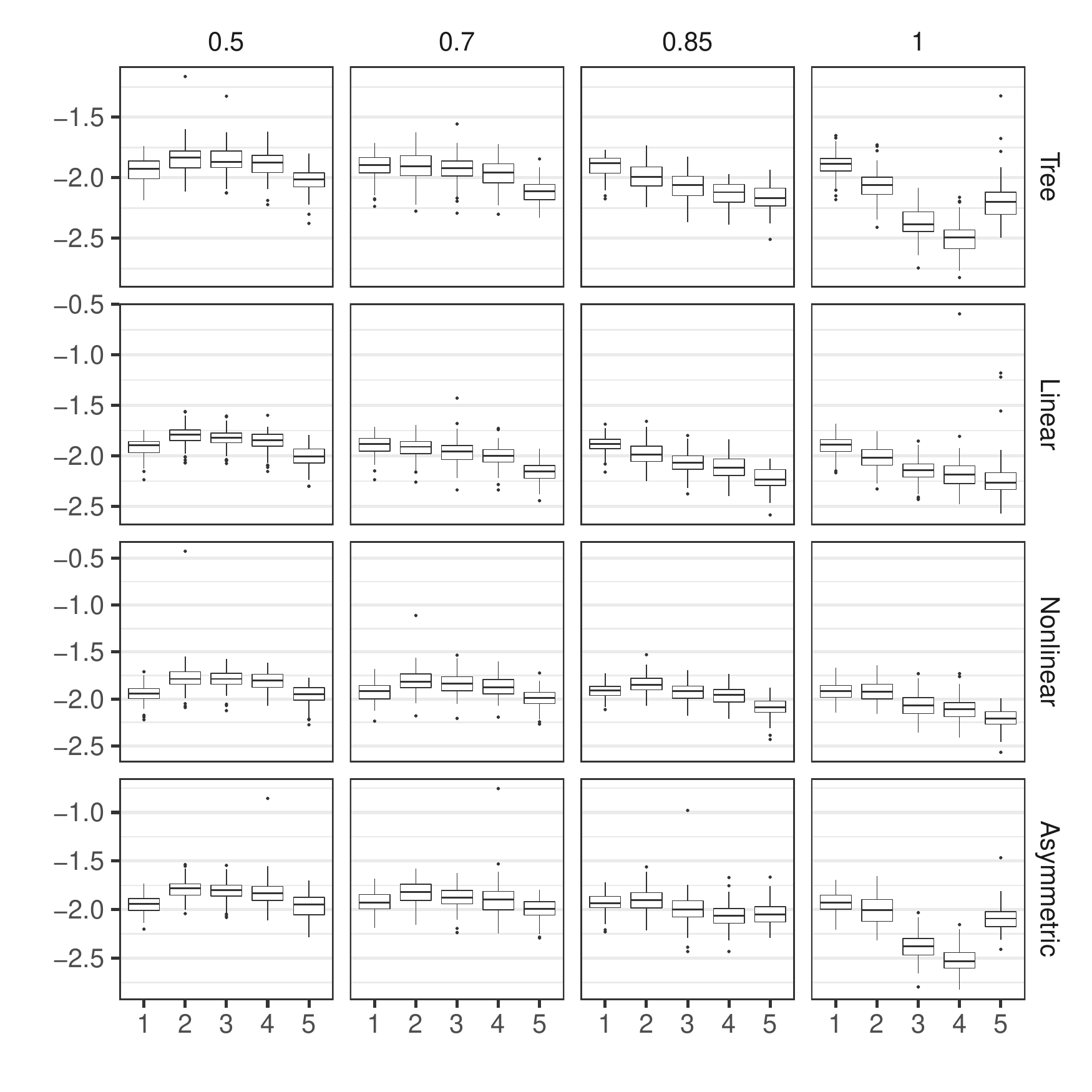}
        \caption[]{{\small Exponential, ISE$_{\text{out}}$}}
    \end{subfigure}
}
\caption{Boxplots of integrated squared error (ISE, both in-sample and out-of-sample) on $\log_{10}$ scale, 
time-varying covariates, N=500, light censoring, Weibull-I, Weibull-D, and Exponential distribution. }
\end{figure}

%--------------------
\newpage
%\subsection{Mean squared error of longitudinal outcome predictions.}

\begin{figure}[h!]
    \vspace*{-0.5cm}
    \makebox[\linewidth][c]{%
    \begin{subfigure}[b]{0.5\textwidth}
        \centering
        \includegraphics[width=\textwidth,height=0.3\textheight,trim=10 8 5 5]{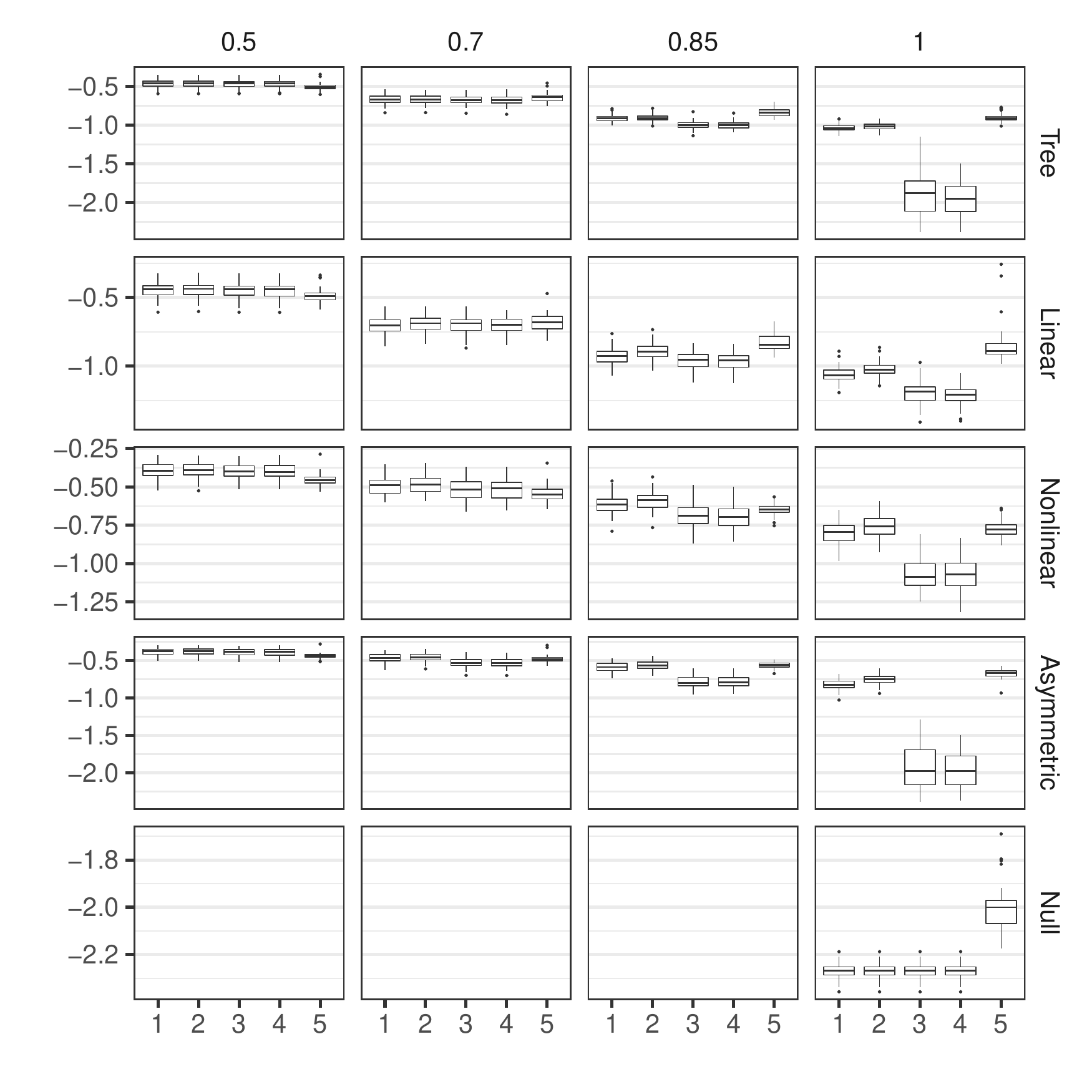}
        \caption[]{{\small Weibull-I, MSE$_{y\text{in}}$}}
    \end{subfigure}\hfill
    \begin{subfigure}[b]{0.5\textwidth}  
        \centering 
        \includegraphics[width=\textwidth,height=0.3\textheight,trim=10 8 5 5]{survvar_Nsub_500_attr_MSEy_test_Weibull-I_Light.pdf}
        \caption[]{{\small Weibull-I, MSE$_{y\text{out}}$}}
    \end{subfigure}
}
%    \vspace\baselineskip
    \makebox[\linewidth][c]{%
    \begin{subfigure}[b]{0.5\textwidth}  
        \centering 
        \includegraphics[width=\textwidth,height=0.3\textheight,trim=10 8 5 5]{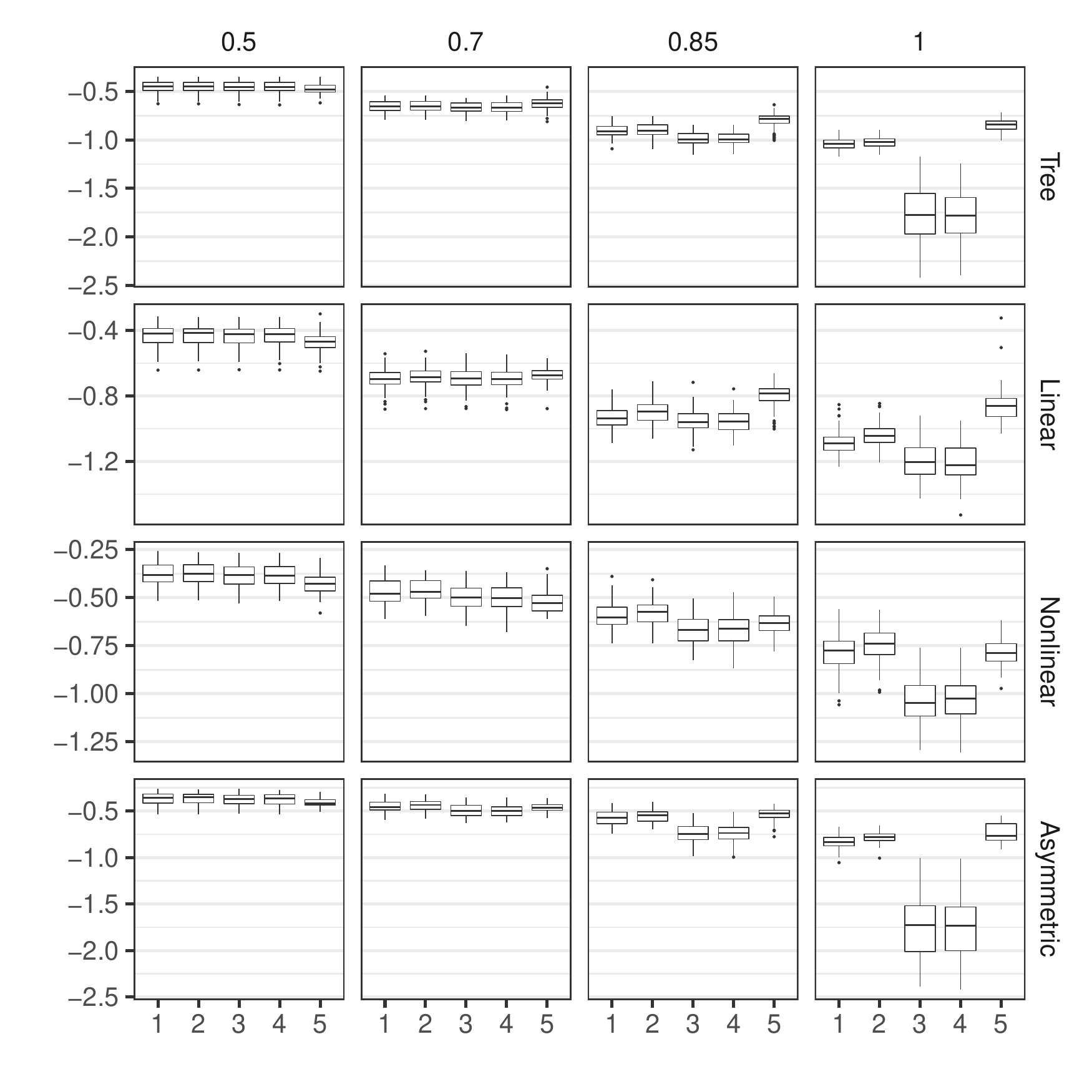}
        \caption[]{{\small Weibull-D, MSE$_{y\text{in}}$}}
    \end{subfigure}\hfill
    \begin{subfigure}[b]{0.5\textwidth}  
        \centering 
        \includegraphics[width=\textwidth,height=0.3\textheight,trim=10 8 5 5 ]{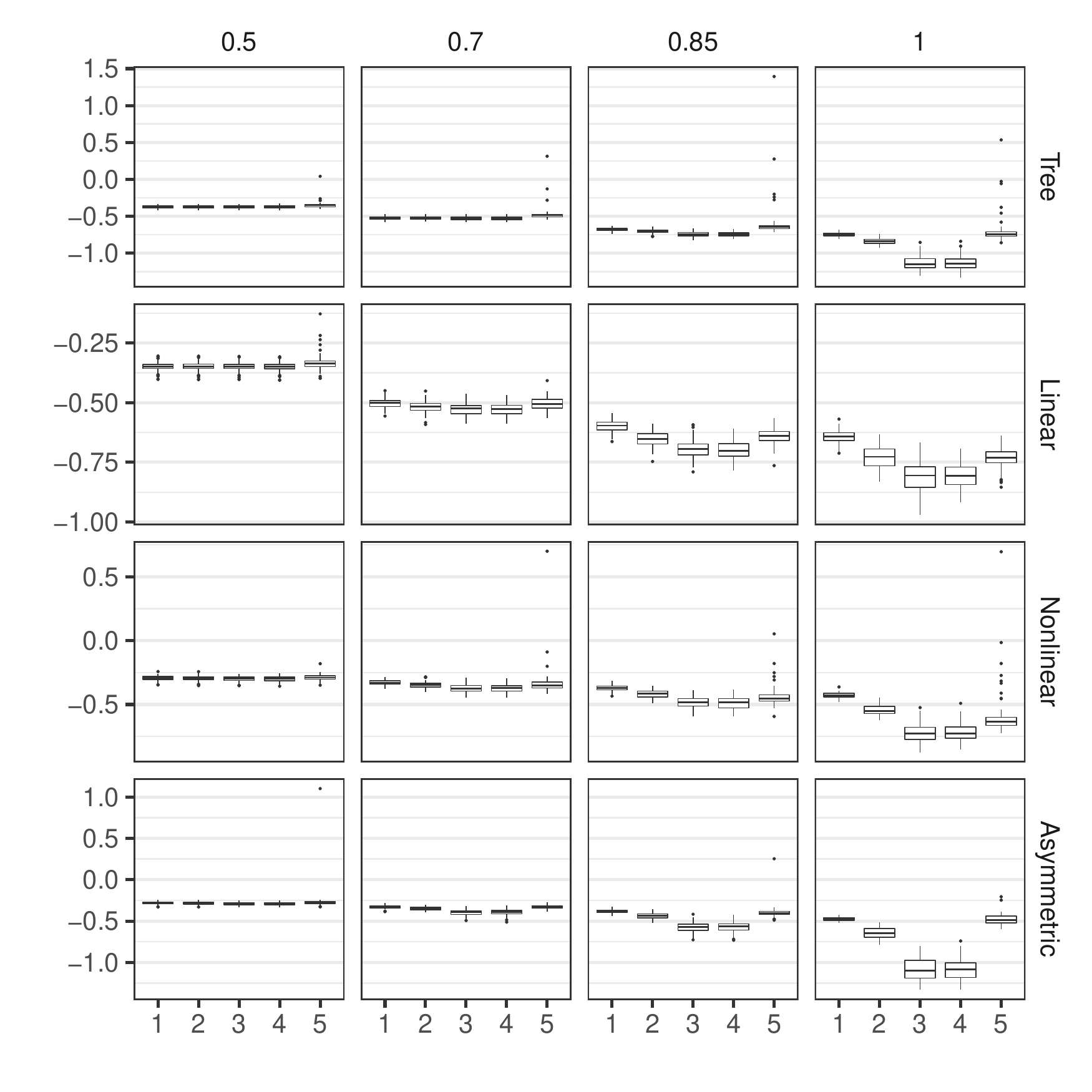}
        \caption[]{{\small Weibull-D, MSE$_{y\text{out}}$}}
    \end{subfigure}
}
\makebox[\linewidth][c]{%
    \begin{subfigure}[b]{0.5\textwidth}  
        \centering 
        \includegraphics[width=\textwidth,height=0.3\textheight,trim=10 8 5 5]{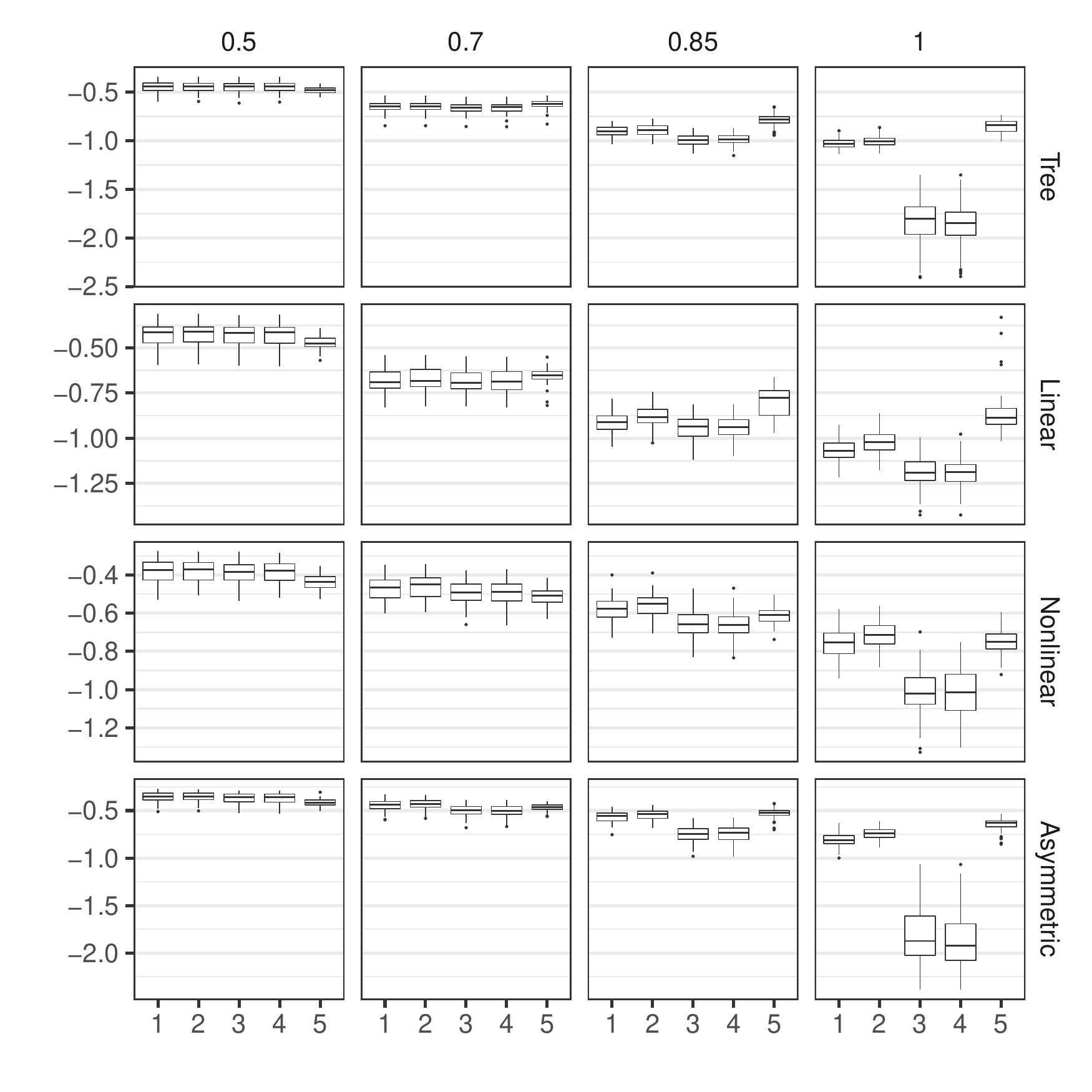}
        \caption[]{{\small Exponential, MSE$_{y\text{in}}$}}
    \end{subfigure}\hfill
    \begin{subfigure}[b]{0.5\textwidth}  
        \centering 
        \includegraphics[width=\textwidth,height=0.3\textheight,trim=10 8 5 5]{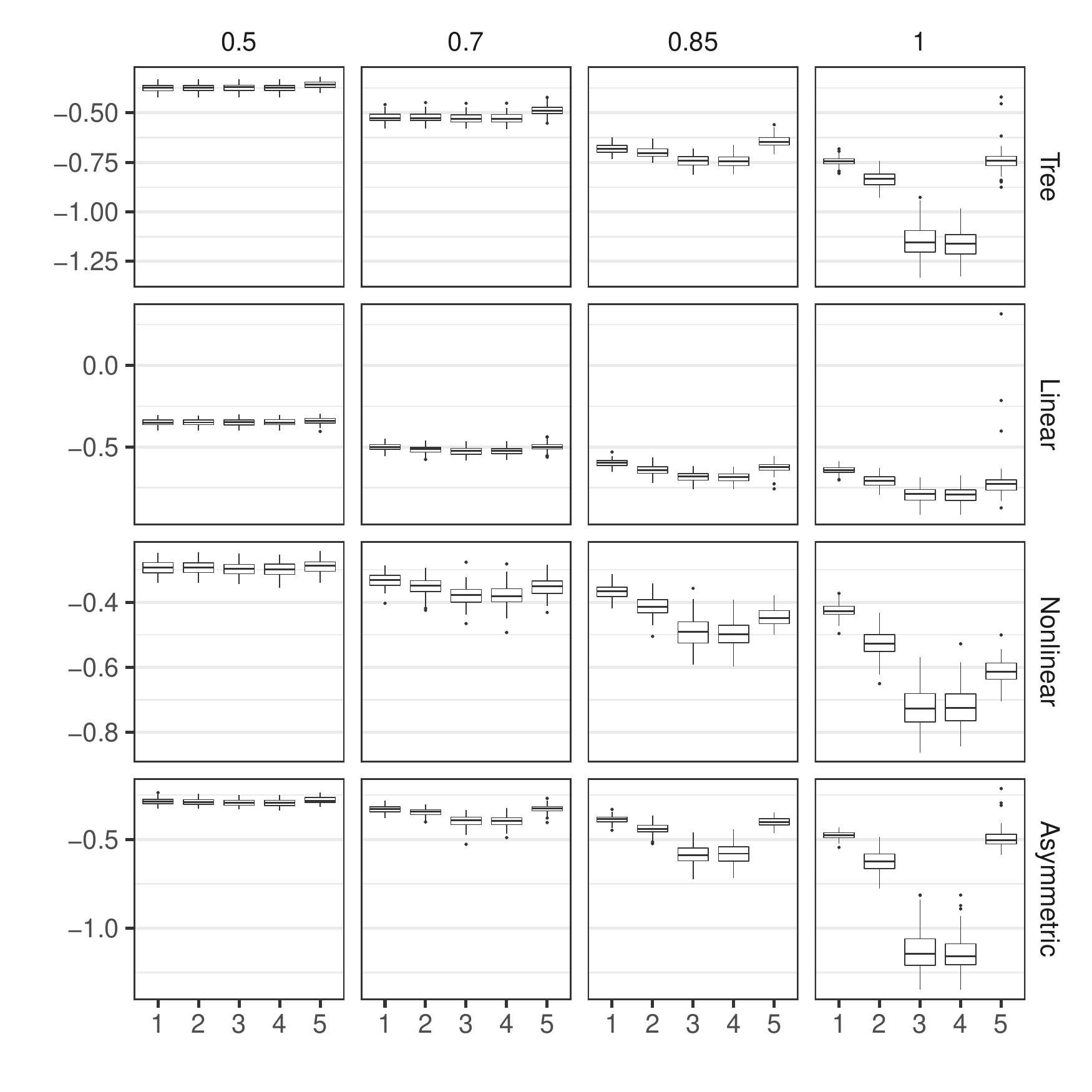}
        \caption[]{{\small Exponential, MSE$_{y\text{out}}$}}
    \end{subfigure}
}
\caption{Boxplots of mean squared error (MSE$_y$, both in-sample and out-of-sample) on $\log_{10}$ scale, 
time-varying covariates, N=500, light censoring, Weibull-I, Weibull-D, and Exponential distribution. }
\end{figure}

%--------------------
\newpage
%\subsection{Mean squared error of estimated Cox PH slope coefficients.}
\begin{figure}[h!]
    \vspace*{-0.5cm}
    \makebox[\linewidth][c]{%
    \begin{subfigure}[b]{0.5\textwidth}
        \centering
        \includegraphics[width=\textwidth,height=0.3\textheight,trim=10 8 5 5]{survvar_Nsub_500_attr_MSEb_Weibull-I_Light.pdf}
        \caption[]{{\small Weibull-I, MSE$_{b}$}}
    \end{subfigure}%\hfill
%    \begin{subfigure}[b]{0.5\textwidth}
%        \centering
%        \includegraphics[width=\textwidth,height=0.3\textheight,trim=10 8 5 5]{figures/survvar_Nsub_500_attr_cover_Weibull-I_Light.pdf}
%        \caption[]{{\small Weibull-I, Cover$_{b}$}}
%    \end{subfigure}
}
%    \vspace\baselineskip
    \makebox[\linewidth][c]{%
    \begin{subfigure}[b]{0.5\textwidth}  
        \centering 
        \includegraphics[width=\textwidth,height=0.3\textheight,trim=10 8 5 5]{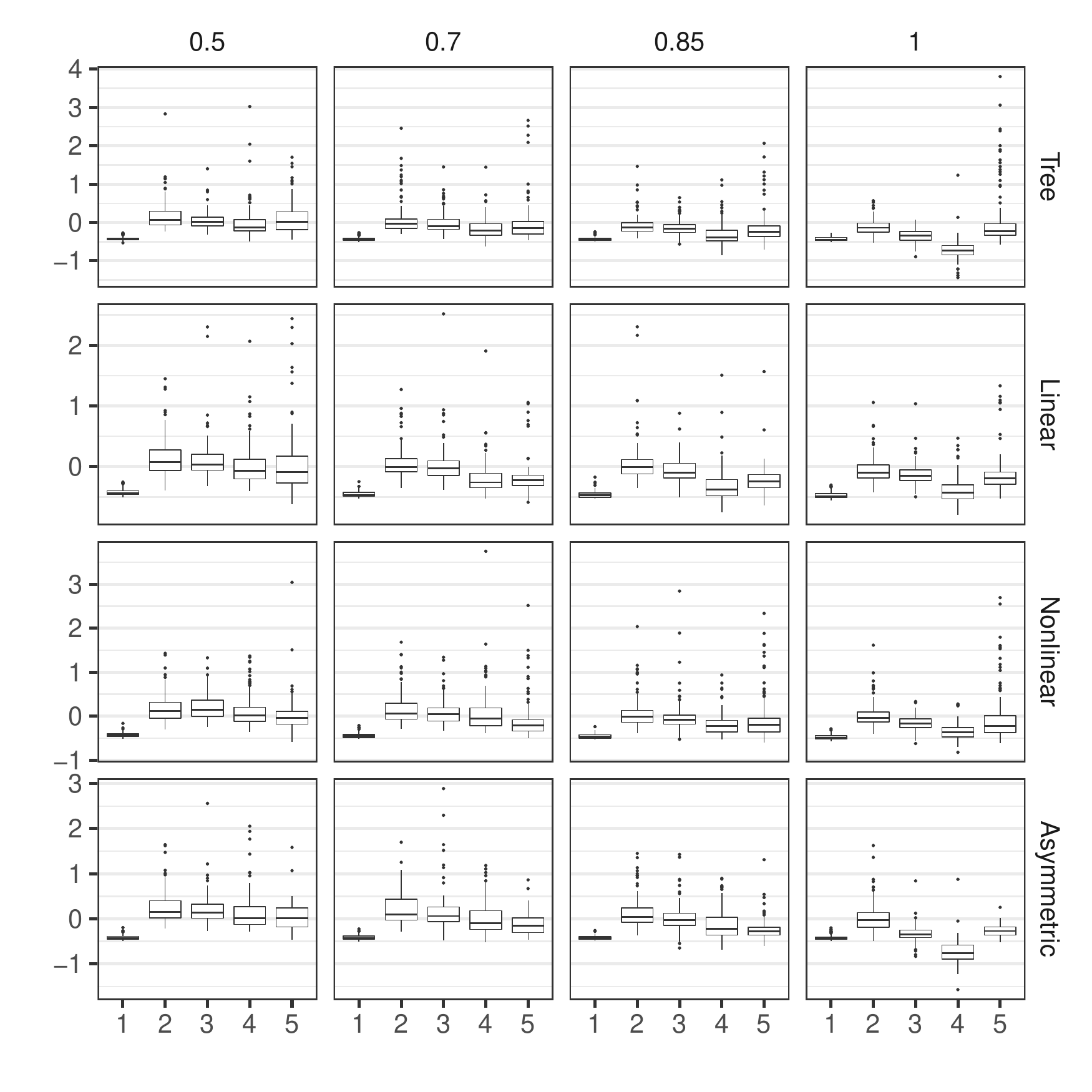}
        \caption[]{{\small Weibull-D, MSE$_{b}$}}
    \end{subfigure}%\hfill
%    \begin{subfigure}[b]{0.5\textwidth}  
%        \centering 
%        \includegraphics[width=\textwidth,height=0.3\textheight,trim=10 8 5 5]{figures/survvar_Nsub_500_attr_cover_Weibull-D_Light.pdf}
%        \caption[]{{\small Weibull-D, Cover$_{b}$}}
%    \end{subfigure}
}
\makebox[\linewidth][c]{%
    \begin{subfigure}[b]{0.5\textwidth}  
        \centering 
        \includegraphics[width=\textwidth,height=0.3\textheight,trim=10 8 5 5]{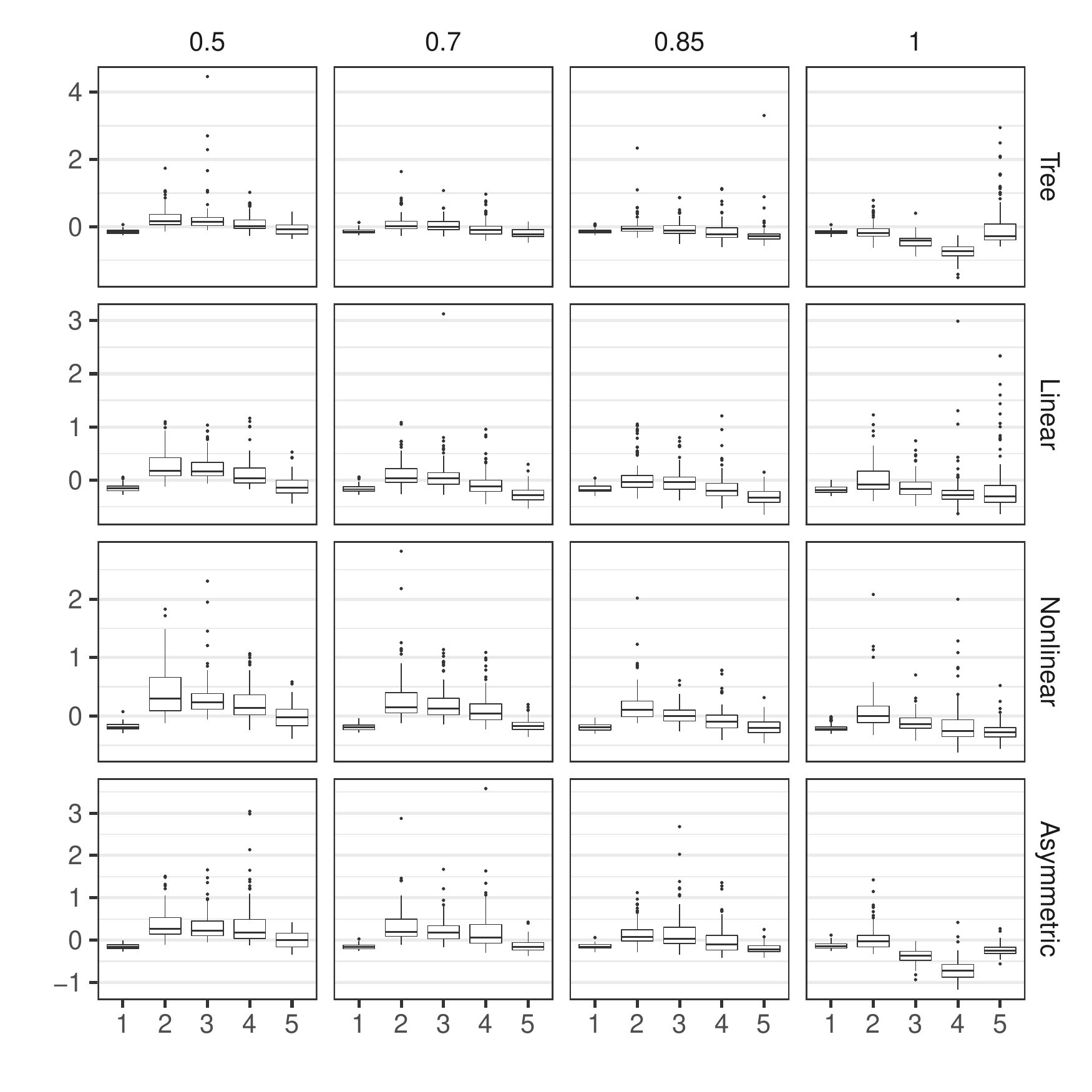}
        \caption[]{{\small Exponential, MSE$_{b}$}}
    \end{subfigure}%\hfill
%    \begin{subfigure}[b]{0.5\textwidth}  
%        \centering 
%        \includegraphics[width=\textwidth,height=0.3\textheight,trim=10 8 5 5]{figures/survvar_Nsub_500_attr_cover_Exponential_Light.pdf}
%        \caption[]{{\small Exponential, Cover$_{b}$}}
%    \end{subfigure}
}
\caption{Boxplots of mean squared error (MSE$_b$) of estimated Cox PH slopes on $\log_{10}$ scale, %and Cover$_b$,
time-varying covariates, N=500, light censoring, Weibull-I, Weibull-D, and Exponential distribution. }
\end{figure}

\section{Comparing various stopping criteria}\label{app:more_stopthre}
\red{In this section, we compare three stopping criteria,
$\TS_{\text{parent}} < \{2.71, 3.84, 6.63\}$,
for Weibull-I distribution and light censoring.
We report results for JLCT$_4$ model in Table~\ref{tb:simulate_models}, 
i.e.\ JLCT with time-varying latent class and survival covariates.
We use the performance measures described in Section~\ref{sec:evaluation}:
ISE$_{\text{in}}$, ISE$_{\text{out}}$, MSE$_{y\text{in}}$,
MSE$_{y\text{out}}$, MSE$_b$, and Acc$_g$.}
\begin{figure}[h!]
    \vspace*{-0.5cm}
    \makebox[\linewidth][c]{%
    \begin{subfigure}[b]{0.5\textwidth}
        \centering
        \includegraphics[width=\textwidth,height=0.3\textheight,trim=10 8 5 5]{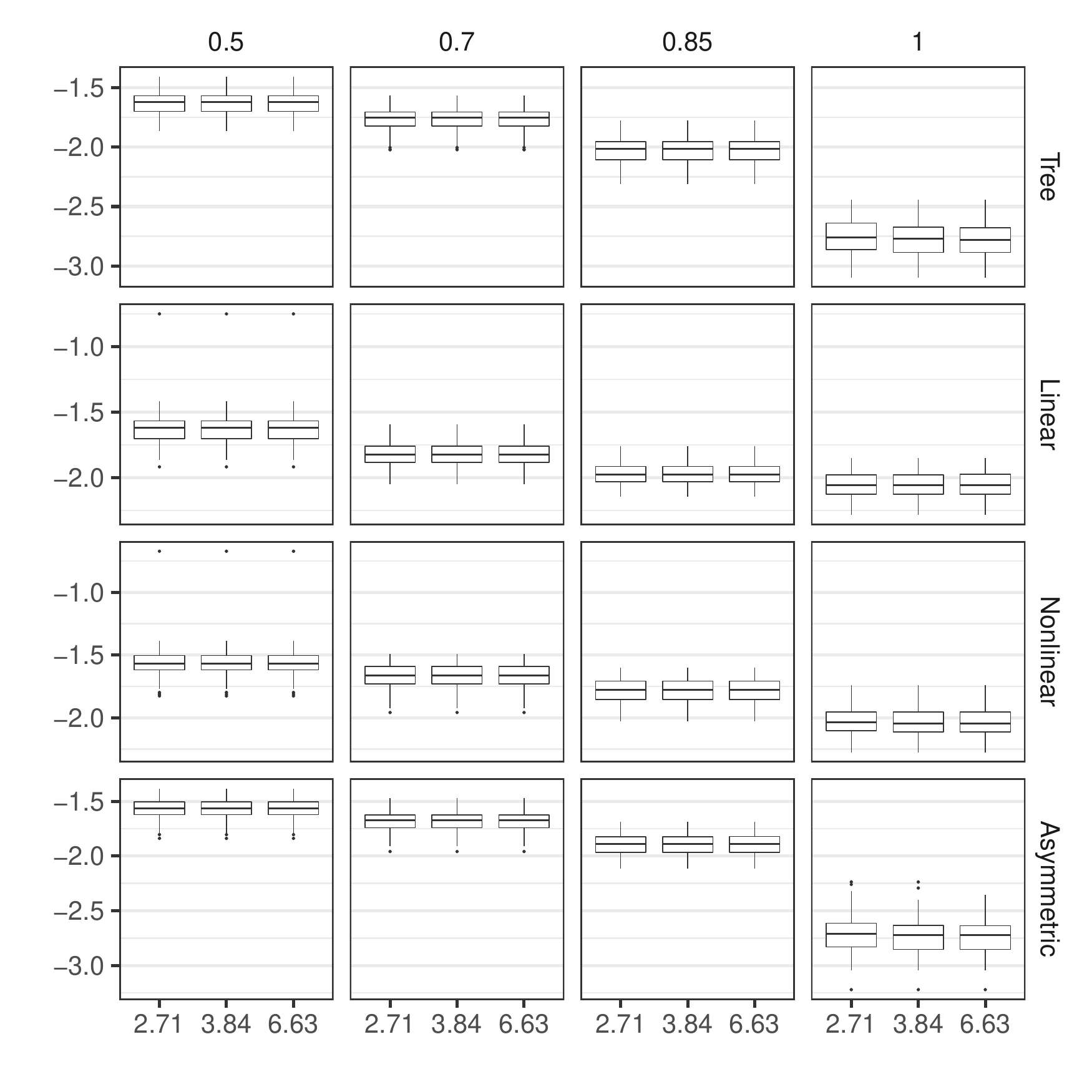}
        \caption[]{{\small $\log_{10}$ ISE$_{\text{in}}$}}
    \end{subfigure}\hfill
    \begin{subfigure}[b]{0.5\textwidth}  
        \centering 
        \includegraphics[width=\textwidth,height=0.3\textheight,trim=10 8 5 5]{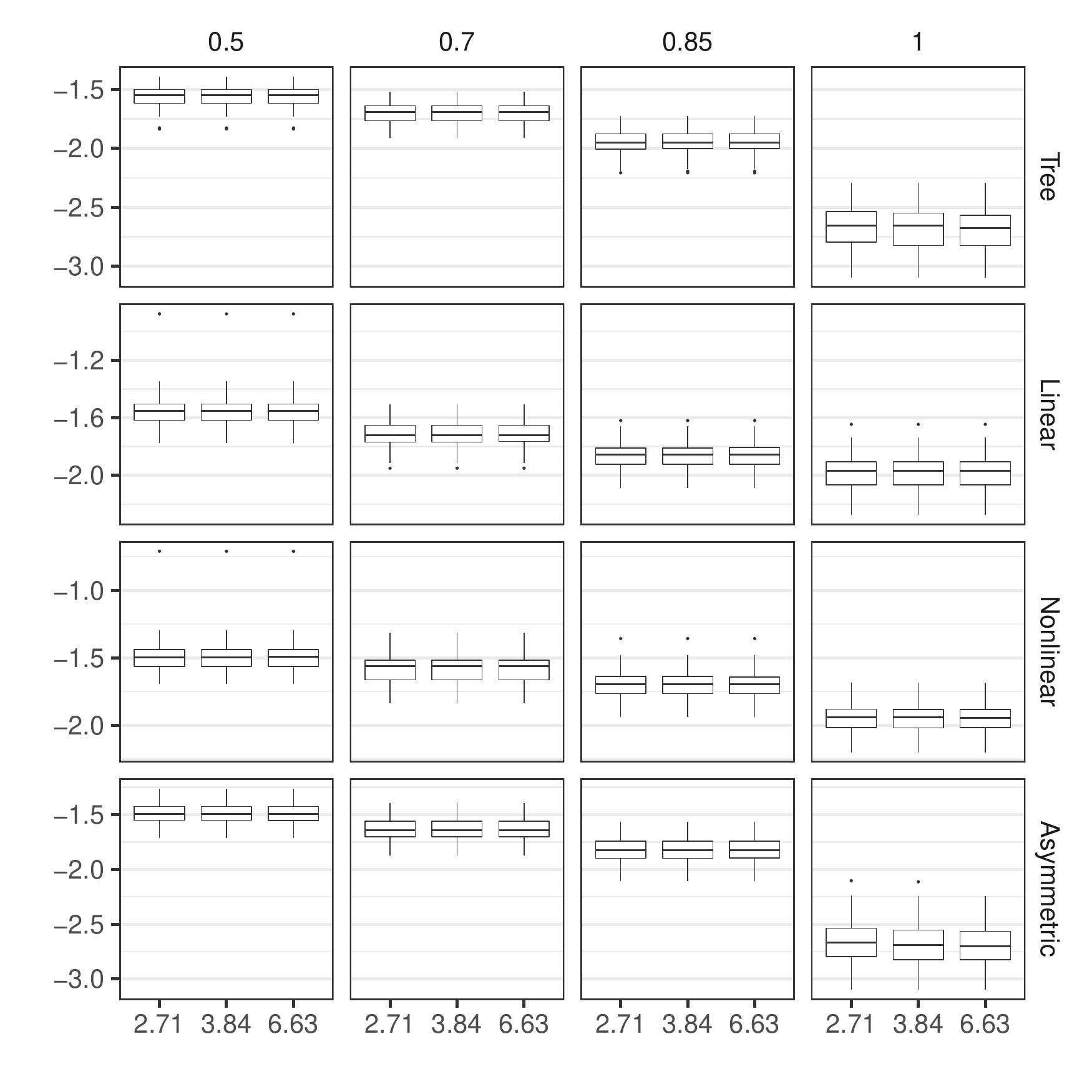}
        \caption[]{{\small $\log_{10}$ ISE$_{\text{out}}$}}
    \end{subfigure}
}
%    \vspace\baselineskip
    \makebox[\linewidth][c]{%
    \begin{subfigure}[b]{0.5\textwidth}  
        \centering 
        \includegraphics[width=\textwidth,height=0.3\textheight,trim=10 8 5 5]{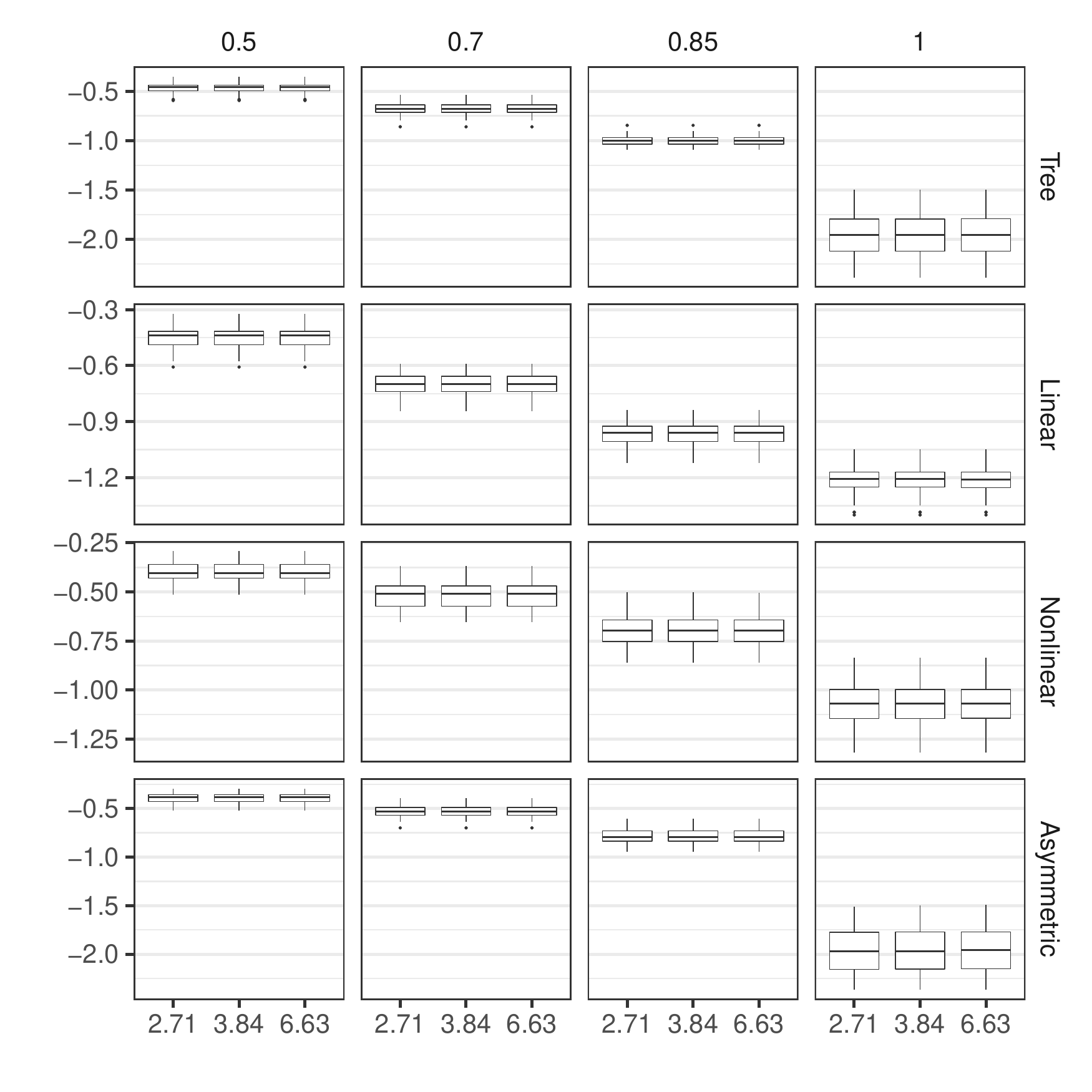}
        \caption[]{{\small $\log_{10}$ MSEy$_{\text{in}}$}}
    \end{subfigure}\hfill
    \begin{subfigure}[b]{0.5\textwidth}  
        \centering 
        \includegraphics[width=\textwidth,height=0.3\textheight,trim=10 8 5 5 ]{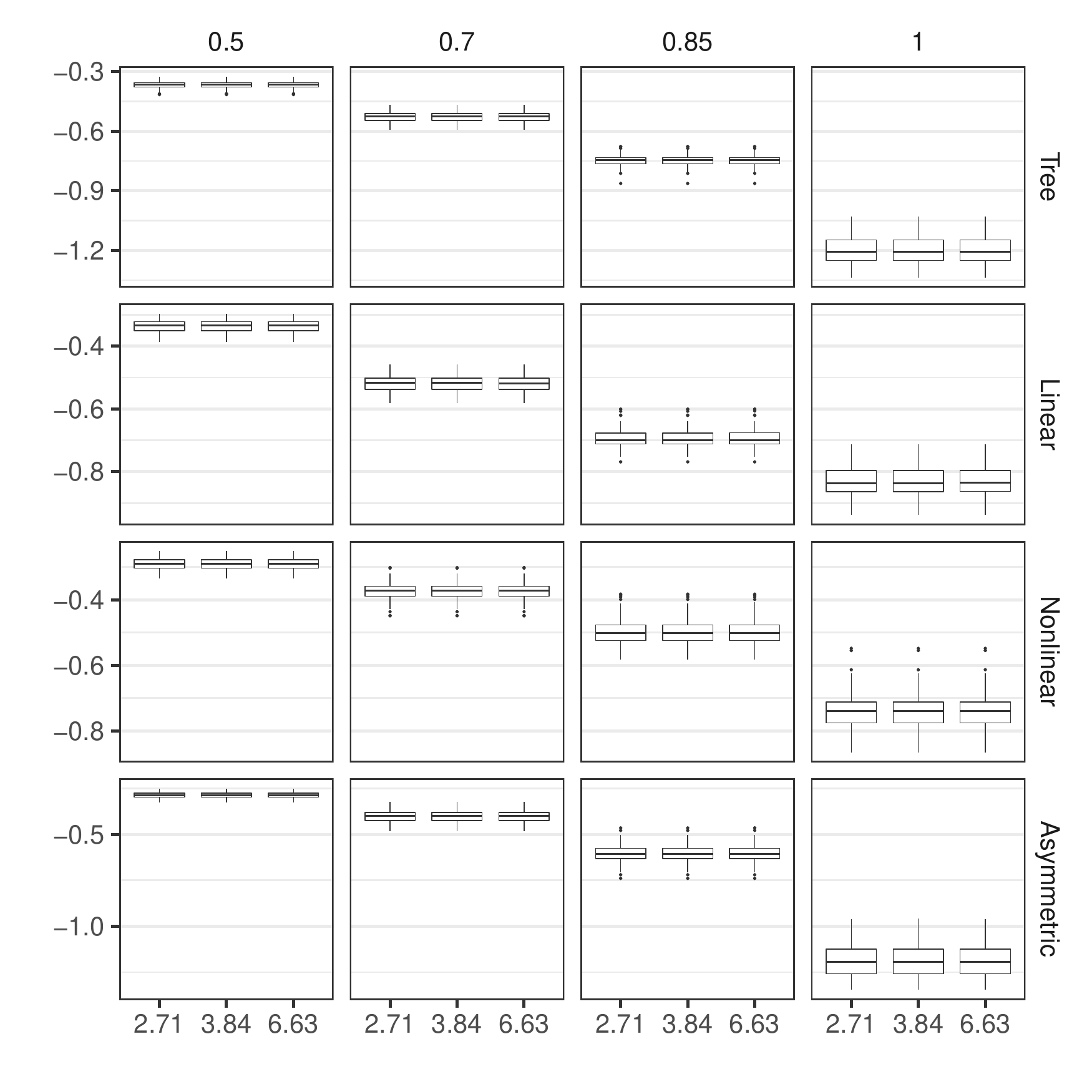}
        \caption[]{{\small $\log_{10}$ MSEy$_{\text{out}}$}}
    \end{subfigure}
}
\makebox[\linewidth][c]{%
    \begin{subfigure}[b]{0.5\textwidth}  
        \centering 
        \includegraphics[width=\textwidth,height=0.3\textheight,trim=10 8 5 5]{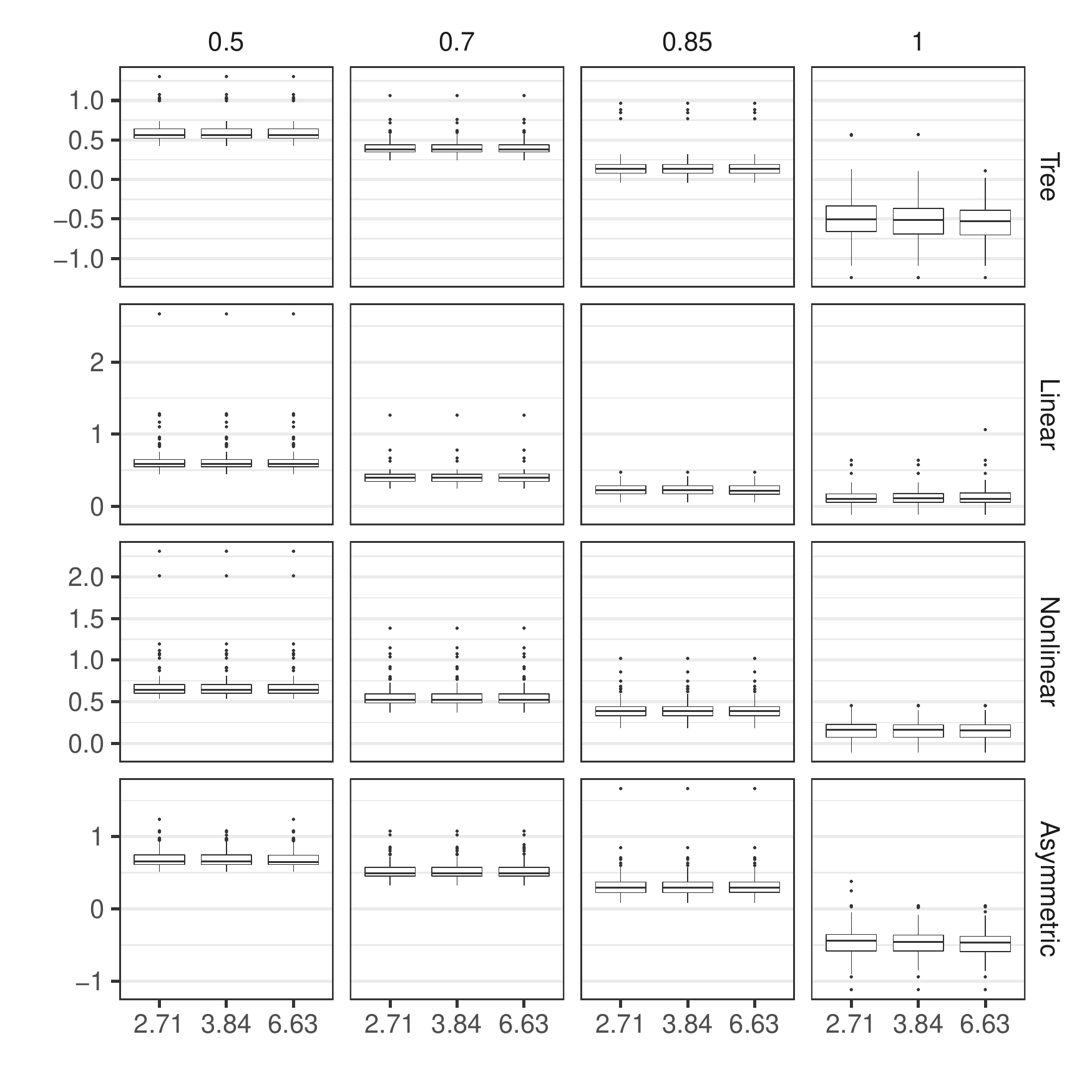}
        \caption[]{{\small $\log_{10}$ MSEb}}
    \end{subfigure}\hfill
    \begin{subfigure}[b]{0.5\textwidth}  
        \centering 
        \includegraphics[width=\textwidth,height=0.3\textheight,trim=10 8 5 5]{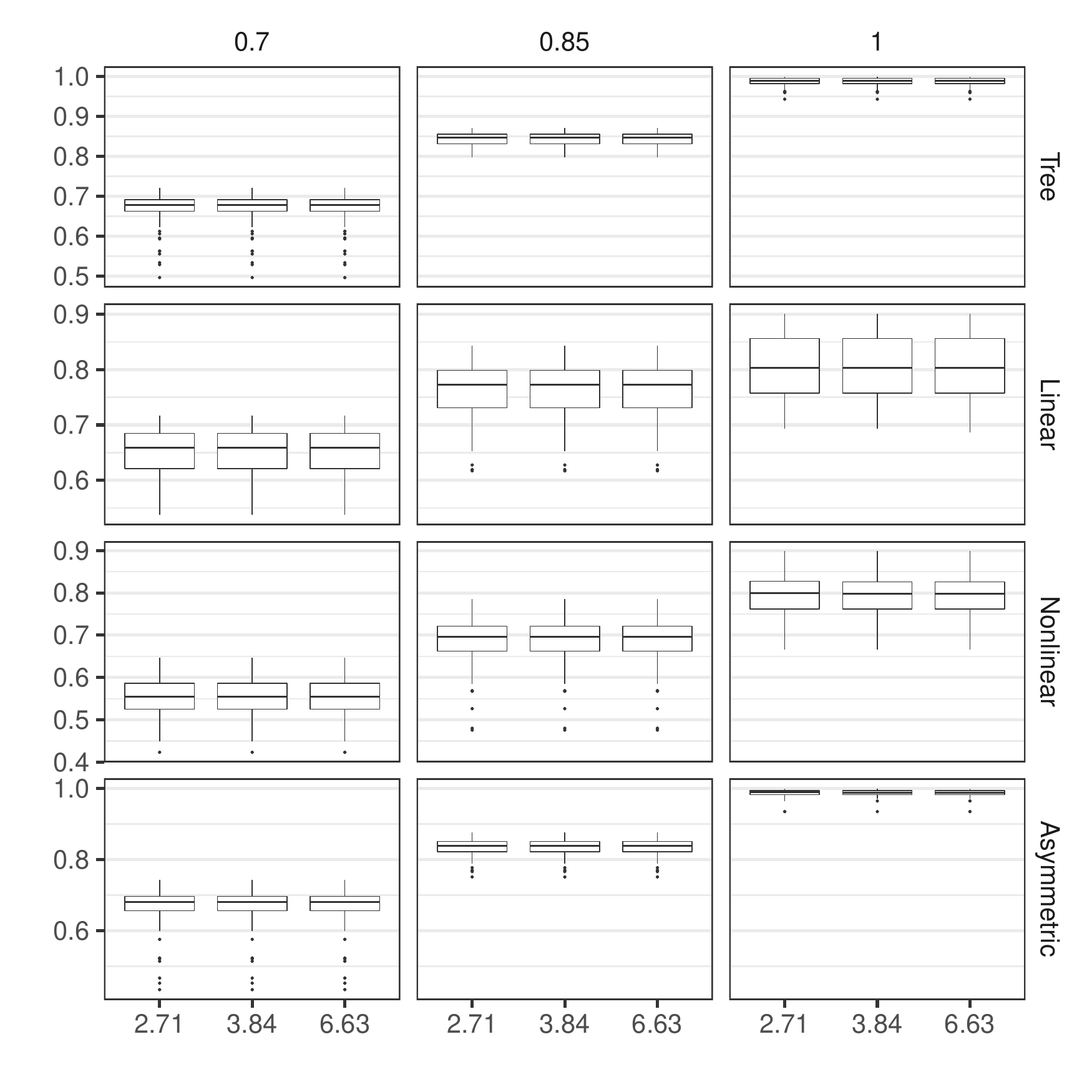}
        \caption[]{{\small Acc$_g$}}
    \end{subfigure}
}
\caption{Boxplots of model performance: ISE$_{\text{in}}$, ISE$_{\text{out}}$, MSE$_{y_\text{in}}$, MSE$_{y_\text{out}}$, and MSE$_b$ 
        on $\log_{10}$ scale, and Acc$_g$, for $N=500$, light censoring, and Weibull-I distributions, stopping 
        criteria $\TS_{\text{parent}} < \{2.71, 3.84, 6.63\}$ for JLCT$_4$ model.}
\end{figure}

\section{Using the \texttt{R} package \texttt{jlctree}}
\label{app:package} In this section, we demonstrate how to use the
\texttt{R} package \texttt{jlctree} to fit JLCT to a simulated
data set and to the PAQUID data set. For more details about the
\texttt{jlctree} package, please refer to the package manual
available at CRAN.

The \texttt{jlctree} package includes an example data set,
\texttt{data\_timevar}, which is generated under the time-varying
setup described in Section~\ref{app:sim_time_var}. The call of
\texttt{jlctree} to fit JLCT (JLCT$_4$ in
Table~\ref{tb:simulate_models}) to \texttt{data\_timevar} is

\noindent\rule{\linewidth}{0.2pt}
\begin{verbatim}
library(jlctree)
data(data_timevar)
tree <- jlctree(survival=Surv(time_L, time_Y, delta)~X3+X4+X5,
            classmb=~X1+X2, fixed=y~X1+X2+X3+X4+X5, random=~1,
            subject='ID',data=data_timevar,
            parms=list(maxng=4))
\end{verbatim}
\noindent\rule{\linewidth}{0.2pt}

Next, we reproduce the results of JLCT on the PAQUID data set. We
first convert the original PAQUID data set (contained in the
\texttt{R} package \texttt{lcmm}) into left-truncated right-censored
(LTRC) format.

\noindent\rule{\linewidth}{0.2pt}
\begin{verbatim}
library(lcmm)
library(NormPsy)
library(data.table)
paquid$normMMSE <- normMMSE(paquid$MMSE)
paquid$age65 <- (paquid$age - 65) / 10
paquidS <- paquid[paquid$agedem > paquid$age_init & paquid$age <= paquid$agedem, ]
paquidS2 <- data.table(paquidS)
paquidS2$age <- paquidS2[,{if(.N==1){age_init}
                          else {c(age_init[1], age[c(1:(.N-1))])}},by=ID][,V1]
temp <- subset(paquidS2, select=c(ID, age_init, agedem, dem, male))
temp <- unique(temp)
data <- tmerge(temp, temp, id=ID, tstart=age_init,
              tstop=agedem, death = event(agedem, dem))
data <- tmerge(data, paquidS2, id=ID, age65 = tdc(age, age65),
               CEP= tdc(age, CEP), normMMSE=tdc(age, normMMSE),
               BVRT=tdc(age, BVRT), IST=tdc(age, IST),
               HIER=tdc(age, HIER), CESD=tdc(age, CESD))
data <- subset(data, !is.na(normMMSE) & !is.na(BVRT)
               & !is.na(IST) & !is.na(HIER) & !is.na(CESD))
\end{verbatim}
\noindent\rule{\linewidth}{0.2pt}
The \texttt{R} code that fits JLCT to the PAQUID dataset using the time-varying covariates,
and plots the obtained tree structure as in Figure~\ref{fig:paquid_tree_var}, is

\noindent\rule{\linewidth}{0.2pt}
\begin{verbatim}
library(jlctree)
library(rpart.plot)
data$age  <- round(10*data$age65+65)
tree_var <- jlctree(survival=Surv(tstart, tstop, death)~CEP+male+
                BVRT+IST+HIER+CESD+poly(age_init, degree=2, raw=TRUE),
                classmb=~CEP+male+age+BVRT+IST+HIER+CESD,
                fixed=normMMSE~CEP+poly(age65, degree=2, raw=TRUE),
                random=~poly(age65, degree=2, raw=TRUE),
                subject='id',data=data,
                parms=list(min.nevents=5, fits=F, fity=F))
rpart.plot(tree_var$tree)
\end{verbatim}
\noindent\rule{\linewidth}{0.2pt}
One can fit JLCT to the PAQUID dataset using the time-invariant covariates
by changing the corresponding arguments in the above \texttt{jlctree} call.

\end{document}